\documentclass[%
 reprint,
superscriptaddress,
floatfix,
nofootinbib,
 amsmath,amssymb,
 aps,
prd,
]{revtex4-2}

\usepackage[dvipsnames]{xcolor}
\usepackage{booktabs}
\usepackage{array}
\usepackage{float}
\usepackage{tabularx}
\usepackage{adjustbox}
\usepackage{amssymb,amsmath,amsfonts}
\usepackage{dsfont} 
\usepackage{graphicx,bm}
\usepackage[hidelinks,colorlinks=true,linkcolor=blue,citecolor=blue]{hyperref}
\newcommand\apjs{ApJS}               % Astrophysical Journal, Supplement
\newcommand\araa{ARA\&A}             % Annual Review of Astronomy and Astrophysics
\newcommand\mnras{MNRAS}             % Monthly Notices of the Royal Astronomical Society
\newcommand\physrep{Phys.~Rep.}      % Physics Reports
\usepackage{tabu}
\usepackage{array}
\usepackage{multirow}
\usepackage{CJK}
\usepackage{mathrsfs}
\usepackage{dcolumn}% Align table columns on decimal point
\usepackage{mathtools}
\usepackage{graphicx}% Include figure files
\usepackage{dcolumn}% Align table columns on decimal point
\usepackage{bm}% bold math
\usepackage[normalem]{ulem}
\begin{document}

%\preprint{APS/123-QED}
%A primer on including dephasing due to environmental effects in Fourier domain gravitational waveforms 
%\title{Dynamical phase transitions in accreting neutron stars}
\title{General-purpose equations of state for hadron--quark matter: first application to accretion-induced phase transitions in neutron stars}

\author{János Takátsy}
\email{janos.takatsy@uni-potsdam.de}
\affiliation{Institut für Physik und Astronomie, Universität Potsdam, Haus 28, Karl-Liebknecht-Str. 24-25, Potsdam, Germany}

\author{Henrique Gieg}
\affiliation{Institut für Physik und Astronomie, Universität Potsdam, Haus 28, Karl-Liebknecht-Str. 24-25, Potsdam, Germany}

\author{Tim Dietrich}
\affiliation{Institut für Physik und Astronomie, Universität Potsdam, Haus 28, Karl-Liebknecht-Str. 24-25, Potsdam, Germany}
\affiliation{Max Planck Institute for Gravitational Physics (Albert Einstein Institute), Am Mühlenberg 1, Potsdam 14476, Germany}


%



\newcommand{\dandelion}{\color{Dandelion}}
\newcommand{\JT}[1]{{\dandelion [JT: #1]}}         

\newcommand{\red}{\color{red}}
\newcommand{\highlight}[1]{{\red #1}}            

\newcommand{\td}[1]{{\color{purple}{[TD: #1]}}}     

\begin{abstract}
Strong first-order phase transitions in the cores of neutron stars (NSs) can have profound implications for astrophysical observables, however, their existence remains an open question. Since most NSs form in binary systems, they are expected to experience prolonged episodes of mass accretion that gradually increase their central density and can potentially trigger such a transition. This might, in turn, generate observable multi-messenger signals, as suggested by previous studies. Motivated by this prospect, we generate a new set of equations of state for hadron--quark matter with first-order phase transitions of various strengths, including neutrino absorption opacities in the quark phase as well. We then perform general-relativistic radiation hydrodynamics simulations of NSs undergoing accretion-induced phase transitions. In contrast to previous findings, our results indicate that these systems do not produce unbound ejecta for the models considered, making accretion-induced phase transitions unlikely sources of observable electromagnetic transients. However, we find that transitions are accompanied by intense neutrino emission, with peak luminosities reaching $\sim10^{52}~\mathrm{erg/s}$ for the strongest transition considered. To facilitate future numerical studies, we make the family of equations of state and neutrino opacities used in this work publicly available, enabling the systematic exploration of the role of the phase transition strength.
\end{abstract}

\maketitle

\section{Introduction}
\label{sec:introduction}

Neutron stars (NSs) provide essential information for testing the properties of strong interaction in the dense, low-temperature region of the phase diagram of quantum chromodynamics (QCD), as nuclear \cite[e.g.,][]{Akmal1998,Wiringa1988,Horowitz:1999fk,Garg:2018uam,Thiel:2019tkm,PREX:2021umo} and heavy-ion experiments \cite{ALICE,PHENIX,STAR,STAR,NA61,CBM,NICA} are unable to probe matter under these conditions. Moreover, despite recent developments in the fields of the complex Langevin method or alternative expansion schemes \cite{Guenther2020,Attanasio2020,Borsanyi2021}, the sign problem still poses a huge challenge for first-principles calculations of QCD \cite{deForcrand2010}, also preventing their application in this regime.

During the past decade, NS observations revealed several breakthrough measurements, strongly constraining the range of possible NS equations of state (EoSs). The discovery of pulsars with masses around $2~M_\odot$ \cite{Demorest:2010bx,Antoniadis:2013pzd,NANOGrav:2017wvv,Cromartie:2019kug,Fonseca:2021wxt,Romani:2022jhd,Saffer:2024tlb,Romani:2025ytn} established a robust lower bound on the maximum mass, ruling out EoSs too soft to support such massive NSs. Additionally, the detection of the binary NS merger GW170817 by the by the LIGO Scientific and Virgo Collaborations \cite{LIGOScientific:2017vwq,LIGOScientific:2018cki,LIGOScientific:2018hze} put constraints on the tidal deformability, while multi-messenger observations of the same event \cite{LIGOScientific:2017ync,LIGOScientific:2017zic} provided further bounds on NS properties. Meanwhile, pulse-profile modeling of X-ray observations with the NICER mission has yielded simultaneous mass and radius measurements for several pulsars, including PSR J0030+0451 \cite{Miller:2019cac,Riley:2019yda,Raaijmakers:2019qny,Vinciguerra2024} and PSR J0740+6620 \cite{Miller:2021qha,Riley:2021pdl,Raaijmakers:2021uju,Salmi2024b,Dittmann2024}, as well as multiple other pulsars \cite{Salmi2024,Choudhury2024,Qi2025,Mauviard2025,Miller2025}. Together, these multi-messenger observations have narrowed down the range of viable EoSs substantially \cite{De:2018uhw, Tews:2018iwm,Rezzolla2017,Annala:2017llu,Dietrich:2020efo,Pang:2021jta,Malik:2022zol,Altiparmak:2022bke,Marczenko:2022jhl,Takatsy:2023xzf,Koehn:2024set}.

Nevertheless, despite continued improvements in measurements of the bulk properties of NSs, these observations fail to resolve the fundamental question of their internal composition. This is both due to the fact that different EoSs can lead to the same bulk NS properties, and that similar EoSs can be obtained from models with different underlying compositions. Several alternative measurements have been suggested recently that tackle this problem by going beyond static properties of NSs in some way. For example, peaks in the gravitational-wave (GW) spectrum of binary NS post-merger signals could encode important information about composition \cite{Bauswein:2013jpa,Bauswein:2015vxa,Sarin:2020gxb,Breschi:2021xrx,Puecher:2022oiz,LIGOScientific:2017fdd,LIGOScientific:2018urg}. However, due to their high frequency, these post-merger signals are only expected to become observable with third-generation GW detectors \cite[e.g.,]{Clark:2015zxa,LIGOScientific:2017fdd,Torres-Rivas:2018svp}. On the other hand, while the measurement of f-mode frequencies through dynamical tides can become feasible using the current infrastructure of GW detectors, they correlate strongly with bulk NS properties \cite{Chan:2014kua,Godzieba:2020bbz,Pradhan:2022vdf,Pradhan:2023zmg}. Another promising probe is the observation of g-mode resonances during the binary inspiral, whose properties are highly sensitive to the composition of dense matter \cite[e.g.,][]{Kantor:2014lja,Passamonti:2015oia,Constantinou:2021hba,Jaikumar:2021jbw,Zhao:2022toc,Counsell:2024pua}. The difficulty for these measurements is posed by the weak coupling of these modes to the tidal field. Yet another class of tidal oscillations is predicted for NSs with a strong phase transition. Recent studies suggest that such interface modes can produce a significant phase shift in the inspiral signal \cite{Miniutti:2002bh,Counsell:2025hcv,Zhu:2022pja,Lau:2020bfq,Rodriguez:2025oes, Pereira:2025xsi}.

In fact, the existence of a QCD phase transition at NS densities has long been a subject of considerable interest. While EoSs featuring a strong phase transition are already somewhat constrained by current mass-radius and tidal deformability measurements \cite{Christian:2019qer,Christian:2020xwz,Christian:2021uhd,Christian:2023hez,Blomqvist:2025cxe}, whether such a phase transition occurs in the cores of NSs remains an open question. Hydrodynamical simulations featuring such phase transitions have been carried out extensively in the context of core-collapse supernovae \cite[e.g.,][]{Sagert:2008ka,Fischer:2010wp,Fischer:2017lag,Zha:2020gjw,Kuroda:2021eiv} and binary mergers \cite[e.g.,][]{Bauswein:2018bma,Most:2019onn,Gieg:2019yzq,Blacker:2020nlq}. These studies investigated whether the effect of the transition can become observable through their GW signals, while for supernovae multi-messenger signals including neutrino emission were also explored. Both of these scenarios correspond to strong density changes on a very short timescale, which then trigger the transition. However, such extreme conditions are not necessarily required for the transition to occur.

It is now well established that the majority of stars reside in binary or multiple systems \cite{Sana:2012,Duchene:2013,Moe:2017}. The main contribution to binary NS mergers is also expected to be given by massive stellar binary evolution \cite{Postnov:2014tza,Tauris:2017omb,Vigna-Gomez:2018dza}. These binaries are expected to go through multiple episodes of mass exchange, including the accretion onto the newly formed compact object from its stellar companion. In fact, the observed population of millisecond or recycled pulsars with extremely short rotational periods is a direct consequence of accretion-driven spin-up phases \cite[e.g.,][]{Bhattacharya:1991}. During these phases NSs can accrete $10^{-3}-10^{-1}~M_\odot$ from their companion over secular timescales \cite[e.g.,][]{Tauris:2012}. However, if a first-order phase transition is triggered during this phase, it can have profound implications for the evolution of the accreting NS, possibly producing an observable multi-messenger transient. Despite the potential importance of these events, only a handful of papers have studied the dynamical evolution of single NSs in the transition regime \cite{Largani:2023kjx,Haque:2026ero,Naseri:2026xrq,Ecker:2026ixk}.

One of these studies investigated the dynamical response of twin stars to perturbations employing a zero-temperature barotropic EoS \cite{Haque:2026ero}, finding that given a sufficiently large perturbation, NSs can migrate between the two stable branches in both directions. Using a finite-temperature EoS, Ref.~\cite{Naseri:2026xrq} has shown that transitions from the hadronic branch to a stable configuration on the hybrid branch are not possible through perturbations, unless unphysically large cooling rates are introduced, due to the thermal pressure arising in the center. One might still make hadronic stars migrate to the hybrid branch by adding sufficient amount of extra material to the system so that no such massive stable hadronic solutions exist. Largani et al.\ \cite{Largani:2023kjx} have already investigated this scenario through one-dimensional simulations including a Boltzmann solver to handle neutrino transport. They found that even for an EoS with a weak phase transition, the onset of the transition can trigger the ejection of $\sim10^{-3}~M_\odot$ of matter from the NS, resulting in an electromagnetic transient. Although~\cite{Largani:2023kjx} did not perform such simulations, this suggests that stronger phase transitions, including those producing twin-star solutions \cite{Gerlach:1968,Glendenning:1998ag}, can potentially produce even more ejecta, enabling the strength of the transition to be probed through observations.

In this paper we study the evolution of accreting NSs with hybrid EoSs right at the onset of the phase transition. We do this by performing general relativistic radiation hydrodynamics simulations using the \texttt{BAM} code. We extend the study of Ref.~\cite{Largani:2023kjx} by employing EoSs with different hadron--quark transition strengths, and implementing neutrino opacities in the quark phase as well. To enable this study, we present a new set of general-purpose EoSs together with neutrino opacities for hadron--quark matter. This set of EoSs was purposefully created in a way that future dynamical studies can investigate the effect of various phase transition strengths in a controlled manner, while keeping the hadronic EoS and the onset of phase transition unchanged.

This paper is structured as follows. In Sec.~\ref{sec:methods}, we describe our hybrid EoSs, including the hadronic and quark models, as well as the construction of the phase transition itself. We also provide a description of our numerical solver, the construction of the initial conditions, and the treatment of neutrino opacities. In Sec.~\ref{sec:results}, we present some of the main properties of our EoSs, as well as results from our simulations. Finally, in Sec.~\ref{sec:conclusion}, we conclude and discuss possible implications. Additionally, in the Appendix we discuss alternative initial conditions, and numerical obstacles arising in simulations of NSs with phase transitions.

\vspace{-0.4cm}
\section{Methods}
\label{sec:methods}

\subsection{Hybrid equation of state}
\label{ssec:eos}

We construct our hybrid EoSs based on the method described in Ref.~\cite{Bastian:2020unt}, by using separate hadronic and quark models, and applying the Gibbs condition to realize the transition between the two phases. In both phases only electrons are included as leptonic degrees of freedom. For the hadronic EoS we choose the relativistic mean field model of Typel et al.\ (DD2) with density-dependent nuclear interactions \cite{Typel2010,Hempel2009}. In addition to being well established within the astrophysics community, this model manages to accurately reproduce nuclear properties, while also being stiff enough to support hybrid star branches with a strong first-order phase transition.

For the quark phase we employ the model described in Ref.~\cite{Bastian:2020unt}, which uses a relativistic mean field approach with two quark flavors and a density-dependent potential between the quarks. The potential is defined through the (iso)scalar and (iso)vector densities and is given by:
\begin{align}
U(n_\mathrm{s},n_\mathrm{v},n_\mathrm{iv}) =& \,\, C n_\mathrm{s}^{4/3} + D(n_\mathrm{v}) n_\mathrm{s}^{2/3} + \omega_4 n_\mathrm{v}^2 \nonumber \\
    &+ \frac{\omega_8}{1+\omega_8'n_\mathrm{v}^2} n_\mathrm{v}^4 + \rho_4 n_\mathrm{iv}^2 \: ,
\end{align}
where $n_\mathrm{s}$, $n_\mathrm{v}$, and $n_\mathrm{iv}$ are the scalar, vector and isovector densities, respectively. The first two terms represent the so-called string-flip model \cite{Kaltenborn:2017hus}, generating the following effective quark mass contribution:
\begin{equation}
    \delta m_Q = \frac{4}{3}C n_\mathrm{s}^{1/3} + \frac{2}{3} D(n_\mathrm{v}) n_\mathrm{s}^{-1/3} \: ,
\end{equation}
which diverges for low values of the scalar density, suppressing quarks in that regime. For our calculations we set $C=0$, while the form of $D(n_\mathrm{v})$ is chosen as:
\begin{align}
    D(n_\mathrm{v}) =
    \begin{cases}
        D_0 \exp[-\alpha(n_\mathrm{v}-n_\mathrm{ref})^2] \: & ,\: n_\mathrm{v} \geq n_\mathrm{ref} \\
        D_0 \: & \:, n_\mathrm{v} < n_\mathrm{ref}
    \end{cases} \: .
\end{align}
Further description of this specific choice and the remaining terms can be found in Ref.~\cite{Bastian:2020unt}. We set $n_\mathrm{ref}$ to zero in all of our models, while values of the EoS parameters are listed in Tab.~\ref{tab:params}.

The third and fourth terms in the potential describe repulsion from four-fermion and higher-order interactions, respectively. The higher-order term provides the necessary stiffness for the EoS to be able to produce massive NSs in the hybrid branch, while the denominator ensures that the EoS stays causal at asymptotically large densities. Finally, the last term represents an isovector interaction, which drives the flavor-asymmetry of quarks.

This potential induces the self-energies, $\Sigma_\mathrm{s}$, $\Sigma_\mathrm{is}$, $\Sigma_\mathrm{v}$, $\Sigma_\mathrm{iv}$ through:
\begin{equation}
    \Sigma_\alpha = \frac{\partial U}{\partial n_\alpha} \: ,
\end{equation}
with $\alpha\in\{\mathrm{s},\mathrm{is},\mathrm{v},\mathrm{iv} \}$. These self-energies enter the effective masses and chemical potentials of the quarks:
\begin{align}
    M_j &= m_j + \Sigma_\mathrm{s} + \tau_j \Sigma_\mathrm{is} \: , \nonumber \\
    \tilde{\mu}_j &= \mu_j - \Sigma_\mathrm{v} - \tau_j \Sigma_\mathrm{iv} \: ,
\end{align}
where $j \in \{ u, d \}$ stands for the quark flavor, and $\tau_j = \pm 1$ represents the isospin of the quarks. The above density-dependent self-energies then enter the definition of the densities themselves. For example, in case of $n_s$, we have:
\begin{equation}
    n_\mathrm{s} = \sum_j n_{\mathrm{s},j} \: ,
\end{equation}
with
\begin{equation}
    n_{\mathrm{s},j} = g_j\int \frac{\mathrm{d}^3p}{(2\pi)^3} \frac{M_j}{\sqrt{p^2 + M_j^2}} \left( f_j + \bar{f}_j \right) \: ,
\end{equation}
where $f_j\equiv f_j(E_j,\tilde{\mu}_j,T)$ and $\bar{f}_j\equiv \bar{f}_j(E_j,\tilde{\mu}_j,T)$ denotes the Fermi-Dirac distribution of particles and anti-particles of species $j$, and $g_j$ is the degeneracy factor. Hence, the definition of the quark potential through the density variables defines a self-consistency problem, whose numerical solution provides the values of the densities at a given chemical potential and temperature, as well as the thermodynamic quantities of the EoS, as described in Ref.~\cite{Bastian:2020unt}.

In addition to the above equations, we impose local charge neutrality between the quarks $(u,d)$ and electrons $(e)$:
\begin{equation}
    \frac{2}{3} n_u - \frac{1}{3} n_d - n_e = 0 \: .
    \label{eq:ch_neutr}
\end{equation}
Since we also want to explore the properties of the model out of beta-equilibrium, we do not impose the usual condition on the chemical potentials. Instead, introducing the chemical potential  $\mu_l$ associated with lepton-number conservation, we write the chemical potentials of the different species as:
\begin{align}
    \mu_u &= \frac{1}{3} \mu_B + \frac{2}{3} \mu_C \: , \nonumber \\
    \mu_d &= \frac{1}{3} \mu_B - \frac{1}{3} \mu_C \: , \nonumber \\
    \mu_e &= \mu_l - \mu_C \: , \nonumber \\
    \mu_{\nu_e} &= \mu_l \: ,
\end{align}
with $\mu_B$ and $\mu_C$ representing the baryon and charge chemical potentials, respectively. This way we can parameterize the EoS through $T$, $\mu_B$ and $\mu_C$, obtaining the EoS at different temperatures, baryon densities and electron fractions. The electron fraction is defined as $Y_e = n_e/n_\mathrm{B}$, with $n_\mathrm{B} = 1/3(n_u + n_d)$ the baryon number density. Note that the lepton chemical potential is not a free parameter and is set by the charge neutrality condition, Eq.~\eqref{eq:ch_neutr}.

In order to find the phase transition surface within this three-dimensional space, we need to apply the Gibbs condition, which involves thermal equilibrium $T^H = T^Q$, mechanical equilibrium $p^H = p^Q$, as well as chemical equilibrium $\mu^H_\mathrm{B,C} = \mu^Q_\mathrm{B,C}$ for the baryon and charge chemical potentials simultaneously, among the baryonic species. In a practical application this means that after subtracting the contribution of leptons to the pressure, we need to find the intersection line between the surfaces described by $p^H(\mu_\mathrm{B},\mu_\mathrm{C})$ and $p^Q(\mu_\mathrm{B},\mu_\mathrm{C})$, at a fixed temperature. Outside of this line, matter is in the phase that maximizes the pressure (see e.g., Fig.~9.5 in Ref~\cite{Schaffner-Bielich:2020psc}).

\begin{table*}[t]
\centering
\caption{Parameters of the quark model, as well as the relative density jump during the phase transition, and the maximum NS mass supported by the model. We use DD2 as the hadronic part, and the values $C=0$ and $m_{u/d}=5$~MeV for all EoSs. The names follow the nomenclature introduced for the EoSs in Ref.~\cite{Bastian:2020unt}.}
\label{tab:params}

\setlength{\tabcolsep}{4pt}
\renewcommand{\arraystretch}{1.1}

\begin{tabular}{ccccccccc}
\hline\hline
\rule{0pt}{3ex}Name & $D_0^{1/2}$ [MeV] & $\alpha$ [fm$^6$] & $\omega_4$ [MeV\,fm$^3$] & $\omega_8$ [MeV\,fm$^9$] & $\omega_8'$ [fm$^6$] & $\rho_4$ [MeV\,fm$^3$] & $\Delta n/n_\mathrm{tr}$ & $M_\mathrm{max}$ $[M_\odot]$ \\[2pt]
\hline
\rule{0pt}{3ex}BBKF 2.1 & 265.0 & 0.37 & -1.0 & 0.8 & 0.05 & 40.0 & 0.79 & 1.98 \\
BBKF 2.2 & 265.0 & 0.45 & 4.7 & 0.28 & 0.05 & 44.0 & 0.68 & 1.99 \\
BBKF 2.3 & 251.0 & 0.495 & 9.3 & 0 & 0 & 47.5 & 0.54 & 2.05 \\
BBKF 2.4 & 235.0 & 0.5 & 12.05 & 0 & 0 & 50.0 & 0.44 & 2.13 \\
BBKF 2.5 & 205.0 & 0.4 & 14.0 & 0 & 0 & 50.0 & 0.29 & 2.17 \\[1pt]
\hline\hline
\end{tabular}

\end{table*}

An additional difficulty is posed by the fact that we want the EoS not as a function of chemical potentials $\mu_\mathrm{B}$ and $\mu_\mathrm{C}$, but as a function of baryon number density and electron fraction. A given intersection point, however, corresponds to different $n_\mathrm{B}$ and $Y_e$ in the two phases (see Fig. 4 in Ref~\cite{Bastian:2020unt}). Therefore, in the phase transition region, different baryon number densities at a fixed electron fraction will correspond to different intersection points in the $\mu_\mathrm{B}-\mu_\mathrm{C}$ plane. We thus need to find the line connecting the $(n_\mathrm{B},Y_e)$ values in the two phases corresponding to the same transition point. For this, we can use the fact that along this line the intensive quantities $T, \mu_\mathrm{B}, \mu_\mathrm{C}, p$ are constant, while the extensive quantities, such as the electron number density or baryon number density mix according to:
\begin{align}
    n_e = (1-\chi)n_e^H(\mu_\mathrm{B}, \mu_\mathrm{C}) + \chi  n_e^Q(\mu_\mathrm{B}, \mu_\mathrm{C}) \: , \nonumber \\
    n_\mathrm{B} = (1-\chi)n_\mathrm{B}^H(\mu_\mathrm{B}, \mu_\mathrm{C}) + \chi n_\mathrm{B}^Q(\mu_\mathrm{B}, \mu_\mathrm{C}) \: ,
\end{align}
parameterized by the mixing factor $\chi$. This defines the set of lines within the phase transition region, which connect the intersection points in the purely hadronic and purely quark phases. Note that other extensive quantities, such as the energy density or entropy density can also be obtained through analogous equations. After this procedure, both the $n_\mathrm{B}$ and $Y_e$ variables, as well as all the other quantities are obtained as functions of a variable that parametrizes where we are along the intersection line, and the mixing factor $\chi$. This can always be inverted to obtain all the quantities as a function of $n_\mathrm{B}$ and $Y_e$. Note that since leptons were not considered to be in chemical equilibrium with the baryonic species, the lepton chemical potential still needs to be recovered in the transition region. This is done by ensuring that the electron chemical potential is consistent with the electron number density at all points. Finally, the contribution of electrons to the extensive quantities, like pressure and energy density, needs to be reintroduced.

\begin{figure}[!t]
    \centering
    \includegraphics[width=\columnwidth]{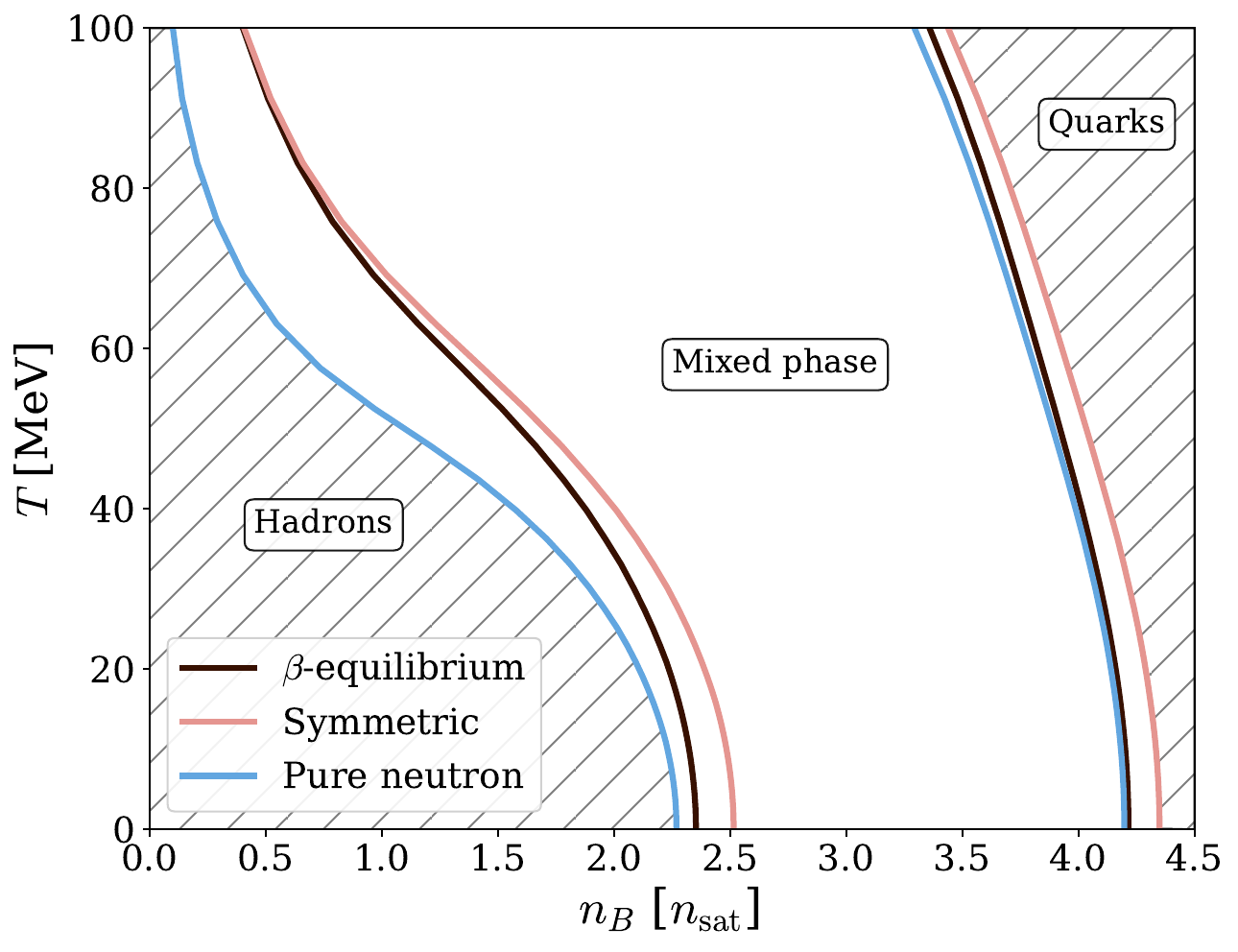}
    \caption{Phase diagram corresponding to the BBKF 2.1 EoS. Density is given in units of saturation density $n_\mathrm{sat}$. Solid lines with different colors represent phase boundaries for matter in neutrinoless beta-equilibrium, symmetric nuclear matter ($Y_e=0.5$), and pure neutron matter ($Y_e=0$). Hatched regions correspond to purely hadronic and quark matter.}
    \label{fig:phase_diagram}
\end{figure}

We calculate five different EoSs in a way that they (i) conform with current astrophysical constraints, e.g., all of the mass--radius relations reach $\sim2~M_\odot$, (ii) display a phase transition at approximately the same NS mass of $\sim1.4~M_\odot$, while (iii) covering the range of phase transition strengths, from mass--radius relations with a small kink to relations having twin star solutions. The parameters of the model provide sufficient freedom to produce EoSs with such properties, and for the cases presented in this work, are summarized in Tab.~\ref{tab:params}. As an example, the phase diagram corresponding to the BBKF 2.1 EoS is shown in Fig.~\ref{fig:phase_diagram} for matter in neutrinoless beta-equilibrium, for symmetric nuclear matter ($Y_e=0.5$), and for pure neutron matter ($Y_e=0$). Note how the phase transition line is shifted to lower densities for higher temperatures and lower electron fractions. This change in the transition line with varying $T$ and $Y_e$ underscores the importance of the proper treatment of phase transition in general purpose EoSs, as nuclear matter during binary NS mergers, for example, can exhibit a large range of electron fractions while temperatures can also easily rise above $\sim30$~MeV.

\subsection{Spacetime and matter evolution}

We perform general-relativistic radiative hydrodynamics simulations using the \texttt{BAM} code~\cite{Bruegmann:2006ulg, Thierfelder:2011yi, Bernuzzi:2016pie, Dietrich:2015iva}, which solves Einstein's field equations numerically in the $3+1$ formulation. Spacetime is evolved through a Z4c formulation with constraint damping terms \cite{Bernuzzi:2009ex,Hilditch:2012fp}, with the gauge set using the $1+\log$ slicing \cite{Bona:1994dr} and $\Gamma$-driver shift \cite{Alcubierre:2002kk} conditions. Matter is evolved as an ideal fluid, with conserved variables governed by the general-relativistic hydrodynamics equations in the Valencia formulation \cite{Banyuls:1997zz}. Additionally, neutrino radiation transport is implemented through an energy-integrated (gray) M1 moment scheme \cite{Schianchi:2023uky}. The time evolution of radiation fields and the coupling with matter is handled through an implicit-explicit scheme \cite{Izquierdo:2022eaz,Jaeger:2026lnh}, using the explicit RK(4,4) (4 stages, 4th order accurate) and the associated implicit I42L (4 stages, 2nd order accurate, L-stable) method. We use a Courant-Friedrichs-Lewy factor of 0.25. The computational grid is structured as a hierarchy of nested refinement levels, with each level containing one or more Cartesian boxes with fixed grid spacing. The resolution between successive levels increases by a factor of two. Because of the symmetry of the considered system, we impose octant symmetry on the spatial grid to reduce the computational costs. For the high-resolution shock-capturing scheme employed in the hydrodynamics code, we use the WENOZ reconstruction \cite{Borges:2008} and the HLLE Riemann solver \cite{Einfeldt:1988,Amiram:2006zjz}. We adopt a cold, static atmosphere to model the vacuum region surrounding the compact objects. The fluid quantities of a grid cell are set to atmosphere values once the rest-mass density $\rho$ falls below a density threshold~\citep{Thierfelder:2011yi}. The atmosphere density is chosen as the minimum rest-mass density covered by the EoS tables ($\sim2\times10^3$~g/cm$^3$).

\subsection{Initial conditions}
\label{ssec:in_cond}

The initial conditions are set by adding a small perturbation to a Tolman--Oppenheimer--Volkoff (TOV) solution, where the initial NS is set to be at the very edge of the purely hadronic branch. We experimented with different ways of triggering the phase transition, such as putting matter outside of the NS with a Bondi profile, as well as adding an extra layer of matter between the different layers of the NS in a smooth way, similarly to Ref.~\cite{Largani:2023kjx}. These approaches are discussed in Appendices~\ref{sec:app_Bondi} and \ref{sec:app_envelope}. These methods either resulted in artificial outflows from the NS, unrelated to the phase transition, as well as heating of the surface layer. In the case of Bondi accretion, the central density in fact even decreased instead of increasing due to the pressure support of the thermally heated outer layers.

\begin{table*}[t]
\centering
\caption{Parameters corresponding to the 3 EoSs we used for our simulations, as well as to the initial conditions. $\rho^{\mathrm{tr}}_L$ and $\rho^{\mathrm{tr}}_U$ denote the rest-mass densities at the lower and upper end of the phase transition, respectively. $\rho_c$ refers to the central density set for the initial TOV star, while $M_b$ denotes the baryonic mass of this NS. Finally, $\Delta M_b$ represents the baryonic mass added to the system by the method described in Sec.~\ref{ssec:in_cond}.}
\label{tab:sim_setup}

\setlength{\tabcolsep}{4pt}
\renewcommand{\arraystretch}{1.1}

\begin{tabular}{ccccccc}
\hline\hline
\rule{0pt}{3ex}EoS & $\rho^{\mathrm{tr}}_L$ [g/cm$^3$] & $\rho^{\mathrm{tr}}_U$ [g/cm$^3$] & $\rho_c$ [g/cm$^3$] & $M_b$ [$M_\odot$] & $\Delta M_b$ [$M_\odot$] & $\Delta M_b / M_b$ \\[2pt]
\hline
\rule{0pt}{3ex}BBKF 2.1 & $5.83\times10^{14}$ & $1.04\times10^{15}$ & $6.18\times10^{14}$ & $1.563$ & $6.25\times10^{-3}$ & $0.4\%$ \\
BBKF 2.3 & $5.82\times10^{14}$ & $8.99\times10^{14}$ & $6.80\times10^{14}$ & $1.558$ & $6.23\times10^{-3}$ & $0.4\%$ \\
BBKF 2.5 & $5.82\times10^{14}$ & $7.41\times10^{14}$ & $6.80\times10^{14}$ & $1.551$ & $6.20\times10^{-3}$ & $0.4\%$ \\[1pt]
\hline\hline
\end{tabular}

\end{table*}

Finally, we chose the following approach. After setting up the spacetime around a static TOV star, we artificially increased the density of the NS uniformly by 0.4\%, without modifying the spacetime solution. This way we ensure that there is additional baryonic mass added to the system, approximately $6\times10^{-3}~M_\odot$ in this case. The obtained setup is not in equilibrium and is, as intended, slightly perturbed. In addition, the setup is constraint-violating, but this violation is quickly damped through the Z4c scheme and constraint violations are propagated away from the numerical grid. During all of our simulations we ensure and monitor that constraint violations are damped after an initial short transient. 
Additionally, even though the star initially compresses due to being out of equilibrium and heats up slightly, this heating is distributed evenly across the NS and amounts to a lower temperature than in the other approaches mentioned above. Moreover, no artificial outflows are generated during this approach.

A summary of our simulated setups is provided in Tab.~\ref{tab:sim_setup}, focusing on the corner cases BBKF 2.1, 2.3 and 2.5. Due to utilizing a Gibbs construction and using EoSs with only a finite resolution in density, the phase transitions obtained numerically are slightly smoothed out, meaning that the baryonic mass corresponding to the TOV solutions increases throughout the transition. Therefore, in order to ensure that the solutions including the additional mass correspond to NSs on the hybrid branch, we initialize our NSs slightly above the lower transition density. For the BBKF 2.5 EoSs this is especially critical, since the difference in baryonic mass between the TOV solutions at the upper and lower phase transition densities is $\sim5\times10^{-2}~M_\odot$.

\subsection{Neutrino Rates}

Weak interactions involving neutrinos are of fundamental importance for determining the thermal and compositional evolution of NS matter. Without phase transitions, relevant reactions include absorption, emission and iso-energetic scattering on free nucleons and heavy nuclei. Thus, due to the large rest-masses of nuclei compared to the energy of the interacting neutrinos, momentum exchange is typically ignored, leading to analytic expressions for the computation of said rates. On the other hand, in dense media, the effective masses of nuclei decrease to energies comparable to those of neutrinos in thermodynamical equilibrium with matter, and single-particle potentials become important for enhancing interaction rates of neutrinos, while shifting the suppression threshold of anti-neutrinos towards higher energies. Altogether, this implies that neglecting momentum exchange is no longer a suitable approximation. Moreover, other microscopic effects come into play, such as momentum-exchange-dependent coupling constants for weak-interactions (form factors), weak-magnetism, and pseudo-scalar corrections due to the presence of pions in dense nuclear environments.

Therefore, in this work we adopt the more realistic description of interaction rates for nucleons provided by the full-kinematics approach~\cite{Guo:2020tgx}, in which analytic expressions for interaction rates are no longer available, as integration over the nucleon phase-space becomes non-trivial.

The electron fraction of beta-equilibrated matter on the two sides of the phase transition is generally different. Therefore, when matter is being compressed into the quark phase, it will end up in a state that is out of beta equilibrium. To return to $\beta$-equilibrium, it will emit (anti)neutrinos. Thus, in order to reproduce this process, one needs to include neutrino rates in the quark phase as well. Here, we take a first step and calculate neutrino opacities for the charged processes:
\begin{align}
    \nu_e + d &\leftrightarrow u + e^- \:,
    \label{eq:nu} \\
    \bar{\nu}_e + u  &\leftrightarrow d + e^+ \:,
    \label{eq:antinu}
\end{align}
according to Refs.~\cite{Steiner:2001rp,Colvero:2014zfa}, using the current-current interaction Lagrangian:
\begin{align}
    \mathcal{L}_\mathrm{int} = &\frac{G_F}{\sqrt{2}} [\bar{u}(\nu)\gamma_\mu(1-\gamma_5)u(4)] \nonumber \\
    & \times[\bar{u}(2)\gamma_\mu(C_A-C_V\gamma_5)u(3)] + \mathrm{h.c.} \: ,
\end{align}
where $G_F$ is the Fermi constant, and $C_V$ and $C_A$ are the appropriate vector and axial coupling constants, with $C_V=\cos\theta_C=0.973$ and $C_A = C_V$ for the neutrino process, and $C_V=\cos\theta_C$ and $C_A = -C_V$ for the anti-neutrino process \cite{Steiner:2001rp,Colvero:2014zfa}. The numbers, 2, 3 and 4 refer to the particles in Eqs.~\eqref{eq:nu}-\eqref{eq:antinu}, in the appropriate order. The precise expressions employed for neutrino absorption on quarks in this work are outlined in Ref.~\cite{Steiner:2001rp}. In principle, neutrino scattering on quarks should be considered for completeness. Due to the low up and down quark masses, such scattering would be inelastic, as shown in Ref.~\cite{Colvero:2014zfa}. Therefore, we highlight this limitation of our method, as appropriate treatment of inelastic processes in M1 energy-integrated schemes is still a challenging endeavor (see, however, Ref.~\cite{Chiesa:2024lnu} for a model of inelastic scattering as an emission/absorption process, Ref.~\cite{Cheong:2024cnb} for applications with spectral M1, Ref.~\cite{Foucart:2026kpd} with Monte Carlo treatment of neutrinos). On the other hand, under the conditions expected to be found in our simulated quark phases, Ref.~\cite{Colvero:2014zfa} argues that scattering contributes up to a few tens of percent to the total opacity, meaning that retaining just absorption reactions as in the present work is a reasonable first approximation. Moreover, since the onset of the quark phase lies at considerably high densities, where radiation should be trapped, we do not expect neutrino-quark reactions to significantly affect the dynamics of low density material, but only to provide the channel for chemical equilibration where quarks are present (see, e.g., Ref.~\cite{Alford:2025jtm} for a study suggesting the possibility of a gas of trapped neutrinos within a quark phase).

The spectral opacities $\kappa(\epsilon_{\nu})$ are then obtained through a double integral over particle energies. In addition, for the M1 moment scheme, the energy-averaged, equilibrium number- and absorption opacities are obtained through the integrals:
\begin{align}
    \label{eq:kappa_n_eq}
    \kappa_n^\mathrm{eq} = \frac{\int_0^\infty \tilde\kappa(\epsilon_{\nu}) \epsilon_\nu^2 f(\epsilon_\nu,T,\mu_\nu) \, \mathrm{d}\epsilon_{\nu}}{\int_0^\infty \epsilon_\nu^2 f(\epsilon_\nu,T,\mu_\nu) \, \mathrm{d}\epsilon_{\nu}} \: , \\
    \kappa_a^\mathrm{eq} = \frac{\int_0^\infty \tilde\kappa(\epsilon_{\nu}) \epsilon_\nu^3 f(\epsilon_\nu,T,\mu_\nu) \, \mathrm{d}\epsilon_{\nu}}{\int_0^\infty \epsilon_\nu^3 f(\epsilon_\nu,T,\mu_\nu) \, \mathrm{d}\epsilon_{\nu}}~\label{eq:kappa-a-avg} \: ,
\end{align}
where $\tilde\kappa(\epsilon) \equiv \kappa(\epsilon)/[1-f(\epsilon_\nu, T,\mu_\nu)]$ is the neutrino opacity corrected for stimulated absorption \cite{Burrows:2004vq}, and $f(\epsilon_\nu, T,\mu_\nu)$ is the Fermi-Dirac distribution assuming matter and radiation to be in local thermodynamical equilibrium. Hence, note that these opacities correspond to neutrinos being in equilibrium with matter. In regions where neutrinos are supposed to be trapped, we compute interaction rates based on equilibrium values $Y_{e,{\rm eq}}$, $T_{\rm eq}$ inferred from energy and lepton-number conservation (see, e.g., \cite{Perego:2019adq} for the original formulation, and~\cite{Radice:2021jtw, Musolino:2023pao, Foucart:2024npn} for applications). More precisely, the trapping conditions are generally set by the equilibration timescale
\begin{equation}
    \tau_\nu = [c\kappa_a^{\rm eq}(\kappa_a^{\rm eq} + \kappa^{\rm eq}_s)]^{-1/2},
\end{equation}
where $\kappa_s^{\rm eq}$ is the iso-energetic scattering opacity, defined as in Eq.~\eqref{eq:kappa-a-avg} for a spectral counterpart $\kappa_s(\epsilon)$~\cite{Burrows:2004vq}. Then, as standard, for $\tau \equiv \min[\tau_{\nu_e},\tau_{\bar\nu_e}]$ smaller than half an evolution timestep, we use $Y_{e,\rm eq},~T_{\rm eq}$ for the rates. For $\tau$ between half and one timestep, we linearly interpolate temperature and electron fraction between equilibrium and fluid values. Finally, for $\tau$ larger than one timestep, we apply the scaling of Ref.~\cite{Foucart:2016rxm}.

It is worth noting that such a procedure has proven important to ensure stability of M1 simulations employing hadronic EoSs, for which $\kappa_s^{\rm eq}$ is well known. For the quark case, however, the absence of $\kappa_s^{\rm eq}$ might artificially increase the equilibration timescale, in particular in regions of high density, low temperature, where iso-energetic scattering dominates over charged-current absorption~\cite{Jaeger:2026lnh}. Moreover, using the inelastic neutrino-electron scattering opacity as a proxy for the inelastic neutrino-quark scattering opacity, it is shown in, e.g., Ref.~\cite{Ng:2023syk}, that such opacities are comparable to iso-energetic elastic scattering on nucleons. Therefore, we might not be able to accurately capture equilibration of neutrinos and quark matter.

Finally, with the above prescription for average opacities, the energy emission rate $Q$ and number emission rate $R$ follow the integrated Kirchhoff's law,
\begin{eqnarray}
    Q &=& c\kappa_a^{\rm eq} \frac{4\pi T^4}{(hc)^3}F_3(\mu_\nu/T)\\
    R &=& c\kappa_n^{\rm eq} \frac{4\pi T^3}{(hc)^3}F_2(\mu_\nu/T),
\end{eqnarray}
where $F_i(\mu_\nu/T)$ is the $i$-th order ultra-relativistic Fermi integral.

\section{Results}
\label{sec:results}

\subsection{Hybrid equations of state}

\begin{figure}[!t]
    \centering
    \includegraphics[width=\columnwidth]{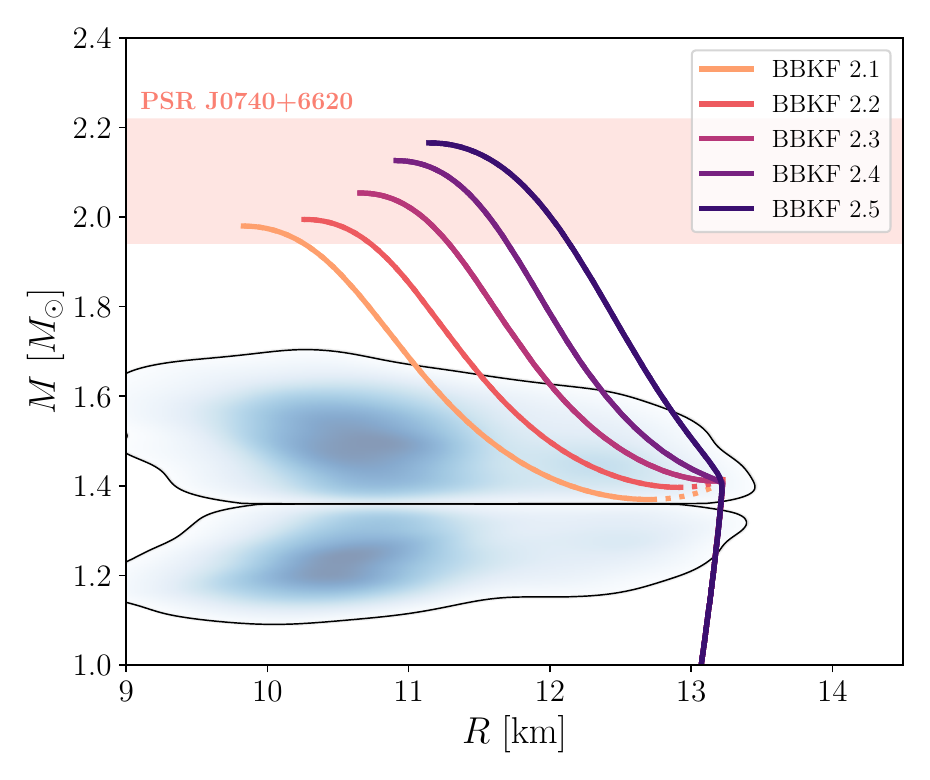}
    \caption{Mass--radius relations calculated with the five new EoSs. The blue shaded area corresponds to the posterior density calculated from GW170817 in Ref.~\cite{LIGOScientific:2018cki} with the black contour denoting the 90\% credible interval, while the light red region denotes the 95\% credible interval on the mass of PSR J0740+6620 \cite{Fonseca:2021wxt}.}
    \label{fig:MR}
\end{figure}

As mentioned in Sec.~\ref{ssec:eos}, we created our set of EoSs along the following principles: (i) the mass--radius relations calculated from the EoSs conform with current astrophysical constraints, (ii) all of them display a phase transition at approximately the same NS mass of $\sim1.4~M_\odot$, while (iii) they cover a broad range of phase transition strengths. This is demonstrated in Fig.~\ref{fig:MR}, which shows the mass--radius relations calculated from the five EoSs. Arguably the biggest obstacle in producing EoSs with strong phase transitions in a way that they conform with astrophysical measurements is that the hadronic branch should be soft enough to satisfy the upper limits on tidal deformability obtained from GW170817, while keeping the EoSs overall stiff enough to support massive NSs of $\sim2~M_\odot$.

Out of the five EoSs BBKF 2.1 and 2.2 showcase twin-star branches, separated by unstable branches from the hadronic parts. BBKF 2.3 is a limiting case between EoSs with and without a twin-star branch, exhibiting a nearly horizontal mass--radius relation after the onset of the phase transition, while the mass--radius relations of BBKF 2.4 and 2.5 do not have a twin-star branch.

\begin{figure*}[!t]
    \centering
    \includegraphics[width=\textwidth]{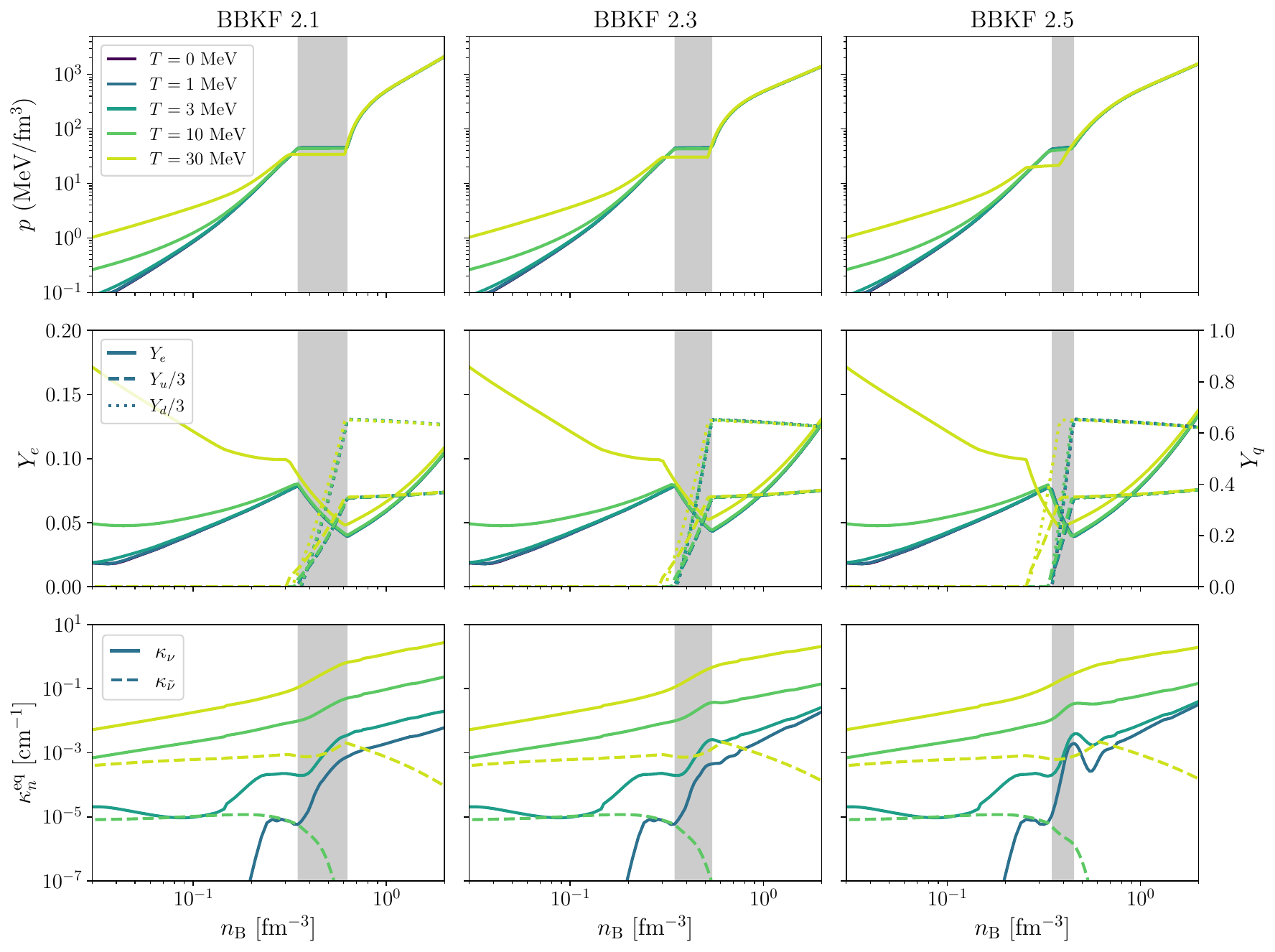}
    \caption{Thermodynamic properties in neutrinoless beta-equilibrium as a function of baryon number density for the three EoSs BBKF 2.1 (left panels), 2.3 (middle panels) and 2.5 (right panel). Different colors correspond to different temperatures. From top to bottom different panels show the pressure (top), the electron and quark fractions (middle), and the number-density averaged electron neutrino and anti-neutrino opacities (bottom). The gray shaded areas are highlighting the phase transition region in the zero-temperature limit.}
    \label{fig:EoS_fig}
\end{figure*}

We also show some of the main thermodynamic properties of the EoSs in neutrinoless beta-equilibrium ($\mu_l=0$) and at different temperatures for the corner cases of BBKF 2.1, 2.3 and 2.5 in Fig.~\ref{fig:EoS_fig}.
The top panels of Fig.~\ref{fig:EoS_fig} show the pressure as a function of baryon number density. One can observe that the change in temperature does not significantly affect the pressure at densities above saturation up to at least $T=10$~MeV, while for $T=30$~MeV both the EoS below saturation density significantly changes and the location of the phase transition is modified. However, the EoS in the quark region remains almost the same due to its high density.

The particle fractions shown in the middle panels are defined as:
\begin{equation}
    Y_i = \frac{n_i}{n_\mathrm{B}} \: ,
\end{equation}
for particle $i$ with $n_i$ the number density of said particle, while $n_\mathrm{B}$ is the baryon number density. Since the baryon number of a single quark is $1/3$, the total quark number density $n_q = n_u + n_d$ in the quark phase is $Y_q = 3$. Note how the electron fraction, except for $T=30$~MeV is monotonically increasing below and above the phase transition line, when crossing the phase transition $Y_e$ drops by $\sim0.04-0.05$. This is mainly due to the appearance of $d$ quarks, which also have a negative charge and can thus replace some of the electrons to make the total charge density vanish. Above the transition line, the ratio of $u$ and $d$ quarks does not change significantly, with $d$ quarks being almost twice as numerous than $u$ quarks.

Finally, the bottom panels of Fig.~\ref{fig:EoS_fig} show the number-density averaged equilibrium electron neutrino and anti-neutrino opacities, as defined in Eq.~\eqref{eq:kappa_n_eq}. For $T=0$ these opacities are naturally zero, since the sharp Fermi surfaces do not allow any reaction to happen. Increasing the temperature first makes the neutrino opacities start increasing. This is because $\mu_p>m_p^\mathrm{eff}$, $\mu_n>m_n^\mathrm{eff}$ and $\mu_e>m_e$, thus even neutrinos slightly above the Fermi surface can create electrons and protons with positive energy. In contrast, the chemical potential of positrons is significantly negative, hence creating a positron with positive energy would require an anti-neutrino or positron substantially above its Fermi surface, which is dynamically not accessible for these temperatures. However, once they become accessible at $T\gtrsim10$~MeV, anti-neutrino opacities also start increasing dramatically and contributing to the total neutrino emissivity. For large temperatures ($T\gtrsim10$~MeV) neutrino opacities scale as $\propto T^2$. Interestingly, although for hadronic matter, absorption opacities are suppressed at high densities due to Pauli blocking of the reaction products~(see, e.g., Ref.~\cite{Chiesa:2024lnu}), absorption opacities calculated for the quark phase grow monotonically (as also depicted in Ref.~\cite{Colvero:2014zfa}).

\subsection{Simulation results}

\begin{figure*}[!t]
    \centering
    \includegraphics[width=\textwidth]{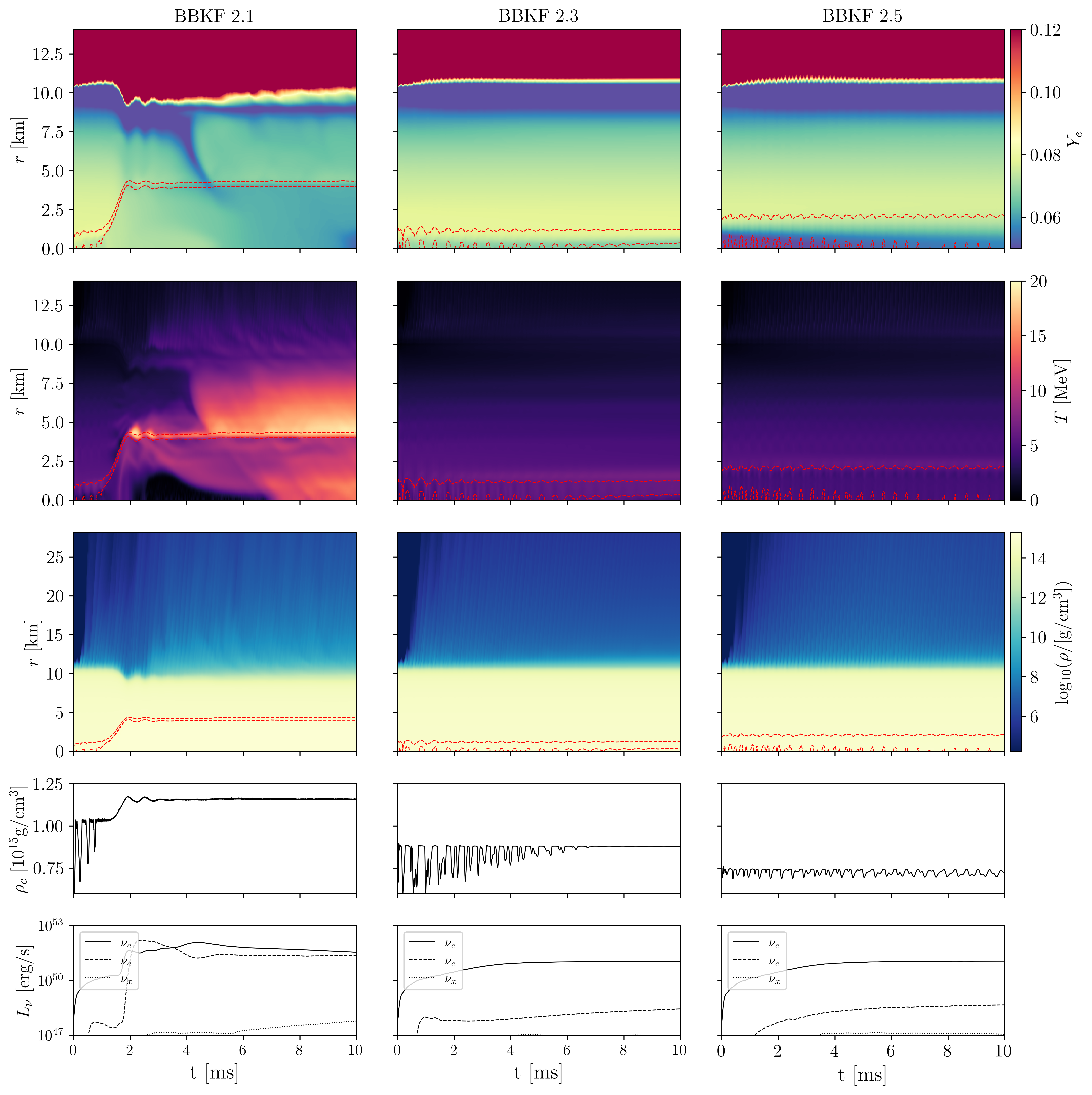}
    \caption{The time evolution of the (angle-averaged) electron fraction, $Y_e$ (top row), temperature (second row), rest-mass density (third row), as a function of radius, as well as the evolution of the central rest-mass density (fourth row), and neutrino luminosities for different neutrino species (bottom row). Luminosities are shown as a function of retarded time. The three columns from left to right show results for the BBKF 2.1, BBKF 2.3 and BBKF 2.5 EoSs, respectively. In the panels showing the angle-averaged variables, dashed red lines denote both the lower and upper boundaries of the phase transitions, calculated using the conditions $Y_q/3=0.05$ and $Y_q/3=0.95$, respectively.}
    \label{fig:PT_evolution}
\end{figure*}

We performed three simulations, employing the EoSs BBKF 2.1,  BBKF 2.3 and  BBKF 2.5, with progressively weaker transition strengths. We first set up static NSs and then added $\sim6\times10^{-3}~M_\odot$ to the system in baryonic mass, as described in Sec.~\ref{ssec:in_cond}. The initial mass of the TOV stars was tuned in a way that the solutions would lie below the upper phase transition threshold, while together with the additional baryonic mass TOV solutions would be on the hybrid branch (see Tab.~\ref{tab:sim_setup}, for the exact numerical values). We used 7 nested levels with 192 grid points per direction on each level, corresponding to a spatial resolution of $\sim0.1$~km on the finest level. We note that even though we did not expect any turbulent behavior that would have required a high resolution to be resolved, a sufficiently high resolution is required to suppress artificial heating due to the discontinuity between the hadronic and quark layers (see Appendix~\ref{sec:app_heating} for further discussion).

The time evolution of diagnostic quantities is shown in Fig.~\ref{fig:PT_evolution} for BBKF 2.1, BBKF 2.3 and BBKF 2.5 from left to right, respectively. From top to bottom, the first three rows represent the (angle-averaged) net electron fraction, temperature and rest-mass density for radial shells. Dashed red lines mark shells at the lower (larger radius) and upper (smaller radius) densities across the phase transition. For these thersholds we used the condition of the quark fraction reaching $Y_q/3=0.05$ and $0.95$, respectively. The last two rows correspond to the time evolution of central rest-mass density $\rho_c$ and neutrino luminosity $L_\nu$, defined in the usual manner as
\begin{equation}
    L_\nu = \oint dS~n^i \sqrt\gamma(\alpha F_i - \beta_i E),
\end{equation}
at a surface of constant coordinate radius $r = 200~M_\odot$, where $n^i$ is the unit radial normal, $\gamma$ is the 3-metric determinant, $\alpha$ the lapse, $\beta_i$ the shift vector, and $E$, $F_i$ are the neutrino energy density and flux, respectively. Luminosities are shown as a function of retarded time.

From the first row we may identify the NS surfaces for all runs at $r \sim 10~{\rm km}$, marked both by the sharp $Y_e$ gradient towards the external $Y_e \geq 0.12$ region, approximately coinciding with the region of $\rho \geq 10^{12}~{\rm g/cm^3}$, depicted in the third row. Note that radius here stands for the coordinate radius, initialized in isotropic coordinates, using 1+log-slicing and the $\Gamma$-driver shift, in contrast to the Schwarzschild radii depicted in Fig.~\ref{fig:MR}.

For BBKF 2.1, we have high amplitude oscillations in $\rho_c$ prompted by the initial density perturbation. This is due to the rather long radially unstable region of the mass-radius diagram, which is crossed as the star migrates to a stable configuration in the hybrid branch. The formation of a pure quark core is observed as a non-oscillatory increase of $\rho_c$ at $\sim 1.3~{\rm ms}$, which we define as the onset time. Consequently, the radius contracts slightly, and the innermost red dashed line, marking the boundary of the pure quark phase, becomes apparent in the first three panels. From $t \sim 2 ~{\rm ms}$ on, we note that the outermost transition line retains high temperatures $T \sim 20~{\rm MeV}$, arguably induced by a sound-speed discontinuity~\cite{Dimmelmeier:2009vw}. Material in the outer layers get progressively hotter over time until $T = 5 - 10~{\rm MeV}$ is reached at densities as low as $\rho \sim 10^{10}~{\rm g~cm^{-3}}$. Then, additional thermal pressure on outer layers makes the radius slightly expand, and overall higher neutrino luminosities (depicted in the lowest, left panel of Fig.~\ref{fig:PT_evolution}) are achieved, with $L_{\bar\nu_e}$ comparable to $L_{\nu_e}$, as usual in sufficiently hot media. Towards the interior of the star, the inner transition line shows no shock-heating, but a gradual increase in the temperature is observed.

For the BBKF 2.3 EoS, the transition happens gradually on a longer timescale, since the mass-radius relation does not have an unstable branch, but the first section of the hybrid branch is nearly horizontal on the diagram. A quark core is formed, but in this case it only covers a small fraction of the overall NS mass. With less violent $\rho_c$ oscillations compared to BBKF 2.1, the heating across the outer transition line is much milder, and, consequently, neutrino emission is dominated by $\nu_e$ with roughly one order of magnitude less luminosity than BBKF 2.1.

Finally, for the BBKF 2.5 EoS, the NS does not completely transition to the hybrid branch, but the central density keeps oscillating between the two phases. Therefore, no relaxed quark core is formed in this case. This is consistent with the findings of Ref.~\cite{Largani:2023kjx}, who used an EoS with a similarly weak transition, and found that the central density keeps oscillating between the two phases; cf.\ Fig.~\ref{fig:PT_evolution}. Moreover, the lack of an unstable branch for this EoS produces $\rho_c$ oscillations with much smaller amplitudes than in the other two cases. Internal heating is substantially less pronounced, yielding neutrino luminosities similar to BBKF 2.3. Neutrino luminosities from heavy lepton flavours, $\nu_x$ are orders of magnitude lower than the respective luminosities from electron (anti)neutrinos in all cases.

For all our simulations, we find negligible amounts of unbound material. Instead, due to the stars' oscillations (more so for BBKF 2.1, which crosses the unstable branch), a cloud of bound matter surrounds the stars.

Note that Ref.~\cite{Largani:2023kjx} reports $\sim 10^{-3}~M_\odot$ of ejected mass. However, while in our simulations a very small amount of baryonic mass was added to the systems ($\sim 6\times10^{-3} M_\odot$ cf. Tab.~\ref{tab:sim_setup}), in their case, the simulated setup is such that an additional $M_{\rm envelope} = 0.018~M_\odot$ (three times more than in our case) is accreted by an isentropic NS. This difference between the simulation setups might be one of the reasons for the disagreement between the two results. In the results of Ref.~\cite{Largani:2023kjx} ejecta also seem to emerge once significant neutrino emission due to the phase transition appears. We do not find such a drastic change in the behaviour of the outermost layers of the NS after the phase transition is triggered, which might be explained by differences in neutrino opacities as well as the different scheme applied for neutrino radiation transport.

Due to the inclusion of neutrino emission in the quark phase, we expect the quark core of the NS to relax into a compositional equilibrium state after the transition with a lower electron fraction compared to its initial state. This is because our hybrid EoSs have the property that in neutrinoless $\beta$-equilibrium the quark phase has a lower electron fraction, with a difference of $\Delta Y_e\approx0.04-0.05$ between the two sides of the transition boundary. This behavior is also explicitly shown in our Fig.~\ref{fig:EoS_fig} and Fig.~7 of Ref.~\cite{Bastian:2020unt}. As seen in the first row of Fig.~\ref{fig:PT_evolution}, we verify to some extent the relaxation of $Y_e$ towards lower values for BBKF 2.1 and BBKF 2.3 comparing the outer and inner transition lines. For BBKF 2.5, the composition $Y_e$ remains mostly unchanged along the evolution.

\begin{figure}
    \centering
    \includegraphics[width=\linewidth]{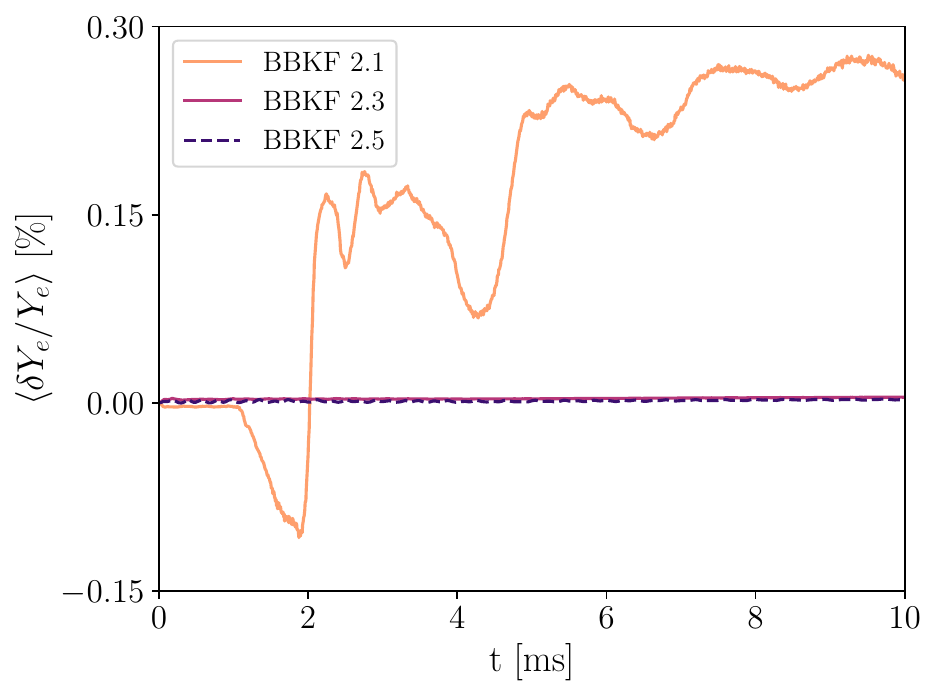}
    \caption{Evolution of the mass-averaged fractional deviation of $Y_e$ with respect to equilibrium with a gas of trapped neutrinos [Eq.~\eqref{eq:avg-dev}] for the EoSs BBKF 2.1, BBKF 2.3, BBKF 2.5.}
    \label{fig:delta-ye}
\end{figure}

Once neutrino emission and absorption by quarks is taken into account, a reasonable expectation is that $Y_e$ should evolve into equilibrium values set by the presence of the trapped gas of neutrinos, instead of the one set by neutrinoless $\beta$-equilibrium. To assess whether this is the case, we follow Ref.~\cite{Gieg:2026beb} and compute the equilibrium net electron fraction $\tilde{Y}_e = \tilde{Y}_e(\rho, T, Y_{Le})$, where
\begin{equation}
    Y_{Le} = Y_e + \frac{n_{\nu_e} - n_{\bar\nu_e}}{n_B},
\end{equation}
is the electron lepton number, which depends on the fluid's net electron fraction $Y_e$, and on the (evolved) electron (anti) neutrino number density $n_{\nu_e}$ ($n_{\bar\nu_e}$). Then, the fractional deviation
\begin{equation}\label{eq:dev-def}
    \delta Y_e/Y_e \equiv 1 - \tilde{Y}_e(\rho,T,Y_{Le})/Y_e,
\end{equation}
allows one to quantify how close the system evolves to a fluid-radiation equilibrium mixture. More precisely, since we are interested in verifying equilibration at high densities, we evaluate the mass-averaged fractional deviation (here restricted to the $x-y$ plane) as
\begin{equation}\label{eq:avg-dev}
\left\langle \delta Y_e/Y_e\right\rangle_{xy} \equiv \frac{\int {\rm d}x{\rm d}y \sqrt{\gamma} D(\delta Y_e/Y_e)}{\int {\rm d}x{\rm d}y \sqrt{\gamma}D},
\end{equation}
where $\gamma$ is the determinant of the spatial metric, and $D$ is the rest-mass density in the Eulerian frame (i.e., measured with respect to an observer at rest with respect to a spatial hypersurface). The time evolution of the quantity defined in Eq.~\eqref{eq:avg-dev} is presented in Fig.~\ref{fig:delta-ye}.

We can observe that matter in the simulation corresponding to the BBKF 2.1 EoS is driven out of equilibrium due to the formation of a pure quark core, as an increased deviation develops around the onset time. As the quark core settles (around $t = 2~{\rm ms}$), the deviation abruptly increases to positive values, and remains so until the end of the simulation, indicating that, in order to attain $\beta$-equilibrium, either (i) $Y_e$ should decrease or (ii) $\tilde{Y}_e$ (and, consequently, $Y_{Le}$) should increase at fixed $(\rho, T)$. Overall, we note that neutrino trapping and compositional equilibration are captured to the sub-percent level. For the BBKF 2.3 and BBKF 2.5 EoSs, the absence of a large core implies that the NSs are not thrown out of compositional equilibrium, leading to much smaller deviations of $\mathcal{O}\left(10^{-5}\right)$.

As a last remark, we note that the persistence of deviations for BBKF 2.1, although small, suggests the importance of neutrino-quark scattering (which, we note again, is an inelastic process). For instance, it is shown in Ref.~\cite{Thompson:2000gv} that neutrino-electron scattering (also inelastic) plays an important role for the trapping and thermalization of heavy lepton neutrinos. Hence, we conjecture that the inclusion of neutrino-quark scattering would allow to better capture the thermalization of neutrinos, possibly contributing to diminishing the observed deviations.

\section{Conclusions}
\label{sec:conclusion}

In this paper we presented a new set of general-purpose EoSs including a first-order phase-transition into two-flavor (up and down) deconfined quark matter, constructed in such a way that the onset density for all cold, beta-equilibrated models is the same, but with varied density gaps across the transition. All models satisfy current astrophysical constraints on maximum TOV masses and radii in the intermediate mass range $1.4 - 1.7~M_\odot$. The BBKF 2.1 and BBKF 2.2 EoSs posses a radially unstable branch connecting the hadronic and hybrid NSs, while for the BBKF 2.3 EoS an almost-flat connection between hadronic and hybrid NSs represents a borderline-case for radial stability. Finally, BBKF 2.4 and BBKF 2.5 are always radially stable. The above set of EoSs was purposefully created in a way to be used by the broader astrophysical community in dynamical simulations, allowing a systematic exploration of the role of the density gap across the phase-transition, while keeping the hadronic EoS and the onset of phase transition the same. In this regard, all five EoSs will be made publicly available through the CompOSE database\footnote{\href{https://compose.obspm.fr/}{https://compose.obspm.fr/}}, while neutrino opacities will be shared through Zenodo. \footnote{\href{https://zenodo.org/}{https://zenodo.org/}}

As a test case for the new EoS models, physically motivated by a large fraction of NSs going through prolonged phases of accretion, and by the previous study of Ref.~\cite{Largani:2023kjx}, we performed a set of general relativistic radiation hydrodynamics simulations of perturbed TOV stars employing the BBKF 2.1, BBKF 2.3 and BBKF 2.5 EoSs, attempting to capture the migration of initially hadronic stars (or stars in a mixed phase) to the hybrid branch, where a quark core is fully formed, through a small increment of baryonic mass. Out of the three simulations, only BBKF 2.1 formed a significant quark core, since along the migration this model crosses the radially unstable branch (for which radial oscillations are amplified), reaching central densities beyond the upper transition density. We also showed that with the inclusion of neutrino-quark absorption reactions, we observe equilibration of quarks and a gas of trapped neutrinos, indicated by a small (within the sub percent level) fractional deviation of the fluid's net electron fraction with respect to the one predicted by equilibrium at the same density, temperature and electron lepton number.

Notably, we find differences between our results and those of Ref.~\cite{Largani:2023kjx} considering the amount of ejected matter. While we did not find any unbound matter in any of our simulations, Ref.~\cite{Largani:2023kjx} reports an amount of $\sim10^{-3}~M_\odot$ being ejected for an EoS that is closest to BBKF 2.5 in terms of phase transition strength. Although the cause of the disagreement between the two results is not completely clear, among the possible reasons one could mention the difference in simulation setup, with Ref.~\cite{Largani:2023kjx} adding an outer envelope of matter with three times as much mass to an isentropic TOV star, while also employing a different radiation transport scheme using different neutrino opacities.

Finally, a number of numerical experiments with different prescriptions to model accretion are presented in the Appendix. Overall, we find that the use of temperature-dependent EoSs poses a major challenge for the numerical modeling of the migration of a hadronic NS to the hybrid branch, as neutrino cooling does not operate on timescales small enough to counteract the rising thermal pressure, preventing the migration to occur on timescales feasibly achievable in numerical relativity setups.

\begin{acknowledgments}

The authors would like to thank Niels-Uwe Bastian for useful input regarding the BBKF EoSs.
J.T. acknowledges support from the Alexander von Humboldt Foundation under the project no. 1240213 - HFST-P.
H.G.\ and T.D.\ acknowledge funding from the EU Horizon under ERC Starting Grant, no.\ SMArt-101076369. Views and opinions expressed are those of the authors only and do not necessarily reflect those of the European Union or the European Research Council. Neither the European Union nor the granting authority can be held responsible for them. Simulations were performed on the DFG-funded research cluster Jarvis at the University of Potsdam (INST 336/173-1; project number: 502227537).

\end{acknowledgments}

\appendix

\section{Bondi accretion}
\label{sec:app_Bondi}

The solution for the relativistic Bondi flow can be found in standard textbooks \cite[e.g.][]{Shapiro:1983du}. Here, we repeat the most important steps. To set up initial conditions for relativistic Bondi accretion on a fixed Schwarzschild background, we need to solve the continuity and Bernoulli equations in usual Schwarzschild coordinates, i.e.,
\begin{align}
    \dot{M} = 4\pi r^2 \rho u^r = \mathrm{const.} \: , \label{eq:continuity} \\
    h u_t = \mathrm{const.} = -h_\infty \: ,
\end{align}
where $h = (\epsilon + p)/\rho$ is the specific enthalpy per baryon, with $\epsilon$ the energy density and $p$ the pressure. Using the metric functions we can express the second equation using $u^r$:
\begin{equation}
    h\left(\rho(r)\right) \left[ \left( 1 - \frac{2M}{r} \right) + \left( u^r \right)^2 \right]^{1/2} = h_\infty \: .
    \label{eq:Bernoulli}
\end{equation}
These equations describe a two-parameter family of solutions, from which the physical solution is selected by requiring a transsonic flow, which is regular everywhere. The sonic point (or critical point) is defined as the point where the fluid velocity observed by a static observer becomes equal to the speed of sound:
\begin{equation}
    v_c^2 = \gamma_{\mu\nu} v_c^\mu v_c^\nu = c_{s}(\rho_c)^2 \: ,
\end{equation}
with
\begin{equation}
    v^\mu = \frac{\gamma^\mu_\nu u^\nu}{-u_\alpha n^\alpha} \: ,
\end{equation}
and the subscript $c$ denoting the value at the sonic point. In addition, we need to require that $\mathrm{d}u^r/\mathrm{d}r$ remains regular at the sonic point. Finally, we end up with the equations:
\begin{figure}[!t]
    \centering
    \includegraphics[width=\columnwidth]{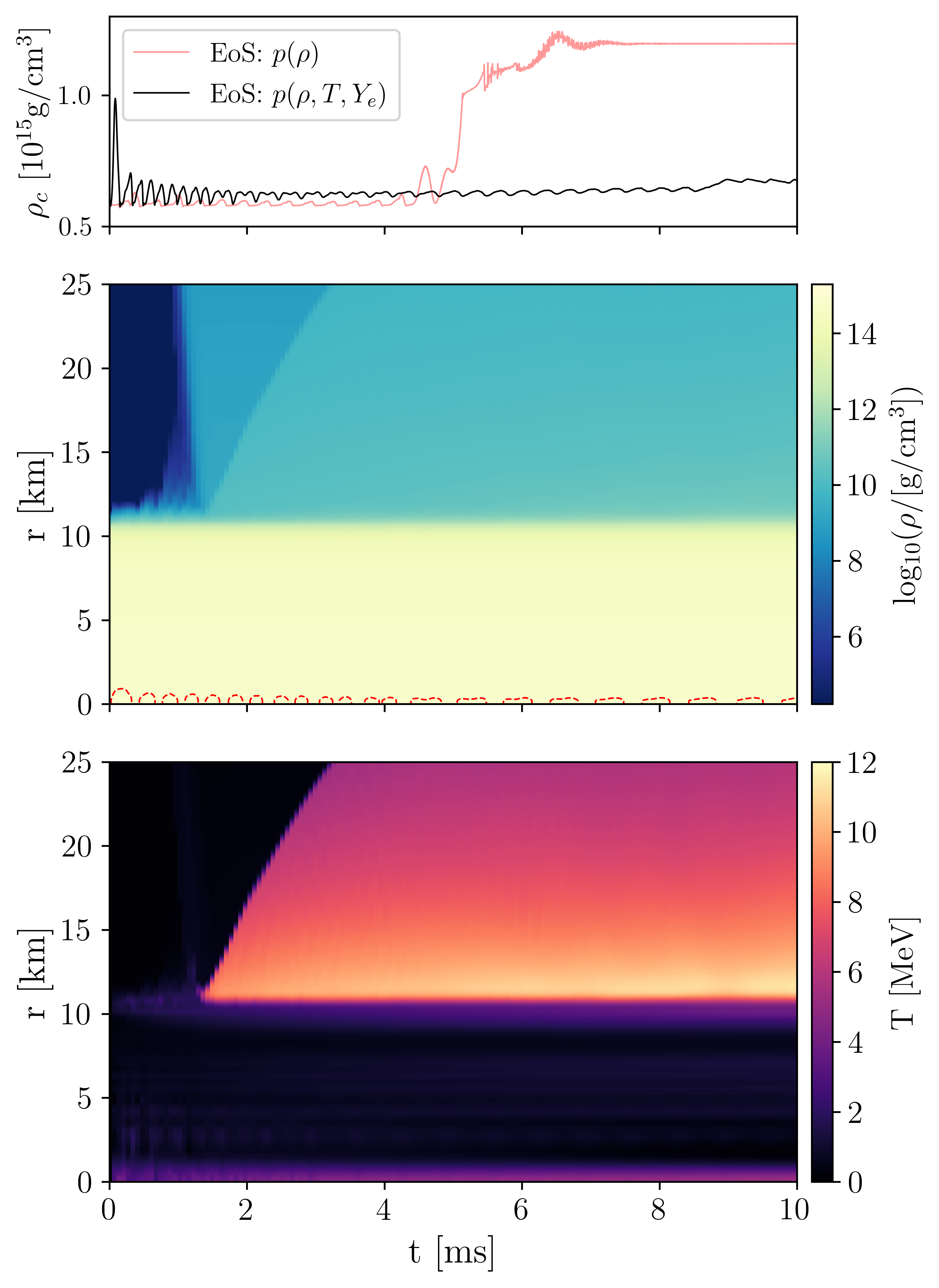}
    \caption{Density and temperature evolution of Bondi accretion on a cold isolated NS with an accretion rate $\dot{M} = 2\times10^{-6}$, in dimensionless units. \textit{Top}: The evolution of the central density. The red line shows results from a simulation where a cold barotropic EoS was used. \textit{Middle}: The angle-averaged density as a function of time and radius. The red dashed line denotes the boundary layer at the lower threshold of the phase transition. \textit{Bottom}: The angle-averaged temperature as a function of time and radius.}
    \label{fig:Bondi}
\end{figure}
\begin{align}
    \frac{M}{r_c} = \frac{2 c_{s}(\rho_c)^2}{1 + 3 c_{s}(\rho_c)^2} \: ,
    \label{eq:sonic_dudr} \\
    (u^r_c)^2 = \frac{c_{s}(\rho_c)^2}{1 + 3c_{s}(\rho_c)^2} \: ,
    \label{eq:ur_cs2}
\end{align}
and the additional constraint
\begin{equation}
    h(\rho_c) \sqrt{\frac{1}{1 + 3c_{s}(\rho_c)^2}} = h_\infty \: .
    \label{eq:hinfty_cond}
\end{equation}
This equation is to be solved numerically for a given $h_\infty$ for a chosen EoS to obtain the density and sound speed at the sonic point, from which the mass flux $\dot{M}$ can also be obtained using the continuity equation with Eqs.~\eqref{eq:sonic_dudr}-\eqref{eq:ur_cs2}. Finally, solving the Bernoulli equation, Eq.~\eqref{eq:Bernoulli} with the continuity equation and the EoS gives us the solution at arbitrary radial coordinate $r$.

Instead of $h_\infty$, it is more practical to use $\dot{M}$ as a parameter for our Bondi flow. The constraint equation then reads
\begin{equation}
    \dot{M} = \pi\rho_c M^2 \left( \frac{c_{s}(\rho_c)^2}{1 + 3c_{s}(\rho_c)^2} \right)^{-3/2} \: ,
\end{equation}
from which we can once again obtain $h_\infty$ and the solution for the Bondi flow at all radial distances.

For our simulations we set $\dot{M} = 2\times10^{-6}$ in dimensionless units ($\sim10^{7}~M_\odot\,\mathrm{yr}^{-1}$), the inner radius of the Bondi flow to $r_\mathrm{in}=50$~km, while the total accreted mass is set to $\Delta M = 8\times10^{-3}~M_\odot$. We perform these simulations without the inclusion of neutrino transport, evolving the momentum and energy equations, as well as the equation for lepton number conservation, together with the full EoS, $p(\rho,T,Y_e)$. The density and temperature evolution is shown in Fig.~\ref{fig:Bondi} for the EoS BBKF 2.1, which has the strongest phase transition. We can see that the impact of the accreted matter after $\sim1.5$~ms does not increase the size of the quark core, or even the central density initially. Instead, it creates a bloated atmosphere, with a temperature $\sim10$~MeV, which occupies an increasingly large region due to the support from thermal pressure. In fact, while the central density starts gradually increasing during later times, a quark core does not appear in the center during the simulated time. Even if a transition did happen, the produced bloated atmosphere would create an unrealistic physical setup. Hence, we conclude that this method makes it unlikely to capture the migration to the hybrid branch due to accretion in relativistic simulations employing temperature-dependent EoSs. For comparison, the top panel of Fig.~\ref{fig:Bondi} also shows the evolution of the central density in a similar simulation employing a cold barotropic EoS, ($p(\rho)$). Note that in the absence of thermal effects, the transition to the hybrid branch can indeed proceed.

In order to understand why it is difficult to capture the transition through simulating a physical accretion, it is instructive to look at the relevant timescales and physical accretion rates. Let us assume that we have the computational resources to accurately simulate the accretion of $\Delta M \sim 10^{-3}~M_\odot$ over a timescale of $\sim1$~s. This would translate to an accretion rate of:
\begin{equation}
    \dot{M}_\mathrm{sim} \approx 3\times 10^4~M_\odot\,\mathrm{yr}^{-1} \: .
\end{equation}
Instead, the Eddington rate corresponding to a $1.4~M_\odot$ NS is:
\begin{equation}
    \dot{M}_\mathrm{Edd} \approx 1.5\times 10^{-8}~M_\odot\,\mathrm{yr}^{-1} \: .
\end{equation}
Therefore, there are $\sim12$ orders of magnitude difference in what is feasible in a numerical simulation, and what would be physically reasonable.

In fact, this heated outer layer would not affect the accretion process significantly, since the physical cooling timescales are much shorter than the accretion timescale. The amount of matter accreted by a NS accreting at the Eddington rate during a single cooling timescale, which is $\sim10-10^3$~s for a NS with a temperature of a few MeV \cite{Prakash:2000jr,Yakovlev:2004iq}, is $\sim10^{-14}-10^{-12}~M_\odot$, meaning that only a thin layer of the NS atmosphere would be affected, which would have a negligible effect on the overall dynamical evolution of the NS.

Finally, we highlight a recent work~\cite{Naseri:2026xrq} in which similar difficulties in capturing a migration between branches, related to excessive thermal support developing in the outer layers of the NS, seem to also be present. There the authors show that the problem may be circumvented by adopting a suitable cooling prescription, based on a cooling timescale comparable to the dynamical one. The authors also anticipate that such a fast cooling could not be provided by neutrinos, whose diffusion timescale exceeds the dynamical timescale by a few orders of magnitude. Our simulations corroborate such an expectation.

\section{Adding matter through envelope}

In an effort to investigate if the difference between our results and those of Ref.~\cite{Largani:2023kjx} are due to the different initial conditions, we tried to reproduce their initial setup. In order to do this, we first calculated the TOV solution for a NS at the transition line, then inserted an additional layer of $\Delta M = 8\times10^{-3}~M_\odot$ with a constant density of $\sim6\times10^{12}$~g/cm$^3$ in-between the layers with higher and lower densities, so that the density remains a continuous function of $r$. Unlike Ref.~\cite{Largani:2023kjx}, we set the initial temperature to be of a cold NS, however, we found that the initial relaxation of the star nevertheless heats up the outer layers to a few MeV, comparable to the initial conditions of~\cite{Largani:2023kjx}.
\label{sec:app_envelope}
\begin{figure*}[!t]
    \centering
    \begin{minipage}[t]{\columnwidth}
        \vspace{0pt}
        \centering
        \includegraphics[width=\linewidth]{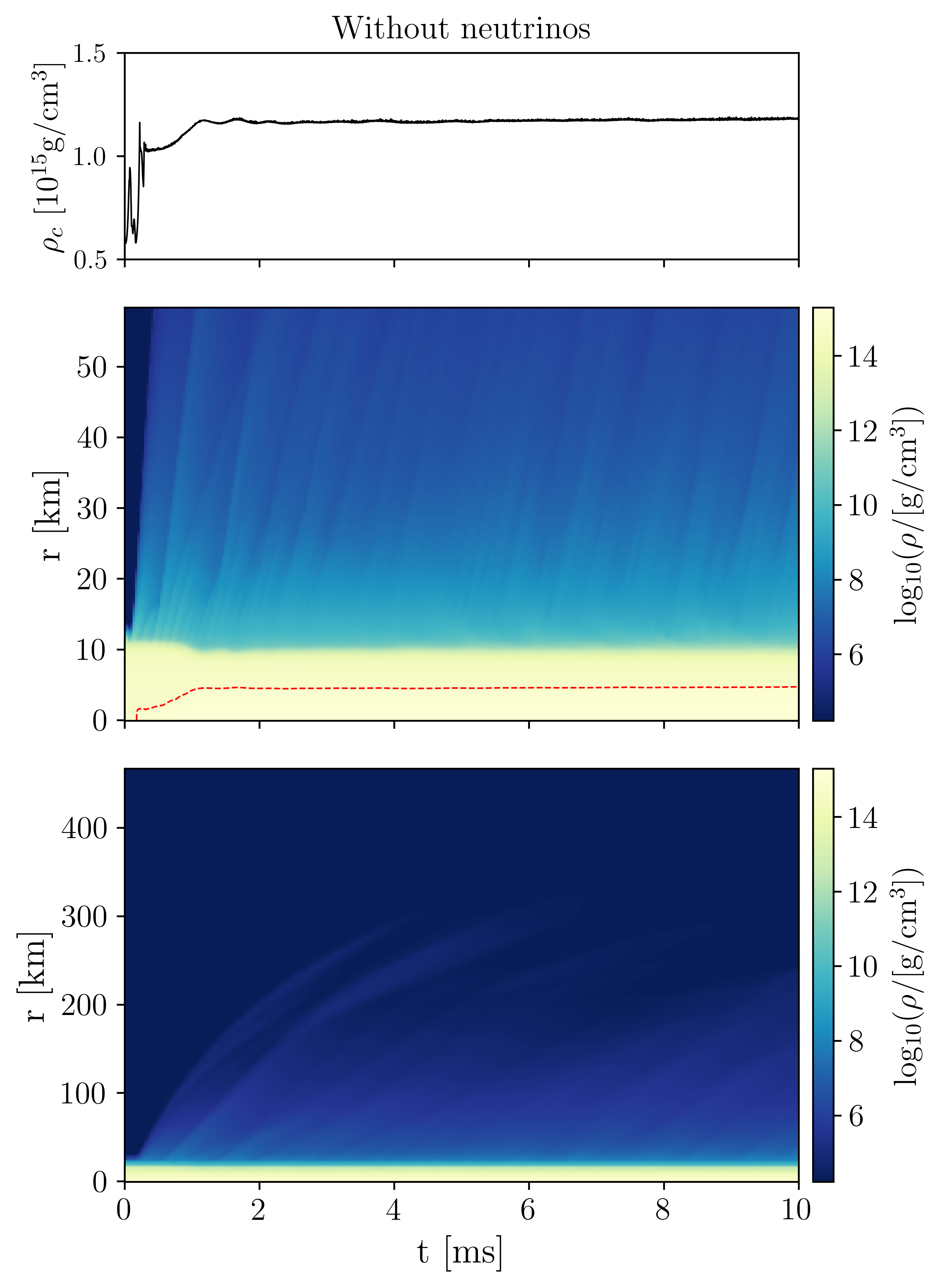}
        \caption{The evolution of the central density (top), and angle-averaged density at different radii (middle and bottom) in a simulation without neutrino transport. The red dashed line denotes the boundary at the upper threshold of the phase transition, within which matter is in a purely quark phase.}
        \label{fig:3D_env}
    \end{minipage}
    \hfill
    \begin{minipage}[t]{\columnwidth}
        \vspace{0pt}
        \centering
        \includegraphics[width=\linewidth]{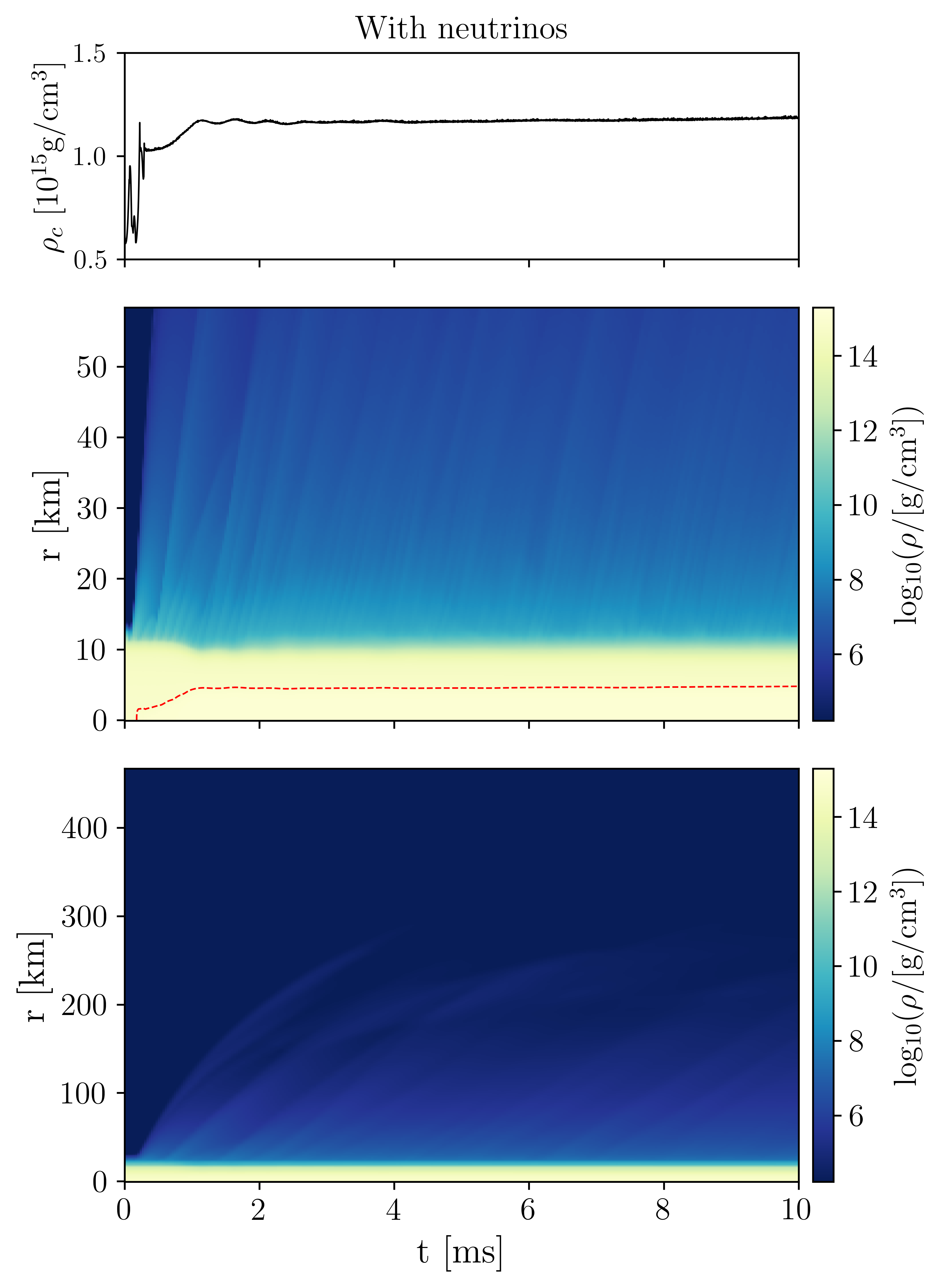}
        \caption{The same as Fig.~\ref{fig:3D_env} but for a simulation with neutrino transport.}
        \label{fig:M1_env}
    \end{minipage}
\end{figure*}

The simulation results using this initial setup are shown in Fig.~\ref{fig:3D_env} and Fig.~\ref{fig:M1_env} without and with neutrino transport, respectively. For both simulations we use the BBKF 2.1 EoS, which has the strongest phase transition. In our simulation without neutrino transport we fail to reproduce the relaxation of the NS without matter being ejected beyond 50~km, as in the reference simulation in Ref.~\cite{Largani:2023kjx}. Even though there is a difference in the utilized EoS, as their reference simulation is performed using an EoS without phase transition, this difference is not because of that but possibly due to different initial conditions, since matter is not ejected due to the transition but because of the additional matter bouncing back from the center of the NS. We also performed simulations setting the density of the added envelope to lower values, but found that this does not reduce the amount of matter escaping the direct vicinity of the NS.

When neutrino transport is also accounted for, we find no substantial difference between the overall evolutions, and do not find that additional matter is being ejected due to neutrinos. In fact, in none of the simulations do we find any matter becoming unbound from the NS, using the Bernoulli criterion
\begin{eqnarray}
    hu_t < -1; ~v^r > 0,
\end{eqnarray}
for $h$ the specific enthalpy, $v^r$ the (Eulerian) radial velocity, both conditions checked at a surface of constant coordinate radius.

\section{Artificial heating by discontinuity}
\label{sec:app_heating}

\begin{figure*}[t]
    \centering
    \includegraphics[width=\textwidth]{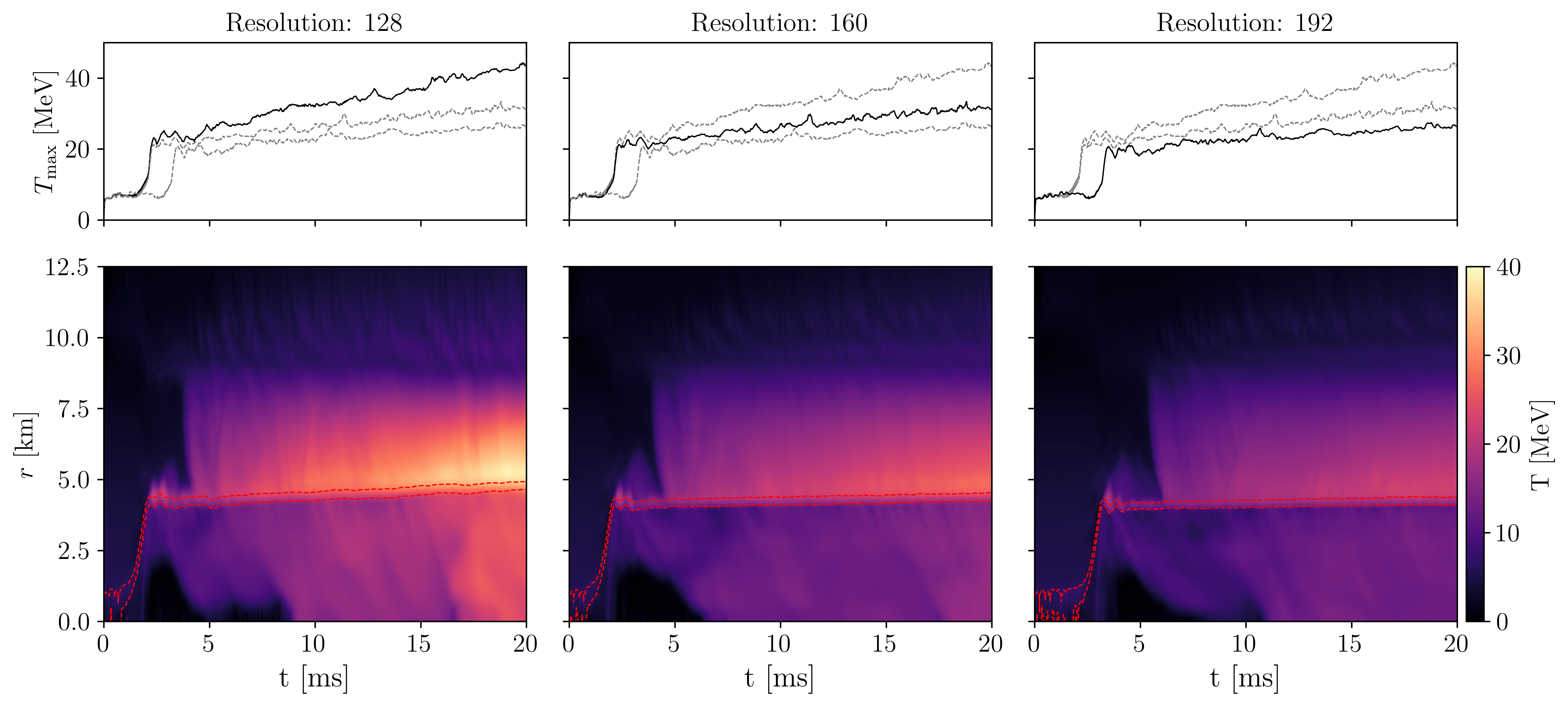}
    \caption{The time evolution of the maximum temperature (top) and the angle-averaged temperature at different radii (bottom). From left to right the different panels correspond to increasing resolution using 128, 160 and 192 grid points per level in each direction, corresponding to spatial resolutions of $\sim0.15$~km, $\sim0.12$~km and $\sim0.1$~km, respectively. The dashed red lines in the bottom panels denote the boundary between the lower and upper boundaries of the phase transition.}
    \label{fig:3D_heating}
\end{figure*}

During the hybrid phase of our simulations we observe that a localized increase in temperature develops near the sharp phase boundary. We attribute this feature primarily to numerical diffusion across the discontinuity, which is broadened over a finite number of grid cells by the shock-capturing scheme. The associated interpolation of conserved states and subsequent primitive-variable recovery can generate an effective thermal contribution. We therefore regard the resulting heating as numerically induced rather than as physical heating associated with the phase transition. This feature reduces as the numerical resolution of the grid is increased, as shown in Fig.~\ref{fig:3D_heating}, where the top panels show the time evolution of the maximum temperature in the NS. Here we performed simulations using the BBKF 2.1 EoS without neutrino transport. The initial temperature increase is constant across all resolutions and can be attributed to the relaxation of the star after the initial conditions. A second jump that also remains relatively constant throughout various resolutions is created during the transition of the hadronic star into a hybrid star. After that, however, the gradual increase in the maximum temperature can be attributed to this spurious heating, which diminishes with higher resolution. The bottom panels show the angle-averaged temperature as a function of time and radial coordinate, with the red dashed lines denoting the boundary between the hadronic and quark phase. These figures show that the heating first appears at the boundary layer and from there it diffuses outwards. Based on these results we use the maximum resolution of 192 points per level in our simulations with neutrino transport, and even then, only evolve the NSs for up to 10 ms.

%\nocite{*}

\bibliography{Bibliography}{}

\begin{thebibliography}{147}%
\makeatletter
\providecommand \@ifxundefined [1]{%
 \@ifx{#1\undefined}
}%
\providecommand \@ifnum [1]{%
 \ifnum #1\expandafter \@firstoftwo
 \else \expandafter \@secondoftwo
 \fi
}%
\providecommand \@ifx [1]{%
 \ifx #1\expandafter \@firstoftwo
 \else \expandafter \@secondoftwo
 \fi
}%
\providecommand \natexlab [1]{#1}%
\providecommand \enquote  [1]{``#1''}%
\providecommand \bibnamefont  [1]{#1}%
\providecommand \bibfnamefont [1]{#1}%
\providecommand \citenamefont [1]{#1}%
\providecommand \href@noop [0]{\@secondoftwo}%
\providecommand \href [0]{\begingroup \@sanitize@url \@href}%
\providecommand \@href[1]{\@@startlink{#1}\@@href}%
\providecommand \@@href[1]{\endgroup#1\@@endlink}%
\providecommand \@sanitize@url [0]{\catcode `\\12\catcode `\$12\catcode `\&12\catcode `\#12\catcode `\^12\catcode `\_12\catcode `\%12\relax}%
\providecommand \@@startlink[1]{}%
\providecommand \@@endlink[0]{}%
\providecommand \url  [0]{\begingroup\@sanitize@url \@url }%
\providecommand \@url [1]{\endgroup\@href {#1}{\urlprefix }}%
\providecommand \urlprefix  [0]{URL }%
\providecommand \Eprint [0]{\href }%
\providecommand \doibase [0]{https://doi.org/}%
\providecommand \selectlanguage [0]{\@gobble}%
\providecommand \bibinfo  [0]{\@secondoftwo}%
\providecommand \bibfield  [0]{\@secondoftwo}%
\providecommand \translation [1]{[#1]}%
\providecommand \BibitemOpen [0]{}%
\providecommand \bibitemStop [0]{}%
\providecommand \bibitemNoStop [0]{.\EOS\space}%
\providecommand \EOS [0]{\spacefactor3000\relax}%
\providecommand \BibitemShut  [1]{\csname bibitem#1\endcsname}%
\let\auto@bib@innerbib\@empty
%</preamble>
\bibitem [{\citenamefont {Akmal}\ \emph {et~al.}(1998)\citenamefont {Akmal}, \citenamefont {Pandharipande},\ and\ \citenamefont {Ravenhall}}]{Akmal1998}%
  \BibitemOpen
  \bibfield  {author} {\bibinfo {author} {\bibfnamefont {A.}~\bibnamefont {Akmal}}, \bibinfo {author} {\bibfnamefont {V.~R.}\ \bibnamefont {Pandharipande}},\ and\ \bibinfo {author} {\bibfnamefont {D.~G.}\ \bibnamefont {Ravenhall}},\ }\bibfield  {title} {\bibinfo {title} {{The Equation of state of nucleon matter and neutron star structure}},\ }\href {https://doi.org/10.1103/PhysRevC.58.1804} {\bibfield  {journal} {\bibinfo  {journal} {Phys. Rev. C}\ }\textbf {\bibinfo {volume} {58}},\ \bibinfo {pages} {1804} (\bibinfo {year} {1998})},\ \Eprint {https://arxiv.org/abs/nucl-th/9804027} {arXiv:nucl-th/9804027} \BibitemShut {NoStop}%
\bibitem [{\citenamefont {Wiringa}\ \emph {et~al.}(1988)\citenamefont {Wiringa}, \citenamefont {Fiks},\ and\ \citenamefont {Fabrocini}}]{Wiringa1988}%
  \BibitemOpen
  \bibfield  {author} {\bibinfo {author} {\bibfnamefont {R.~B.}\ \bibnamefont {Wiringa}}, \bibinfo {author} {\bibfnamefont {V.}~\bibnamefont {Fiks}},\ and\ \bibinfo {author} {\bibfnamefont {A.}~\bibnamefont {Fabrocini}},\ }\bibfield  {title} {\bibinfo {title} {{Equation of state for dense nucleon matter}},\ }\href {https://doi.org/10.1103/PhysRevC.38.1010} {\bibfield  {journal} {\bibinfo  {journal} {Phys. Rev. C}\ }\textbf {\bibinfo {volume} {38}},\ \bibinfo {pages} {1010} (\bibinfo {year} {1988})}\BibitemShut {NoStop}%
\bibitem [{\citenamefont {Horowitz}\ \emph {et~al.}(2001)\citenamefont {Horowitz}, \citenamefont {Pollock}, \citenamefont {Souder},\ and\ \citenamefont {Michaels}}]{Horowitz:1999fk}%
  \BibitemOpen
  \bibfield  {author} {\bibinfo {author} {\bibfnamefont {C.~J.}\ \bibnamefont {Horowitz}}, \bibinfo {author} {\bibfnamefont {S.~J.}\ \bibnamefont {Pollock}}, \bibinfo {author} {\bibfnamefont {P.~A.}\ \bibnamefont {Souder}},\ and\ \bibinfo {author} {\bibfnamefont {R.}~\bibnamefont {Michaels}},\ }\bibfield  {title} {\bibinfo {title} {{Parity violating measurements of neutron densities}},\ }\href {https://doi.org/10.1103/PhysRevC.63.025501} {\bibfield  {journal} {\bibinfo  {journal} {Phys. Rev. C}\ }\textbf {\bibinfo {volume} {63}},\ \bibinfo {pages} {025501} (\bibinfo {year} {2001})},\ \Eprint {https://arxiv.org/abs/nucl-th/9912038} {arXiv:nucl-th/9912038} \BibitemShut {NoStop}%
\bibitem [{\citenamefont {Garg}\ and\ \citenamefont {Col{\`o}}(2018)}]{Garg:2018uam}%
  \BibitemOpen
  \bibfield  {author} {\bibinfo {author} {\bibfnamefont {U.}~\bibnamefont {Garg}}\ and\ \bibinfo {author} {\bibfnamefont {G.}~\bibnamefont {Col{\`o}}},\ }\bibfield  {title} {\bibinfo {title} {{The compression-mode giant resonances and nuclear incompressibility}},\ }\href {https://doi.org/10.1016/j.ppnp.2018.03.001} {\bibfield  {journal} {\bibinfo  {journal} {Prog. Part. Nucl. Phys.}\ }\textbf {\bibinfo {volume} {101}},\ \bibinfo {pages} {55} (\bibinfo {year} {2018})},\ \Eprint {https://arxiv.org/abs/1801.03672} {arXiv:1801.03672 [nucl-ex]} \BibitemShut {NoStop}%
\bibitem [{\citenamefont {Thiel}\ \emph {et~al.}(2019)\citenamefont {Thiel}, \citenamefont {Sfienti}, \citenamefont {Piekarewicz}, \citenamefont {Horowitz},\ and\ \citenamefont {Vanderhaeghen}}]{Thiel:2019tkm}%
  \BibitemOpen
  \bibfield  {author} {\bibinfo {author} {\bibfnamefont {M.}~\bibnamefont {Thiel}}, \bibinfo {author} {\bibfnamefont {C.}~\bibnamefont {Sfienti}}, \bibinfo {author} {\bibfnamefont {J.}~\bibnamefont {Piekarewicz}}, \bibinfo {author} {\bibfnamefont {C.~J.}\ \bibnamefont {Horowitz}},\ and\ \bibinfo {author} {\bibfnamefont {M.}~\bibnamefont {Vanderhaeghen}},\ }\bibfield  {title} {\bibinfo {title} {{Neutron skins of atomic nuclei: per aspera ad astra}},\ }\href {https://doi.org/10.1088/1361-6471/ab2c6d} {\bibfield  {journal} {\bibinfo  {journal} {J. Phys. G}\ }\textbf {\bibinfo {volume} {46}},\ \bibinfo {pages} {093003} (\bibinfo {year} {2019})},\ \Eprint {https://arxiv.org/abs/1904.12269} {arXiv:1904.12269 [nucl-ex]} \BibitemShut {NoStop}%
\bibitem [{\citenamefont {Adhikari}\ \emph {et~al.}(2021)\citenamefont {Adhikari} \emph {et~al.}}]{PREX:2021umo}%
  \BibitemOpen
  \bibfield  {author} {\bibinfo {author} {\bibfnamefont {D.}~\bibnamefont {Adhikari}} \emph {et~al.} (\bibinfo {collaboration} {PREX}),\ }\bibfield  {title} {\bibinfo {title} {{Accurate Determination of the Neutron Skin Thickness of $^{208}$Pb through Parity-Violation in Electron Scattering}},\ }\href {https://doi.org/10.1103/PhysRevLett.126.172502} {\bibfield  {journal} {\bibinfo  {journal} {Phys. Rev. Lett.}\ }\textbf {\bibinfo {volume} {126}},\ \bibinfo {pages} {172502} (\bibinfo {year} {2021})},\ \Eprint {https://arxiv.org/abs/2102.10767} {arXiv:2102.10767 [nucl-ex]} \BibitemShut {NoStop}%
\bibitem [{\citenamefont {Grelli}(2018)}]{ALICE}%
  \BibitemOpen
  \bibfield  {author} {\bibinfo {author} {\bibfnamefont {A.}~\bibnamefont {Grelli}} (\bibinfo {collaboration} {ALICE}),\ }\bibfield  {title} {\bibinfo {title} {{ALICE Overview}},\ }\bibfield  {booktitle} {\emph {\bibinfo {booktitle} {{Proceedings, 17th International Conference on Strangeness in Quark Matter (SQM 2017): Utrecht, The Netherlands, July 10-15, 2017}}},\ }\href {https://doi.org/10.1051/epjconf/201817101005} {\bibfield  {journal} {\bibinfo  {journal} {EPJ Web Conf.}\ }\textbf {\bibinfo {volume} {17}},\ \bibinfo {pages} {10100} (\bibinfo {year} {2018})}\BibitemShut {NoStop}%
%%CITATION = ARXIV:1711.03369;%%
\bibitem [{\citenamefont {Sakaguchi}(2019)}]{PHENIX}%
  \BibitemOpen
  \bibfield  {author} {\bibinfo {author} {\bibfnamefont {T.}~\bibnamefont {Sakaguchi}} (\bibinfo {collaboration} {PHENIX}),\ }\bibfield  {title} {\bibinfo {title} {{Overview of latest results from PHENIX}},\ }\bibfield  {booktitle} {\emph {\bibinfo {booktitle} {{Proceedings, 9th International Conference on Hard and Electromagnetic Probes of High-Energy Nuclear Collisions: Hard Probes 2018 (HP2018): Aix-Les-Bains, France, October 1-5, 2018}}},\ }\href {https://doi.org/10.22323/1.345.0035} {\bibfield  {journal} {\bibinfo  {journal} {PoS}\ }\textbf {\bibinfo {volume} {HardProbes2018}},\ \bibinfo {pages} {035} (\bibinfo {year} {2019})}\BibitemShut {NoStop}%
%%CITATION = ARXIV:1901.01951;%%
\bibitem [{\citenamefont {Tlusty}(2018)}]{STAR}%
  \BibitemOpen
  \bibfield  {author} {\bibinfo {author} {\bibfnamefont {D.}~\bibnamefont {Tlusty}},\ }\bibfield  {title} {\bibinfo {title} {{The RHIC Beam Energy Scan Phase II: Physics and Upgrades}},\ }in\ \href@noop {} {\emph {\bibinfo {booktitle} {{13th Conference on the Intersections of Particle and Nuclear Physics (CIPANP 2018) Palm Springs, California, USA, May 29-June 3}}}}\ (\bibinfo {year} {2018})\BibitemShut {NoStop}%
%%CITATION = ARXIV:1810.04767;%%
\bibitem [{\citenamefont {Larsen}(2018)}]{NA61}%
  \BibitemOpen
  \bibfield  {author} {\bibinfo {author} {\bibfnamefont {D.~T.}\ \bibnamefont {Larsen}} (\bibinfo {collaboration} {NA61/SHINE}),\ }\bibfield  {title} {\bibinfo {title} {{Latest Results from the NA61/SHINE Experiment}},\ }\bibfield  {booktitle} {\emph {\bibinfo {booktitle} {{Proceedings, 3rd International Conference on Particle Physics and Astrophysics (ICPPA 2017): Moscow, Russia, October 2-5, 2017}}},\ }\href {https://doi.org/10.18502/ken.v3i1.1742} {\bibfield  {journal} {\bibinfo  {journal} {KnE Energ. Phys.}\ }\textbf {\bibinfo {volume} {3}},\ \bibinfo {pages} {188} (\bibinfo {year} {2018})}\BibitemShut {NoStop}%
%%CITATION = ARXIV:1810.02775;%%
\bibitem [{\citenamefont {Ablyazimov}\ \emph {et~al.}(2017)\citenamefont {Ablyazimov} \emph {et~al.}}]{CBM}%
  \BibitemOpen
  \bibfield  {author} {\bibinfo {author} {\bibfnamefont {T.}~\bibnamefont {Ablyazimov}} \emph {et~al.} (\bibinfo {collaboration} {CBM}),\ }\bibfield  {title} {\bibinfo {title} {{Challenges in QCD matter physics --The scientific programme of the Compressed Baryonic Matter experiment at FAIR}},\ }\href {https://doi.org/10.1140/epja/i2017-12248-y} {\bibfield  {journal} {\bibinfo  {journal} {Eur. Phys. J.}\ }\textbf {\bibinfo {volume} {A53}},\ \bibinfo {pages} {60} (\bibinfo {year} {2017})}\BibitemShut {NoStop}%
%%CITATION = ARXIV:1607.01487;%%
\bibitem [{\citenamefont {Kekelidze}(2017)}]{NICA}%
  \BibitemOpen
  \bibfield  {author} {\bibinfo {author} {\bibfnamefont {V.~D.}\ \bibnamefont {Kekelidze}} (\bibinfo {collaboration} {NICA}),\ }\bibfield  {title} {\bibinfo {title} {{NICA project at JINR: status and prospects}},\ }\bibfield  {booktitle} {\emph {\bibinfo {booktitle} {{Proceedings, International Conference on Instrumentation for Colliding Beam Physics (INSTR17): Novosibirsk, Russia}}},\ }\href {https://doi.org/10.1088/1748-0221/12/06/C06012} {\bibfield  {journal} {\bibinfo  {journal} {JINST}\ }\textbf {\bibinfo {volume} {12}}\bibinfo  {number} { (06)},\ \bibinfo {pages} {C06012}}\BibitemShut {NoStop}%
%%CITATION = JINST,12,C06012;%%
\bibitem [{\citenamefont {Guenther}(2021)}]{Guenther2020}%
  \BibitemOpen
\bibfield  {number} {  }\bibfield  {author} {\bibinfo {author} {\bibfnamefont {J.~N.}\ \bibnamefont {Guenther}},\ }\bibfield  {title} {\bibinfo {title} {{Overview of the QCD phase diagram: Recent progress from the lattice}},\ }\href {https://doi.org/10.1140/epja/s10050-021-00354-6} {\bibfield  {journal} {\bibinfo  {journal} {Eur. Phys. J. A}\ }\textbf {\bibinfo {volume} {57}},\ \bibinfo {pages} {136} (\bibinfo {year} {2021})},\ \Eprint {https://arxiv.org/abs/2010.15503} {arXiv:2010.15503 [hep-lat]} \BibitemShut {NoStop}%
\bibitem [{\citenamefont {Attanasio}\ \emph {et~al.}(2020)\citenamefont {Attanasio}, \citenamefont {J\"ager},\ and\ \citenamefont {Ziegler}}]{Attanasio2020}%
  \BibitemOpen
  \bibfield  {author} {\bibinfo {author} {\bibfnamefont {F.}~\bibnamefont {Attanasio}}, \bibinfo {author} {\bibfnamefont {B.}~\bibnamefont {J\"ager}},\ and\ \bibinfo {author} {\bibfnamefont {F.~P.~G.}\ \bibnamefont {Ziegler}},\ }\bibfield  {title} {\bibinfo {title} {{Complex Langevin simulations and the QCD phase diagram: Recent developments}},\ }\href {https://doi.org/10.1140/epja/s10050-020-00256-z} {\bibfield  {journal} {\bibinfo  {journal} {Eur. Phys. J. A}\ }\textbf {\bibinfo {volume} {56}},\ \bibinfo {pages} {251} (\bibinfo {year} {2020})},\ \Eprint {https://arxiv.org/abs/2006.00476} {arXiv:2006.00476 [hep-lat]} \BibitemShut {NoStop}%
\bibitem [{\citenamefont {Bors\'anyi}\ \emph {et~al.}(2021)\citenamefont {Bors\'anyi}, \citenamefont {Fodor}, \citenamefont {Guenther}, \citenamefont {Kara}, \citenamefont {Katz}, \citenamefont {Parotto}, \citenamefont {P\'asztor}, \citenamefont {Ratti},\ and\ \citenamefont {Szab\'o}}]{Borsanyi2021}%
  \BibitemOpen
  \bibfield  {author} {\bibinfo {author} {\bibfnamefont {S.}~\bibnamefont {Bors\'anyi}}, \bibinfo {author} {\bibfnamefont {Z.}~\bibnamefont {Fodor}}, \bibinfo {author} {\bibfnamefont {J.~N.}\ \bibnamefont {Guenther}}, \bibinfo {author} {\bibfnamefont {R.}~\bibnamefont {Kara}}, \bibinfo {author} {\bibfnamefont {S.~D.}\ \bibnamefont {Katz}}, \bibinfo {author} {\bibfnamefont {P.}~\bibnamefont {Parotto}}, \bibinfo {author} {\bibfnamefont {A.}~\bibnamefont {P\'asztor}}, \bibinfo {author} {\bibfnamefont {C.}~\bibnamefont {Ratti}},\ and\ \bibinfo {author} {\bibfnamefont {K.~K.}\ \bibnamefont {Szab\'o}},\ }\bibfield  {title} {\bibinfo {title} {{Lattice QCD equation of state at finite chemical potential from an alternative expansion scheme}},\ }\href {https://doi.org/10.1103/PhysRevLett.126.232001} {\bibfield  {journal} {\bibinfo  {journal} {Phys. Rev. Lett.}\ }\textbf {\bibinfo {volume} {126}},\ \bibinfo {pages} {232001} (\bibinfo {year} {2021})},\ \Eprint {https://arxiv.org/abs/2102.06660} {arXiv:2102.06660
  [hep-lat]} \BibitemShut {NoStop}%
\bibitem [{\citenamefont {de~Forcrand}(2009)}]{deForcrand2010}%
  \BibitemOpen
  \bibfield  {author} {\bibinfo {author} {\bibfnamefont {P.}~\bibnamefont {de~Forcrand}},\ }\bibfield  {title} {\bibinfo {title} {{Simulating QCD at finite density}},\ }\href {https://doi.org/10.22323/1.091.0010} {\bibfield  {journal} {\bibinfo  {journal} {PoS}\ }\textbf {\bibinfo {volume} {LAT2009}},\ \bibinfo {pages} {010} (\bibinfo {year} {2009})}\BibitemShut {NoStop}%
\bibitem [{\citenamefont {Demorest}\ \emph {et~al.}(2010)\citenamefont {Demorest}, \citenamefont {Pennucci}, \citenamefont {Ransom}, \citenamefont {Roberts},\ and\ \citenamefont {Hessels}}]{Demorest:2010bx}%
  \BibitemOpen
  \bibfield  {author} {\bibinfo {author} {\bibfnamefont {P.}~\bibnamefont {Demorest}}, \bibinfo {author} {\bibfnamefont {T.}~\bibnamefont {Pennucci}}, \bibinfo {author} {\bibfnamefont {S.}~\bibnamefont {Ransom}}, \bibinfo {author} {\bibfnamefont {M.}~\bibnamefont {Roberts}},\ and\ \bibinfo {author} {\bibfnamefont {J.}~\bibnamefont {Hessels}},\ }\bibfield  {title} {\bibinfo {title} {{Shapiro Delay Measurement of A Two Solar Mass Neutron Star}},\ }\href {https://doi.org/10.1038/nature09466} {\bibfield  {journal} {\bibinfo  {journal} {Nature}\ }\textbf {\bibinfo {volume} {467}},\ \bibinfo {pages} {1081} (\bibinfo {year} {2010})},\ \Eprint {https://arxiv.org/abs/1010.5788} {arXiv:1010.5788 [astro-ph.HE]} \BibitemShut {NoStop}%
%%CITATION = ARXIV:1010.5788;%%
\bibitem [{\citenamefont {Antoniadis}\ \emph {et~al.}(2013)\citenamefont {Antoniadis}, \citenamefont {Freire}, \citenamefont {Wex}, \citenamefont {Tauris}, \citenamefont {Lynch}, \citenamefont {van Kerkwijk}, \citenamefont {Kramer}, \citenamefont {Bassa}, \citenamefont {Dhillon}, \citenamefont {Driebe}, \citenamefont {Hessels}, \citenamefont {Kaspi}, \citenamefont {Kondratiev}, \citenamefont {Langer}, \citenamefont {Marsh}, \citenamefont {McLaughlin}, \citenamefont {Pennucci}, \citenamefont {Ransom}, \citenamefont {Stairs}, \citenamefont {van Leeuwen}, \citenamefont {Verbiest},\ and\ \citenamefont {Whelan}}]{Antoniadis:2013pzd}%
  \BibitemOpen
  \bibfield  {author} {\bibinfo {author} {\bibfnamefont {J.}~\bibnamefont {Antoniadis}}, \bibinfo {author} {\bibfnamefont {P.~C.}\ \bibnamefont {Freire}}, \bibinfo {author} {\bibfnamefont {N.}~\bibnamefont {Wex}}, \bibinfo {author} {\bibfnamefont {T.~M.}\ \bibnamefont {Tauris}}, \bibinfo {author} {\bibfnamefont {R.~S.}\ \bibnamefont {Lynch}}, \bibinfo {author} {\bibfnamefont {M.~H.}\ \bibnamefont {van Kerkwijk}}, \bibinfo {author} {\bibfnamefont {M.}~\bibnamefont {Kramer}}, \bibinfo {author} {\bibfnamefont {C.}~\bibnamefont {Bassa}}, \bibinfo {author} {\bibfnamefont {V.~S.}\ \bibnamefont {Dhillon}}, \bibinfo {author} {\bibfnamefont {T.}~\bibnamefont {Driebe}}, \bibinfo {author} {\bibfnamefont {J.~W.~T.}\ \bibnamefont {Hessels}}, \bibinfo {author} {\bibfnamefont {V.~M.}\ \bibnamefont {Kaspi}}, \bibinfo {author} {\bibfnamefont {V.~I.}\ \bibnamefont {Kondratiev}}, \bibinfo {author} {\bibfnamefont {N.}~\bibnamefont {Langer}}, \bibinfo {author} {\bibfnamefont {T.~R.}\ \bibnamefont {Marsh}}, \bibinfo {author}
  {\bibfnamefont {M.~A.}\ \bibnamefont {McLaughlin}}, \bibinfo {author} {\bibfnamefont {T.~T.}\ \bibnamefont {Pennucci}}, \bibinfo {author} {\bibfnamefont {S.~M.}\ \bibnamefont {Ransom}}, \bibinfo {author} {\bibfnamefont {I.~H.}\ \bibnamefont {Stairs}}, \bibinfo {author} {\bibfnamefont {J.}~\bibnamefont {van Leeuwen}}, \bibinfo {author} {\bibfnamefont {J.~P.~W.}\ \bibnamefont {Verbiest}},\ and\ \bibinfo {author} {\bibfnamefont {D.~G.}\ \bibnamefont {Whelan}},\ }\bibfield  {title} {\bibinfo {title} {{A Massive Pulsar in a Compact Relativistic Binary}},\ }\href {https://doi.org/10.1126/science.1233232} {\bibfield  {journal} {\bibinfo  {journal} {Science}\ }\textbf {\bibinfo {volume} {340}},\ \bibinfo {pages} {6131} (\bibinfo {year} {2013})},\ \Eprint {https://arxiv.org/abs/1304.6875} {arXiv:1304.6875 [astro-ph.HE]} \BibitemShut {NoStop}%
%%CITATION = ARXIV:1304.6875;%%
\bibitem [{\citenamefont {Arzoumanian}\ \emph {et~al.}(2018)\citenamefont {Arzoumanian} \emph {et~al.}}]{NANOGrav:2017wvv}%
  \BibitemOpen
  \bibfield  {author} {\bibinfo {author} {\bibfnamefont {Z.}~\bibnamefont {Arzoumanian}} \emph {et~al.} (\bibinfo {collaboration} {NANOGrav}),\ }\bibfield  {title} {\bibinfo {title} {{The NANOGrav 11-year Data Set: High-precision timing of 45 Millisecond Pulsars}},\ }\href {https://doi.org/10.3847/1538-4365/aab5b0} {\bibfield  {journal} {\bibinfo  {journal} {Astrophys. J. Suppl.}\ }\textbf {\bibinfo {volume} {235}},\ \bibinfo {pages} {37} (\bibinfo {year} {2018})},\ \Eprint {https://arxiv.org/abs/1801.01837} {arXiv:1801.01837 [astro-ph.HE]} \BibitemShut {NoStop}%
\bibitem [{\citenamefont {Cromartie}\ \emph {et~al.}(2019)\citenamefont {Cromartie}, \citenamefont {E.~Fonseca},\ and\ \citenamefont {P.~B.~Demorest}}]{Cromartie:2019kug}%
  \BibitemOpen
  \bibfield  {author} {\bibinfo {author} {\bibfnamefont {H.~T.}\ \bibnamefont {Cromartie}}, \bibinfo {author} {\bibfnamefont {S.~M.~R.}\ \bibnamefont {E.~Fonseca}},\ and\ \bibinfo {author} {\bibfnamefont {e.~a.}\ \bibnamefont {P.~B.~Demorest}},\ }\bibfield  {title} {\bibinfo {title} {{Relativistic Shapiro delay measurements of an extremely massive millisecond pulsar}},\ }\href {https://doi.org/10.1038/s41550-019-0880-2} {\bibfield  {journal} {\bibinfo  {journal} {Nat. Astron.}\ }\textbf {\bibinfo {volume} {4}},\ \bibinfo {pages} {72} (\bibinfo {year} {2019})},\ \Eprint {https://arxiv.org/abs/1904.06759} {arXiv:1904.06759 [astro-ph.HE]} \BibitemShut {NoStop}%
\bibitem [{\citenamefont {Fonseca}\ \emph {et~al.}(2021)\citenamefont {Fonseca} \emph {et~al.}}]{Fonseca:2021wxt}%
  \BibitemOpen
  \bibfield  {author} {\bibinfo {author} {\bibfnamefont {E.}~\bibnamefont {Fonseca}} \emph {et~al.},\ }\bibfield  {title} {\bibinfo {title} {{Refined Mass and Geometric Measurements of the High-mass PSR J0740+6620}},\ }\href {https://doi.org/10.3847/2041-8213/ac03b8} {\bibfield  {journal} {\bibinfo  {journal} {Astrophys. J. Lett.}\ }\textbf {\bibinfo {volume} {915}},\ \bibinfo {pages} {L12} (\bibinfo {year} {2021})},\ \Eprint {https://arxiv.org/abs/2104.00880} {arXiv:2104.00880 [astro-ph.HE]} \BibitemShut {NoStop}%
\bibitem [{\citenamefont {Romani}\ \emph {et~al.}(2022)\citenamefont {Romani}, \citenamefont {Kandel}, \citenamefont {Filippenko}, \citenamefont {Brink},\ and\ \citenamefont {Zheng}}]{Romani:2022jhd}%
  \BibitemOpen
  \bibfield  {author} {\bibinfo {author} {\bibfnamefont {R.~W.}\ \bibnamefont {Romani}}, \bibinfo {author} {\bibfnamefont {D.}~\bibnamefont {Kandel}}, \bibinfo {author} {\bibfnamefont {A.~V.}\ \bibnamefont {Filippenko}}, \bibinfo {author} {\bibfnamefont {T.~G.}\ \bibnamefont {Brink}},\ and\ \bibinfo {author} {\bibfnamefont {W.}~\bibnamefont {Zheng}},\ }\bibfield  {title} {\bibinfo {title} {{PSR J0952\ensuremath{-}0607: The Fastest and Heaviest Known Galactic Neutron Star}},\ }\href {https://doi.org/10.3847/2041-8213/ac8007} {\bibfield  {journal} {\bibinfo  {journal} {Astrophys. J. Lett.}\ }\textbf {\bibinfo {volume} {934}},\ \bibinfo {pages} {L18} (\bibinfo {year} {2022})},\ \Eprint {https://arxiv.org/abs/2207.05124} {arXiv:2207.05124 [astro-ph.HE]} \BibitemShut {NoStop}%
\bibitem [{\citenamefont {Saffer}\ \emph {et~al.}(2025)\citenamefont {Saffer} \emph {et~al.}}]{Saffer:2024tlb}%
  \BibitemOpen
  \bibfield  {author} {\bibinfo {author} {\bibfnamefont {A.}~\bibnamefont {Saffer}} \emph {et~al.},\ }\bibfield  {title} {\bibinfo {title} {{A Lower Mass Estimate for PSR J0348+0432 Based on CHIME/Pulsar Precision Timing}},\ }\href {https://doi.org/10.3847/2041-8213/adc25e} {\bibfield  {journal} {\bibinfo  {journal} {Astrophys. J. Lett.}\ }\textbf {\bibinfo {volume} {983}},\ \bibinfo {pages} {L20} (\bibinfo {year} {2025})},\ \Eprint {https://arxiv.org/abs/2412.02850} {arXiv:2412.02850 [astro-ph.HE]} \BibitemShut {NoStop}%
\bibitem [{\citenamefont {Romani}\ \emph {et~al.}(2026)\citenamefont {Romani}, \citenamefont {Beleznay}, \citenamefont {Filippenko}, \citenamefont {Brink},\ and\ \citenamefont {Zheng}}]{Romani:2025ytn}%
  \BibitemOpen
  \bibfield  {author} {\bibinfo {author} {\bibfnamefont {R.~W.}\ \bibnamefont {Romani}}, \bibinfo {author} {\bibfnamefont {M.}~\bibnamefont {Beleznay}}, \bibinfo {author} {\bibfnamefont {A.~V.}\ \bibnamefont {Filippenko}}, \bibinfo {author} {\bibfnamefont {T.~G.}\ \bibnamefont {Brink}},\ and\ \bibinfo {author} {\bibfnamefont {W.}~\bibnamefont {Zheng}},\ }\bibfield  {title} {\bibinfo {title} {{PSR J0952-0607: Tightening a Record-high Neutron Star Mass}},\ }\href {https://doi.org/10.3847/1538-4357/ae28c5} {\bibfield  {journal} {\bibinfo  {journal} {Astrophys. J.}\ }\textbf {\bibinfo {volume} {996}},\ \bibinfo {pages} {101} (\bibinfo {year} {2026})},\ \Eprint {https://arxiv.org/abs/2512.05099} {arXiv:2512.05099 [astro-ph.HE]} \BibitemShut {NoStop}%
\bibitem [{\citenamefont {Abbott}\ \emph {et~al.}(2017{\natexlab{a}})\citenamefont {Abbott} \emph {et~al.}}]{LIGOScientific:2017vwq}%
  \BibitemOpen
  \bibfield  {author} {\bibinfo {author} {\bibfnamefont {B.~P.}\ \bibnamefont {Abbott}} \emph {et~al.} (\bibinfo {collaboration} {LIGO Scientific, Virgo}),\ }\bibfield  {title} {\bibinfo {title} {{GW170817: Observation of Gravitational Waves from a Binary Neutron Star Inspiral}},\ }\href {https://doi.org/10.1103/PhysRevLett.119.161101} {\bibfield  {journal} {\bibinfo  {journal} {Phys. Rev. Lett.}\ }\textbf {\bibinfo {volume} {119}},\ \bibinfo {pages} {161101} (\bibinfo {year} {2017}{\natexlab{a}})},\ \Eprint {https://arxiv.org/abs/1710.05832} {arXiv:1710.05832 [gr-qc]} \BibitemShut {NoStop}%
\bibitem [{\citenamefont {Abbott}\ \emph {et~al.}(2018)\citenamefont {Abbott} \emph {et~al.}}]{LIGOScientific:2018cki}%
  \BibitemOpen
  \bibfield  {author} {\bibinfo {author} {\bibfnamefont {B.~P.}\ \bibnamefont {Abbott}} \emph {et~al.} (\bibinfo {collaboration} {LIGO Scientific, Virgo}),\ }\bibfield  {title} {\bibinfo {title} {{GW170817: Measurements of neutron star radii and equation of state}},\ }\href {https://doi.org/10.1103/PhysRevLett.121.161101} {\bibfield  {journal} {\bibinfo  {journal} {Phys. Rev. Lett.}\ }\textbf {\bibinfo {volume} {121}},\ \bibinfo {pages} {161101} (\bibinfo {year} {2018})},\ \Eprint {https://arxiv.org/abs/1805.11581} {arXiv:1805.11581 [gr-qc]} \BibitemShut {NoStop}%
\bibitem [{\citenamefont {Abbott}\ \emph {et~al.}(2019{\natexlab{a}})\citenamefont {Abbott} \emph {et~al.}}]{LIGOScientific:2018hze}%
  \BibitemOpen
  \bibfield  {author} {\bibinfo {author} {\bibfnamefont {B.~P.}\ \bibnamefont {Abbott}} \emph {et~al.} (\bibinfo {collaboration} {LIGO Scientific, Virgo}),\ }\bibfield  {title} {\bibinfo {title} {{Properties of the binary neutron star merger GW170817}},\ }\href {https://doi.org/10.1103/PhysRevX.9.011001} {\bibfield  {journal} {\bibinfo  {journal} {Phys. Rev. X}\ }\textbf {\bibinfo {volume} {9}},\ \bibinfo {pages} {011001} (\bibinfo {year} {2019}{\natexlab{a}})},\ \Eprint {https://arxiv.org/abs/1805.11579} {arXiv:1805.11579 [gr-qc]} \BibitemShut {NoStop}%
\bibitem [{\citenamefont {Abbott}\ \emph {et~al.}(2017{\natexlab{b}})\citenamefont {Abbott} \emph {et~al.}}]{LIGOScientific:2017ync}%
  \BibitemOpen
  \bibfield  {author} {\bibinfo {author} {\bibfnamefont {B.~P.}\ \bibnamefont {Abbott}} \emph {et~al.} (\bibinfo {collaboration} {LIGO Scientific, Virgo, Fermi GBM, INTEGRAL, IceCube, AstroSat Cadmium Zinc Telluride Imager Team, IPN, Insight-Hxmt, ANTARES, Swift, AGILE Team, 1M2H Team, Dark Energy Camera GW-EM, DES, DLT40, GRAWITA, Fermi-LAT, ATCA, ASKAP, Las Cumbres Observatory Group, OzGrav, DWF (Deeper Wider Faster Program), AST3, CAASTRO, VINROUGE, MASTER, J-GEM, GROWTH, JAGWAR, CaltechNRAO, TTU-NRAO, NuSTAR, Pan-STARRS, MAXI Team, TZAC Consortium, KU, Nordic Optical Telescope, ePESSTO, GROND, Texas Tech University, SALT Group, TOROS, BOOTES, MWA, CALET, IKI-GW Follow-up, H.E.S.S., LOFAR, LWA, HAWC, Pierre Auger, ALMA, Euro VLBI Team, Pi of Sky, Chandra Team at McGill University, DFN, ATLAS Telescopes, High Time Resolution Universe Survey, RIMAS, RATIR, SKA South Africa/MeerKAT}),\ }\bibfield  {title} {\bibinfo {title} {{Multi-messenger Observations of a Binary Neutron Star Merger}},\ }\href
  {https://doi.org/10.3847/2041-8213/aa91c9} {\bibfield  {journal} {\bibinfo  {journal} {Astrophys. J. Lett.}\ }\textbf {\bibinfo {volume} {848}},\ \bibinfo {pages} {L12} (\bibinfo {year} {2017}{\natexlab{b}})},\ \Eprint {https://arxiv.org/abs/1710.05833} {arXiv:1710.05833 [astro-ph.HE]} \BibitemShut {NoStop}%
\bibitem [{\citenamefont {Abbott}\ \emph {et~al.}(2017{\natexlab{c}})\citenamefont {Abbott} \emph {et~al.}}]{LIGOScientific:2017zic}%
  \BibitemOpen
  \bibfield  {author} {\bibinfo {author} {\bibfnamefont {B.~P.}\ \bibnamefont {Abbott}} \emph {et~al.} (\bibinfo {collaboration} {LIGO Scientific, Virgo, Fermi-GBM, INTEGRAL}),\ }\bibfield  {title} {\bibinfo {title} {{Gravitational Waves and Gamma-rays from a Binary Neutron Star Merger: GW170817 and GRB 170817A}},\ }\href {https://doi.org/10.3847/2041-8213/aa920c} {\bibfield  {journal} {\bibinfo  {journal} {Astrophys. J. Lett.}\ }\textbf {\bibinfo {volume} {848}},\ \bibinfo {pages} {L13} (\bibinfo {year} {2017}{\natexlab{c}})},\ \Eprint {https://arxiv.org/abs/1710.05834} {arXiv:1710.05834 [astro-ph.HE]} \BibitemShut {NoStop}%
\bibitem [{\citenamefont {Miller}\ \emph {et~al.}(2019)\citenamefont {Miller}, \citenamefont {Lamb}, \citenamefont {Dittmann} \emph {et~al.}}]{Miller:2019cac}%
  \BibitemOpen
  \bibfield  {author} {\bibinfo {author} {\bibfnamefont {M.~C.}\ \bibnamefont {Miller}}, \bibinfo {author} {\bibfnamefont {F.~K.}\ \bibnamefont {Lamb}}, \bibinfo {author} {\bibfnamefont {A.~J.}\ \bibnamefont {Dittmann}}, \emph {et~al.},\ }\bibfield  {title} {\bibinfo {title} {{PSR J0030+0451 Mass and Radius from NICER Data and Implications for the Properties of Neutron Star Matter}},\ }\href@noop {} {\bibfield  {journal} {\bibinfo  {journal} {Astrophys. J. Lett.}\ }\textbf {\bibinfo {volume} {887}},\ \bibinfo {pages} {L24} (\bibinfo {year} {2019})},\ \Eprint {https://arxiv.org/abs/1912.05705} {arXiv:1912.05705 [astro-ph.HE]} \BibitemShut {NoStop}%
%%CITATION = ARXIV:1912.05705;%%
\bibitem [{\citenamefont {Riley}\ \emph {et~al.}(2019)\citenamefont {Riley}, \citenamefont {Watts}, \citenamefont {Bogdanov} \emph {et~al.}}]{Riley:2019yda}%
  \BibitemOpen
  \bibfield  {author} {\bibinfo {author} {\bibfnamefont {T.~E.}\ \bibnamefont {Riley}}, \bibinfo {author} {\bibfnamefont {A.~L.}\ \bibnamefont {Watts}}, \bibinfo {author} {\bibfnamefont {S.}~\bibnamefont {Bogdanov}}, \emph {et~al.},\ }\bibfield  {title} {\bibinfo {title} {{A NICER View of PSR J0030+0451: Millisecond Pulsar Parameter Estimation}},\ }\href {https://doi.org/10.3847/2041-8213/ab481c} {\bibfield  {journal} {\bibinfo  {journal} {Astrophys. J. Lett.}\ }\textbf {\bibinfo {volume} {887}},\ \bibinfo {pages} {L21} (\bibinfo {year} {2019})},\ \Eprint {https://arxiv.org/abs/1912.05702} {arXiv:1912.05702 [astro-ph.HE]} \BibitemShut {NoStop}%
%%CITATION = ARXIV:1912.05702;%%
\bibitem [{\citenamefont {Raaijmakers}\ \emph {et~al.}(2019)\citenamefont {Raaijmakers}, \citenamefont {Riley}, \citenamefont {Watts} \emph {et~al.}}]{Raaijmakers:2019qny}%
  \BibitemOpen
  \bibfield  {author} {\bibinfo {author} {\bibfnamefont {G.}~\bibnamefont {Raaijmakers}}, \bibinfo {author} {\bibfnamefont {T.~E.}\ \bibnamefont {Riley}}, \bibinfo {author} {\bibfnamefont {A.~L.}\ \bibnamefont {Watts}}, \emph {et~al.},\ }\bibfield  {title} {\bibinfo {title} {{A NICER view of PSR J0030+0451: Implications for the dense matter equation of state}},\ }\href {https://doi.org/10.3847/2041-8213/ab451a} {\bibfield  {journal} {\bibinfo  {journal} {Astrophys. J. Lett.}\ }\textbf {\bibinfo {volume} {887}},\ \bibinfo {pages} {L22} (\bibinfo {year} {2019})},\ \Eprint {https://arxiv.org/abs/1912.05703} {arXiv:1912.05703 [astro-ph.HE]} \BibitemShut {NoStop}%
%%CITATION = ARXIV:1912.05703;%%
\bibitem [{\citenamefont {Vinciguerra}(2024)}]{Vinciguerra2024}%
  \BibitemOpen
  \bibfield  {author} {\bibinfo {author} {\bibfnamefont {S.~e.~a.}\ \bibnamefont {Vinciguerra}},\ }\bibfield  {title} {\bibinfo {title} {{An Updated Mass–Radius Analysis of the 2017–2018 NICER Data Set of PSR J0030+0451}},\ }\href {https://doi.org/10.3847/1538-4357/acfb83} {\bibfield  {journal} {\bibinfo  {journal} {The Astrophysical Journal}\ }\textbf {\bibinfo {volume} {961}},\ \bibinfo {pages} {62} (\bibinfo {year} {2024})},\ \Eprint {https://arxiv.org/abs/2407.06789} {arXiv:2407.06789 [astro-ph.HE]} \BibitemShut {NoStop}%
\bibitem [{\citenamefont {Miller}\ \emph {et~al.}(2021)\citenamefont {Miller} \emph {et~al.}}]{Miller:2021qha}%
  \BibitemOpen
  \bibfield  {author} {\bibinfo {author} {\bibfnamefont {M.~C.}\ \bibnamefont {Miller}} \emph {et~al.},\ }\bibfield  {title} {\bibinfo {title} {{The Radius of PSR J0740+6620 from NICER and XMM-Newton Data}},\ }\href {https://doi.org/10.3847/2041-8213/ac089b} {\bibfield  {journal} {\bibinfo  {journal} {Astrophys. J. Lett.}\ }\textbf {\bibinfo {volume} {918}},\ \bibinfo {pages} {L28} (\bibinfo {year} {2021})},\ \Eprint {https://arxiv.org/abs/2105.06979} {arXiv:2105.06979 [astro-ph.HE]} \BibitemShut {NoStop}%
\bibitem [{\citenamefont {Riley}\ \emph {et~al.}(2021)\citenamefont {Riley} \emph {et~al.}}]{Riley:2021pdl}%
  \BibitemOpen
  \bibfield  {author} {\bibinfo {author} {\bibfnamefont {T.~E.}\ \bibnamefont {Riley}} \emph {et~al.},\ }\bibfield  {title} {\bibinfo {title} {{A NICER View of the Massive Pulsar PSR J0740+6620 Informed by Radio Timing and XMM-Newton Spectroscopy}},\ }\href {https://doi.org/10.3847/2041-8213/ac0a81} {\bibfield  {journal} {\bibinfo  {journal} {Astrophys. J. Lett.}\ }\textbf {\bibinfo {volume} {918}},\ \bibinfo {pages} {L27} (\bibinfo {year} {2021})},\ \Eprint {https://arxiv.org/abs/2105.06980} {arXiv:2105.06980 [astro-ph.HE]} \BibitemShut {NoStop}%
\bibitem [{\citenamefont {Raaijmakers}\ \emph {et~al.}(2021)\citenamefont {Raaijmakers}, \citenamefont {Greif}, \citenamefont {Hebeler}, \citenamefont {Hinderer}, \citenamefont {Nissanke}, \citenamefont {Schwenk}, \citenamefont {Riley}, \citenamefont {Watts}, \citenamefont {Lattimer},\ and\ \citenamefont {Ho}}]{Raaijmakers:2021uju}%
  \BibitemOpen
  \bibfield  {author} {\bibinfo {author} {\bibfnamefont {G.}~\bibnamefont {Raaijmakers}}, \bibinfo {author} {\bibfnamefont {S.~K.}\ \bibnamefont {Greif}}, \bibinfo {author} {\bibfnamefont {K.}~\bibnamefont {Hebeler}}, \bibinfo {author} {\bibfnamefont {T.}~\bibnamefont {Hinderer}}, \bibinfo {author} {\bibfnamefont {S.}~\bibnamefont {Nissanke}}, \bibinfo {author} {\bibfnamefont {A.}~\bibnamefont {Schwenk}}, \bibinfo {author} {\bibfnamefont {T.~E.}\ \bibnamefont {Riley}}, \bibinfo {author} {\bibfnamefont {A.~L.}\ \bibnamefont {Watts}}, \bibinfo {author} {\bibfnamefont {J.~M.}\ \bibnamefont {Lattimer}},\ and\ \bibinfo {author} {\bibfnamefont {W.~C.~G.}\ \bibnamefont {Ho}},\ }\bibfield  {title} {\bibinfo {title} {{Constraints on the Dense Matter Equation of State and Neutron Star Properties from NICER\textquoteright{}s Mass\textendash{}Radius Estimate of PSR J0740+6620 and Multimessenger Observations}},\ }\href {https://doi.org/10.3847/2041-8213/ac089a} {\bibfield  {journal} {\bibinfo  {journal} {Astrophys. J.
  Lett.}\ }\textbf {\bibinfo {volume} {918}},\ \bibinfo {pages} {L29} (\bibinfo {year} {2021})},\ \Eprint {https://arxiv.org/abs/2105.06981} {arXiv:2105.06981 [astro-ph.HE]} \BibitemShut {NoStop}%
\bibitem [{\citenamefont {Salmi}\ \emph {et~al.}(2024)\citenamefont {Salmi}, \citenamefont {Choudhury}, \citenamefont {Kini}, \citenamefont {Riley}, \citenamefont {Vinciguerra}, \citenamefont {Watts}, \citenamefont {Wolff}, \citenamefont {Arzoumanian}, \citenamefont {Bogdanov}, \citenamefont {Chakrabarty}, \citenamefont {Gendreau}, \citenamefont {Guillot}, \citenamefont {Ho}, \citenamefont {Huppenkothen}, \citenamefont {Ludlam}, \citenamefont {Morsink},\ and\ \citenamefont {Ray}}]{Salmi2024b}%
  \BibitemOpen
  \bibfield  {author} {\bibinfo {author} {\bibfnamefont {T.}~\bibnamefont {Salmi}}, \bibinfo {author} {\bibfnamefont {D.}~\bibnamefont {Choudhury}}, \bibinfo {author} {\bibfnamefont {Y.}~\bibnamefont {Kini}}, \bibinfo {author} {\bibfnamefont {T.~E.}\ \bibnamefont {Riley}}, \bibinfo {author} {\bibfnamefont {S.}~\bibnamefont {Vinciguerra}}, \bibinfo {author} {\bibfnamefont {A.~L.}\ \bibnamefont {Watts}}, \bibinfo {author} {\bibfnamefont {M.~T.}\ \bibnamefont {Wolff}}, \bibinfo {author} {\bibfnamefont {Z.}~\bibnamefont {Arzoumanian}}, \bibinfo {author} {\bibfnamefont {S.}~\bibnamefont {Bogdanov}}, \bibinfo {author} {\bibfnamefont {D.}~\bibnamefont {Chakrabarty}}, \bibinfo {author} {\bibfnamefont {K.}~\bibnamefont {Gendreau}}, \bibinfo {author} {\bibfnamefont {S.}~\bibnamefont {Guillot}}, \bibinfo {author} {\bibfnamefont {W.~C.~G.}\ \bibnamefont {Ho}}, \bibinfo {author} {\bibfnamefont {D.}~\bibnamefont {Huppenkothen}}, \bibinfo {author} {\bibfnamefont {R.~M.}\ \bibnamefont {Ludlam}}, \bibinfo {author}
  {\bibfnamefont {S.~M.}\ \bibnamefont {Morsink}},\ and\ \bibinfo {author} {\bibfnamefont {P.~S.}\ \bibnamefont {Ray}},\ }\bibfield  {title} {\bibinfo {title} {{The Radius of the High-mass Pulsar PSR J0740+6620 with 3.6 yr of NICER Data}},\ }\href {https://doi.org/10.3847/1538-4357/ad5f1f} {\bibfield  {journal} {\bibinfo  {journal} {The Astrophysical Journal}\ }\textbf {\bibinfo {volume} {974}},\ \bibinfo {pages} {294} (\bibinfo {year} {2024})},\ \Eprint {https://arxiv.org/abs/2406.14466} {arXiv:2406.14466 [astro-ph.HE]} \BibitemShut {NoStop}%
\bibitem [{\citenamefont {Dittmann}\ \emph {et~al.}(2024)\citenamefont {Dittmann} \emph {et~al.}}]{Dittmann2024}%
  \BibitemOpen
  \bibfield  {author} {\bibinfo {author} {\bibfnamefont {A.~J.}\ \bibnamefont {Dittmann}} \emph {et~al.},\ }\bibfield  {title} {\bibinfo {title} {A more precise measurement of the radius of psr j0740+6620 using updated nicer data},\ }\href {https://doi.org/10.3847/1538-4357/ad5f1e} {\bibfield  {journal} {\bibinfo  {journal} {Astrophys. J.}\ }\textbf {\bibinfo {volume} {974}},\ \bibinfo {pages} {295} (\bibinfo {year} {2024})},\ \Eprint {https://arxiv.org/abs/2406.14467} {arXiv:2406.14467 [astro-ph.HE]} \BibitemShut {NoStop}%
\bibitem [{\citenamefont {Salmi}(2024)}]{Salmi2024}%
  \BibitemOpen
  \bibfield  {author} {\bibinfo {author} {\bibfnamefont {T.~e.~a.}\ \bibnamefont {Salmi}},\ }\bibfield  {title} {\bibinfo {title} {{A NICER View of PSR J1231-1411: A Complex Case}},\ }\href {https://doi.org/10.3847/1538-4357/ad81d2} {\bibfield  {journal} {\bibinfo  {journal} {The Astrophysical Journal}\ }\textbf {\bibinfo {volume} {976}},\ \bibinfo {pages} {58} (\bibinfo {year} {2024})},\ \Eprint {https://arxiv.org/abs/2409.14923} {arXiv:2409.14923 [astro-ph.HE]} \BibitemShut {NoStop}%
\bibitem [{\citenamefont {Choudhury}(2024)}]{Choudhury2024}%
  \BibitemOpen
  \bibfield  {author} {\bibinfo {author} {\bibfnamefont {D.~e.~a.}\ \bibnamefont {Choudhury}},\ }\bibfield  {title} {\bibinfo {title} {{A NICER View of the Nearest and Brightest Millisecond Pulsar: PSR J0437–4715}},\ }\href {https://doi.org/10.3847/2041-8213/ad5a6f} {\bibfield  {journal} {\bibinfo  {journal} {The Astrophysical Journal Letters}\ }\textbf {\bibinfo {volume} {971}},\ \bibinfo {pages} {L20} (\bibinfo {year} {2024})},\ \Eprint {https://arxiv.org/abs/2407.06789} {arXiv:2407.06789 [astro-ph.HE]} \BibitemShut {NoStop}%
\bibitem [{\citenamefont {Qi}\ \emph {et~al.}(2025)\citenamefont {Qi} \emph {et~al.}}]{Qi2025}%
  \BibitemOpen
  \bibfield  {author} {\bibinfo {author} {\bibfnamefont {L.}~\bibnamefont {Qi}} \emph {et~al.},\ }\bibfield  {title} {\bibinfo {title} {{PSR J1231–1411 Revisited: Pulse Profile Analysis of X-Ray Observation}},\ }\href {https://doi.org/10.3847/1538-4357/adb42f} {\bibfield  {journal} {\bibinfo  {journal} {Astrophys. J.}\ }\textbf {\bibinfo {volume} {981}},\ \bibinfo {pages} {99} (\bibinfo {year} {2025})},\ \Eprint {https://arxiv.org/abs/2502.09147} {arXiv:2502.09147 [astro-ph.HE]} \BibitemShut {NoStop}%
\bibitem [{\citenamefont {Mauviard}\ \emph {et~al.}(2025)\citenamefont {Mauviard}, \citenamefont {Guillot}, \citenamefont {Salmi}, \citenamefont {Choudhury}, \citenamefont {Dorsman}, \citenamefont {González-Caniulef}, \citenamefont {Hoogkamer}, \citenamefont {Huppenkothen}, \citenamefont {Kazantsev}, \citenamefont {Kini}, \citenamefont {Olive}, \citenamefont {Stammler}, \citenamefont {Watts}, \citenamefont {Mendes}, \citenamefont {Rutherford}, \citenamefont {Schwenk}, \citenamefont {Svensson}, \citenamefont {Bogdanov}, \citenamefont {Kerr}, \citenamefont {Ray}, \citenamefont {Guillemot}, \citenamefont {Cognard},\ and\ \citenamefont {Theureau}}]{Mauviard2025}%
  \BibitemOpen
  \bibfield  {author} {\bibinfo {author} {\bibfnamefont {L.}~\bibnamefont {Mauviard}}, \bibinfo {author} {\bibfnamefont {S.}~\bibnamefont {Guillot}}, \bibinfo {author} {\bibfnamefont {T.}~\bibnamefont {Salmi}}, \bibinfo {author} {\bibfnamefont {D.}~\bibnamefont {Choudhury}}, \bibinfo {author} {\bibfnamefont {B.}~\bibnamefont {Dorsman}}, \bibinfo {author} {\bibfnamefont {D.}~\bibnamefont {González-Caniulef}}, \bibinfo {author} {\bibfnamefont {M.}~\bibnamefont {Hoogkamer}}, \bibinfo {author} {\bibfnamefont {D.}~\bibnamefont {Huppenkothen}}, \bibinfo {author} {\bibfnamefont {C.}~\bibnamefont {Kazantsev}}, \bibinfo {author} {\bibfnamefont {Y.}~\bibnamefont {Kini}}, \bibinfo {author} {\bibfnamefont {J.-F.}\ \bibnamefont {Olive}}, \bibinfo {author} {\bibfnamefont {P.}~\bibnamefont {Stammler}}, \bibinfo {author} {\bibfnamefont {A.~L.}\ \bibnamefont {Watts}}, \bibinfo {author} {\bibfnamefont {M.}~\bibnamefont {Mendes}}, \bibinfo {author} {\bibfnamefont {N.}~\bibnamefont {Rutherford}}, \bibinfo {author}
  {\bibfnamefont {A.}~\bibnamefont {Schwenk}}, \bibinfo {author} {\bibfnamefont {I.}~\bibnamefont {Svensson}}, \bibinfo {author} {\bibfnamefont {S.}~\bibnamefont {Bogdanov}}, \bibinfo {author} {\bibfnamefont {M.}~\bibnamefont {Kerr}}, \bibinfo {author} {\bibfnamefont {P.~S.}\ \bibnamefont {Ray}}, \bibinfo {author} {\bibfnamefont {L.}~\bibnamefont {Guillemot}}, \bibinfo {author} {\bibfnamefont {I.}~\bibnamefont {Cognard}},\ and\ \bibinfo {author} {\bibfnamefont {G.}~\bibnamefont {Theureau}},\ }\bibfield  {title} {\bibinfo {title} {{A NICER View of the 1.4 $M\odot$ Edge-on Pulsar PSR J0614-3329}},\ }\href {https://doi.org/10.3847/1538-4357/ae145d} {\bibfield  {journal} {\bibinfo  {journal} {The Astrophysical Journal}\ }\textbf {\bibinfo {volume} {995}},\ \bibinfo {pages} {60} (\bibinfo {year} {2025})},\ \Eprint {https://arxiv.org/abs/2506.14883} {arXiv:2506.14883 [astro-ph.HE]} \BibitemShut {NoStop}%
\bibitem [{\citenamefont {Miller}\ \emph {et~al.}(2025)\citenamefont {Miller} \emph {et~al.}}]{Miller2025}%
  \BibitemOpen
  \bibfield  {author} {\bibinfo {author} {\bibfnamefont {M.~C.}\ \bibnamefont {Miller}} \emph {et~al.},\ }\bibfield  {title} {\bibinfo {title} {{The Radius of PSR J0437-4715 from NICER Data}},\ }\bibfield  {journal} {\bibinfo  {journal} {arXiv e-prints}\ }\href {https://doi.org/10.48550/arXiv.2512.08790} {10.48550/arXiv.2512.08790} (\bibinfo {year} {2025}),\ \Eprint {https://arxiv.org/abs/2512.08790} {arXiv:2512.08790 [astro-ph.HE]} \BibitemShut {NoStop}%
\bibitem [{\citenamefont {De}\ \emph {et~al.}(2018)\citenamefont {De}, \citenamefont {Finstad}, \citenamefont {Lattimer}, \citenamefont {Brown}, \citenamefont {Berger},\ and\ \citenamefont {Biwer}}]{De:2018uhw}%
  \BibitemOpen
  \bibfield  {author} {\bibinfo {author} {\bibfnamefont {S.}~\bibnamefont {De}}, \bibinfo {author} {\bibfnamefont {D.}~\bibnamefont {Finstad}}, \bibinfo {author} {\bibfnamefont {J.~M.}\ \bibnamefont {Lattimer}}, \bibinfo {author} {\bibfnamefont {D.~A.}\ \bibnamefont {Brown}}, \bibinfo {author} {\bibfnamefont {E.}~\bibnamefont {Berger}},\ and\ \bibinfo {author} {\bibfnamefont {C.~M.}\ \bibnamefont {Biwer}},\ }\bibfield  {title} {\bibinfo {title} {{Tidal Deformabilities and Radii of Neutron Stars from the Observation of GW170817}},\ }\href {https://doi.org/10.1103/PhysRevLett.121.091102} {\bibfield  {journal} {\bibinfo  {journal} {Phys. Rev. Lett.}\ }\textbf {\bibinfo {volume} {121}},\ \bibinfo {pages} {091102} (\bibinfo {year} {2018})},\ \bibinfo {note} {[Erratum: Phys.Rev.Lett. 121, 259902 (2018)]},\ \Eprint {https://arxiv.org/abs/1804.08583} {arXiv:1804.08583 [astro-ph.HE]} \BibitemShut {NoStop}%
\bibitem [{\citenamefont {Tews}\ \emph {et~al.}(2018)\citenamefont {Tews}, \citenamefont {Margueron},\ and\ \citenamefont {Reddy}}]{Tews:2018iwm}%
  \BibitemOpen
  \bibfield  {author} {\bibinfo {author} {\bibfnamefont {I.}~\bibnamefont {Tews}}, \bibinfo {author} {\bibfnamefont {J.}~\bibnamefont {Margueron}},\ and\ \bibinfo {author} {\bibfnamefont {S.}~\bibnamefont {Reddy}},\ }\bibfield  {title} {\bibinfo {title} {{Critical examination of constraints on the equation of state of dense matter obtained from GW170817}},\ }\href {https://doi.org/10.1103/PhysRevC.98.045804} {\bibfield  {journal} {\bibinfo  {journal} {Phys. Rev. C}\ }\textbf {\bibinfo {volume} {98}},\ \bibinfo {pages} {045804} (\bibinfo {year} {2018})},\ \Eprint {https://arxiv.org/abs/1804.02783} {arXiv:1804.02783 [nucl-th]} \BibitemShut {NoStop}%
\bibitem [{\citenamefont {Rezzolla}\ \emph {et~al.}(2018)\citenamefont {Rezzolla}, \citenamefont {Most},\ and\ \citenamefont {Weih}}]{Rezzolla2017}%
  \BibitemOpen
  \bibfield  {author} {\bibinfo {author} {\bibfnamefont {L.}~\bibnamefont {Rezzolla}}, \bibinfo {author} {\bibfnamefont {E.~R.}\ \bibnamefont {Most}},\ and\ \bibinfo {author} {\bibfnamefont {L.~R.}\ \bibnamefont {Weih}},\ }\bibfield  {title} {\bibinfo {title} {{Using gravitational-wave observations and quasi-universal relations to constrain the maximum mass of neutron stars}},\ }\href {https://doi.org/10.3847/2041-8213/aaa401} {\bibfield  {journal} {\bibinfo  {journal} {Astrophys. J. Lett.}\ }\textbf {\bibinfo {volume} {852}},\ \bibinfo {pages} {L25} (\bibinfo {year} {2018})},\ \Eprint {https://arxiv.org/abs/1711.00314} {arXiv:1711.00314 [astro-ph.HE]} \BibitemShut {NoStop}%
\bibitem [{\citenamefont {Annala}\ \emph {et~al.}(2018)\citenamefont {Annala}, \citenamefont {Gorda}, \citenamefont {Kurkela},\ and\ \citenamefont {Vuorinen}}]{Annala:2017llu}%
  \BibitemOpen
  \bibfield  {author} {\bibinfo {author} {\bibfnamefont {E.}~\bibnamefont {Annala}}, \bibinfo {author} {\bibfnamefont {T.}~\bibnamefont {Gorda}}, \bibinfo {author} {\bibfnamefont {A.}~\bibnamefont {Kurkela}},\ and\ \bibinfo {author} {\bibfnamefont {A.}~\bibnamefont {Vuorinen}},\ }\bibfield  {title} {\bibinfo {title} {{Gravitational-wave constraints on the neutron-star-matter Equation of State}},\ }\href {https://doi.org/10.1103/PhysRevLett.120.172703} {\bibfield  {journal} {\bibinfo  {journal} {Phys. Rev. Lett.}\ }\textbf {\bibinfo {volume} {120}},\ \bibinfo {pages} {172703} (\bibinfo {year} {2018})},\ \Eprint {https://arxiv.org/abs/1711.02644} {arXiv:1711.02644 [astro-ph.HE]} \BibitemShut {NoStop}%
\bibitem [{\citenamefont {Dietrich}\ \emph {et~al.}(2020)\citenamefont {Dietrich}, \citenamefont {Coughlin}, \citenamefont {Pang}, \citenamefont {Bulla}, \citenamefont {Heinzel}, \citenamefont {Issa}, \citenamefont {Tews},\ and\ \citenamefont {Antier}}]{Dietrich:2020efo}%
  \BibitemOpen
  \bibfield  {author} {\bibinfo {author} {\bibfnamefont {T.}~\bibnamefont {Dietrich}}, \bibinfo {author} {\bibfnamefont {M.~W.}\ \bibnamefont {Coughlin}}, \bibinfo {author} {\bibfnamefont {P.~T.~H.}\ \bibnamefont {Pang}}, \bibinfo {author} {\bibfnamefont {M.}~\bibnamefont {Bulla}}, \bibinfo {author} {\bibfnamefont {J.}~\bibnamefont {Heinzel}}, \bibinfo {author} {\bibfnamefont {L.}~\bibnamefont {Issa}}, \bibinfo {author} {\bibfnamefont {I.}~\bibnamefont {Tews}},\ and\ \bibinfo {author} {\bibfnamefont {S.}~\bibnamefont {Antier}},\ }\bibfield  {title} {\bibinfo {title} {{Multimessenger constraints on the neutron-star equation of state and the Hubble constant}},\ }\href {https://doi.org/10.1126/science.abb4317} {\bibfield  {journal} {\bibinfo  {journal} {Science}\ }\textbf {\bibinfo {volume} {370}},\ \bibinfo {pages} {1450} (\bibinfo {year} {2020})},\ \Eprint {https://arxiv.org/abs/2002.11355} {arXiv:2002.11355 [astro-ph.HE]} \BibitemShut {NoStop}%
\bibitem [{\citenamefont {Pang}\ \emph {et~al.}(2021)\citenamefont {Pang}, \citenamefont {Tews}, \citenamefont {Coughlin}, \citenamefont {Bulla}, \citenamefont {Van Den~Broeck},\ and\ \citenamefont {Dietrich}}]{Pang:2021jta}%
  \BibitemOpen
  \bibfield  {author} {\bibinfo {author} {\bibfnamefont {P.~T.~H.}\ \bibnamefont {Pang}}, \bibinfo {author} {\bibfnamefont {I.}~\bibnamefont {Tews}}, \bibinfo {author} {\bibfnamefont {M.~W.}\ \bibnamefont {Coughlin}}, \bibinfo {author} {\bibfnamefont {M.}~\bibnamefont {Bulla}}, \bibinfo {author} {\bibfnamefont {C.}~\bibnamefont {Van Den~Broeck}},\ and\ \bibinfo {author} {\bibfnamefont {T.}~\bibnamefont {Dietrich}},\ }\bibfield  {title} {\bibinfo {title} {{Nuclear Physics Multimessenger Astrophysics Constraints on the Neutron Star Equation of State: Adding NICER{\textquoteright}s PSR J0740+6620 Measurement}},\ }\href {https://doi.org/10.3847/1538-4357/ac19ab} {\bibfield  {journal} {\bibinfo  {journal} {Astrophys. J.}\ }\textbf {\bibinfo {volume} {922}},\ \bibinfo {pages} {14} (\bibinfo {year} {2021})},\ \Eprint {https://arxiv.org/abs/2105.08688} {arXiv:2105.08688 [astro-ph.HE]} \BibitemShut {NoStop}%
\bibitem [{\citenamefont {Malik}\ \emph {et~al.}(2022)\citenamefont {Malik}, \citenamefont {Ferreira}, \citenamefont {Agrawal},\ and\ \citenamefont {Provid{\^e}ncia}}]{Malik:2022zol}%
  \BibitemOpen
  \bibfield  {author} {\bibinfo {author} {\bibfnamefont {T.}~\bibnamefont {Malik}}, \bibinfo {author} {\bibfnamefont {M.}~\bibnamefont {Ferreira}}, \bibinfo {author} {\bibfnamefont {B.~K.}\ \bibnamefont {Agrawal}},\ and\ \bibinfo {author} {\bibfnamefont {C.}~\bibnamefont {Provid{\^e}ncia}},\ }\bibfield  {title} {\bibinfo {title} {{Relativistic Description of Dense Matter Equation of State and Compatibility with Neutron Star Observables: A Bayesian Approach}},\ }\href {https://doi.org/10.3847/1538-4357/ac5d3c} {\bibfield  {journal} {\bibinfo  {journal} {Astrophys. J.}\ }\textbf {\bibinfo {volume} {930}},\ \bibinfo {pages} {17} (\bibinfo {year} {2022})},\ \Eprint {https://arxiv.org/abs/2201.12552} {arXiv:2201.12552 [nucl-th]} \BibitemShut {NoStop}%
\bibitem [{\citenamefont {Altiparmak}\ \emph {et~al.}(2022)\citenamefont {Altiparmak}, \citenamefont {Ecker},\ and\ \citenamefont {Rezzolla}}]{Altiparmak:2022bke}%
  \BibitemOpen
  \bibfield  {author} {\bibinfo {author} {\bibfnamefont {S.}~\bibnamefont {Altiparmak}}, \bibinfo {author} {\bibfnamefont {C.}~\bibnamefont {Ecker}},\ and\ \bibinfo {author} {\bibfnamefont {L.}~\bibnamefont {Rezzolla}},\ }\bibfield  {title} {\bibinfo {title} {{On the Sound Speed in Neutron Stars}},\ }\href {https://doi.org/10.3847/2041-8213/ac9b2a} {\bibfield  {journal} {\bibinfo  {journal} {Astrophys. J. Lett.}\ }\textbf {\bibinfo {volume} {939}},\ \bibinfo {pages} {L34} (\bibinfo {year} {2022})},\ \Eprint {https://arxiv.org/abs/2203.14974} {arXiv:2203.14974 [astro-ph.HE]} \BibitemShut {NoStop}%
\bibitem [{\citenamefont {Marczenko}\ \emph {et~al.}(2023)\citenamefont {Marczenko}, \citenamefont {McLerran}, \citenamefont {Redlich},\ and\ \citenamefont {Sasaki}}]{Marczenko:2022jhl}%
  \BibitemOpen
  \bibfield  {author} {\bibinfo {author} {\bibfnamefont {M.}~\bibnamefont {Marczenko}}, \bibinfo {author} {\bibfnamefont {L.}~\bibnamefont {McLerran}}, \bibinfo {author} {\bibfnamefont {K.}~\bibnamefont {Redlich}},\ and\ \bibinfo {author} {\bibfnamefont {C.}~\bibnamefont {Sasaki}},\ }\bibfield  {title} {\bibinfo {title} {{Reaching percolation and conformal limits in neutron stars}},\ }\href {https://doi.org/10.1103/PhysRevC.107.025802} {\bibfield  {journal} {\bibinfo  {journal} {Phys. Rev. C}\ }\textbf {\bibinfo {volume} {107}},\ \bibinfo {pages} {025802} (\bibinfo {year} {2023})},\ \Eprint {https://arxiv.org/abs/2207.13059} {arXiv:2207.13059 [nucl-th]} \BibitemShut {NoStop}%
\bibitem [{\citenamefont {Takatsy}\ \emph {et~al.}(2023)\citenamefont {Takatsy}, \citenamefont {Kovacs}, \citenamefont {Wolf},\ and\ \citenamefont {Schaffner-Bielich}}]{Takatsy:2023xzf}%
  \BibitemOpen
  \bibfield  {author} {\bibinfo {author} {\bibfnamefont {J.}~\bibnamefont {Takatsy}}, \bibinfo {author} {\bibfnamefont {P.}~\bibnamefont {Kovacs}}, \bibinfo {author} {\bibfnamefont {G.}~\bibnamefont {Wolf}},\ and\ \bibinfo {author} {\bibfnamefont {J.}~\bibnamefont {Schaffner-Bielich}},\ }\bibfield  {title} {\bibinfo {title} {{What neutron stars tell about the hadron-quark phase transition: A Bayesian study}},\ }\href {https://doi.org/10.1103/PhysRevD.108.043002} {\bibfield  {journal} {\bibinfo  {journal} {Phys. Rev. D}\ }\textbf {\bibinfo {volume} {108}},\ \bibinfo {pages} {043002} (\bibinfo {year} {2023})},\ \Eprint {https://arxiv.org/abs/2303.00013} {arXiv:2303.00013 [astro-ph.HE]} \BibitemShut {NoStop}%
\bibitem [{\citenamefont {Koehn}\ \emph {et~al.}(2025)\citenamefont {Koehn} \emph {et~al.}}]{Koehn:2024set}%
  \BibitemOpen
  \bibfield  {author} {\bibinfo {author} {\bibfnamefont {H.}~\bibnamefont {Koehn}} \emph {et~al.},\ }\bibfield  {title} {\bibinfo {title} {{From existing and new nuclear and astrophysical constraints to stringent limits on the equation of state of neutron-rich dense matter}},\ }\href {https://doi.org/10.1103/PhysRevX.15.021014} {\bibfield  {journal} {\bibinfo  {journal} {Phys. Rev. X}\ }\textbf {\bibinfo {volume} {15}},\ \bibinfo {pages} {021014} (\bibinfo {year} {2025})},\ \Eprint {https://arxiv.org/abs/2402.04172} {arXiv:2402.04172 [astro-ph.HE]} \BibitemShut {NoStop}%
\bibitem [{\citenamefont {Bauswein}\ \emph {et~al.}(2013)\citenamefont {Bauswein}, \citenamefont {Baumgarte},\ and\ \citenamefont {Janka}}]{Bauswein:2013jpa}%
  \BibitemOpen
  \bibfield  {author} {\bibinfo {author} {\bibfnamefont {A.}~\bibnamefont {Bauswein}}, \bibinfo {author} {\bibfnamefont {T.~W.}\ \bibnamefont {Baumgarte}},\ and\ \bibinfo {author} {\bibfnamefont {H.~T.}\ \bibnamefont {Janka}},\ }\bibfield  {title} {\bibinfo {title} {{Prompt merger collapse and the maximum mass of neutron stars}},\ }\href {https://doi.org/10.1103/PhysRevLett.111.131101} {\bibfield  {journal} {\bibinfo  {journal} {Phys. Rev. Lett.}\ }\textbf {\bibinfo {volume} {111}},\ \bibinfo {pages} {131101} (\bibinfo {year} {2013})},\ \Eprint {https://arxiv.org/abs/1307.5191} {arXiv:1307.5191 [astro-ph.SR]} \BibitemShut {NoStop}%
\bibitem [{\citenamefont {Bauswein}\ \emph {et~al.}(2016)\citenamefont {Bauswein}, \citenamefont {Stergioulas},\ and\ \citenamefont {Janka}}]{Bauswein:2015vxa}%
  \BibitemOpen
  \bibfield  {author} {\bibinfo {author} {\bibfnamefont {A.}~\bibnamefont {Bauswein}}, \bibinfo {author} {\bibfnamefont {N.}~\bibnamefont {Stergioulas}},\ and\ \bibinfo {author} {\bibfnamefont {H.-T.}\ \bibnamefont {Janka}},\ }\bibfield  {title} {\bibinfo {title} {{Exploring properties of high-density matter through remnants of neutron-star mergers}},\ }\href {https://doi.org/10.1140/epja/i2016-16056-7} {\bibfield  {journal} {\bibinfo  {journal} {Eur. Phys. J. A}\ }\textbf {\bibinfo {volume} {52}},\ \bibinfo {pages} {56} (\bibinfo {year} {2016})},\ \Eprint {https://arxiv.org/abs/1508.05493} {arXiv:1508.05493 [astro-ph.HE]} \BibitemShut {NoStop}%
\bibitem [{\citenamefont {Sarin}\ and\ \citenamefont {Lasky}(2021)}]{Sarin:2020gxb}%
  \BibitemOpen
  \bibfield  {author} {\bibinfo {author} {\bibfnamefont {N.}~\bibnamefont {Sarin}}\ and\ \bibinfo {author} {\bibfnamefont {P.~D.}\ \bibnamefont {Lasky}},\ }\bibfield  {title} {\bibinfo {title} {{The evolution of binary neutron star post-merger remnants: a review}},\ }\href {https://doi.org/10.1007/s10714-021-02831-1} {\bibfield  {journal} {\bibinfo  {journal} {Gen. Rel. Grav.}\ }\textbf {\bibinfo {volume} {53}},\ \bibinfo {pages} {59} (\bibinfo {year} {2021})},\ \Eprint {https://arxiv.org/abs/2012.08172} {arXiv:2012.08172 [astro-ph.HE]} \BibitemShut {NoStop}%
\bibitem [{\citenamefont {Breschi}\ \emph {et~al.}(2022)\citenamefont {Breschi}, \citenamefont {Bernuzzi}, \citenamefont {Godzieba}, \citenamefont {Perego},\ and\ \citenamefont {Radice}}]{Breschi:2021xrx}%
  \BibitemOpen
  \bibfield  {author} {\bibinfo {author} {\bibfnamefont {M.}~\bibnamefont {Breschi}}, \bibinfo {author} {\bibfnamefont {S.}~\bibnamefont {Bernuzzi}}, \bibinfo {author} {\bibfnamefont {D.}~\bibnamefont {Godzieba}}, \bibinfo {author} {\bibfnamefont {A.}~\bibnamefont {Perego}},\ and\ \bibinfo {author} {\bibfnamefont {D.}~\bibnamefont {Radice}},\ }\bibfield  {title} {\bibinfo {title} {{Constraints on the Maximum Densities of Neutron Stars from Postmerger Gravitational Waves with Third-Generation Observations}},\ }\href {https://doi.org/10.1103/PhysRevLett.128.161102} {\bibfield  {journal} {\bibinfo  {journal} {Phys. Rev. Lett.}\ }\textbf {\bibinfo {volume} {128}},\ \bibinfo {pages} {161102} (\bibinfo {year} {2022})},\ \Eprint {https://arxiv.org/abs/2110.06957} {arXiv:2110.06957 [gr-qc]} \BibitemShut {NoStop}%
\bibitem [{\citenamefont {Puecher}\ \emph {et~al.}(2023)\citenamefont {Puecher}, \citenamefont {Dietrich}, \citenamefont {Tsang}, \citenamefont {Kalaghatgi}, \citenamefont {Roy}, \citenamefont {Setyawati},\ and\ \citenamefont {Van Den~Broeck}}]{Puecher:2022oiz}%
  \BibitemOpen
  \bibfield  {author} {\bibinfo {author} {\bibfnamefont {A.}~\bibnamefont {Puecher}}, \bibinfo {author} {\bibfnamefont {T.}~\bibnamefont {Dietrich}}, \bibinfo {author} {\bibfnamefont {K.~W.}\ \bibnamefont {Tsang}}, \bibinfo {author} {\bibfnamefont {C.}~\bibnamefont {Kalaghatgi}}, \bibinfo {author} {\bibfnamefont {S.}~\bibnamefont {Roy}}, \bibinfo {author} {\bibfnamefont {Y.}~\bibnamefont {Setyawati}},\ and\ \bibinfo {author} {\bibfnamefont {C.}~\bibnamefont {Van Den~Broeck}},\ }\bibfield  {title} {\bibinfo {title} {{Unraveling information about supranuclear-dense matter from the complete binary neutron star coalescence process using future gravitational-wave detector networks}},\ }\href {https://doi.org/10.1103/PhysRevD.107.124009} {\bibfield  {journal} {\bibinfo  {journal} {Phys. Rev. D}\ }\textbf {\bibinfo {volume} {107}},\ \bibinfo {pages} {124009} (\bibinfo {year} {2023})},\ \Eprint {https://arxiv.org/abs/2210.09259} {arXiv:2210.09259 [gr-qc]} \BibitemShut {NoStop}%
\bibitem [{\citenamefont {Abbott}\ \emph {et~al.}(2017{\natexlab{d}})\citenamefont {Abbott} \emph {et~al.}}]{LIGOScientific:2017fdd}%
  \BibitemOpen
  \bibfield  {author} {\bibinfo {author} {\bibfnamefont {B.~P.}\ \bibnamefont {Abbott}} \emph {et~al.} (\bibinfo {collaboration} {LIGO Scientific, Virgo}),\ }\bibfield  {title} {\bibinfo {title} {{Search for Post-merger Gravitational Waves from the Remnant of the Binary Neutron Star Merger GW170817}},\ }\href {https://doi.org/10.3847/2041-8213/aa9a35} {\bibfield  {journal} {\bibinfo  {journal} {Astrophys. J. Lett.}\ }\textbf {\bibinfo {volume} {851}},\ \bibinfo {pages} {L16} (\bibinfo {year} {2017}{\natexlab{d}})},\ \Eprint {https://arxiv.org/abs/1710.09320} {arXiv:1710.09320 [astro-ph.HE]} \BibitemShut {NoStop}%
\bibitem [{\citenamefont {Abbott}\ \emph {et~al.}(2019{\natexlab{b}})\citenamefont {Abbott} \emph {et~al.}}]{LIGOScientific:2018urg}%
  \BibitemOpen
  \bibfield  {author} {\bibinfo {author} {\bibfnamefont {B.~P.}\ \bibnamefont {Abbott}} \emph {et~al.} (\bibinfo {collaboration} {LIGO Scientific, Virgo}),\ }\bibfield  {title} {\bibinfo {title} {{Search for gravitational waves from a long-lived remnant of the binary neutron star merger GW170817}},\ }\href {https://doi.org/10.3847/1538-4357/ab0f3d} {\bibfield  {journal} {\bibinfo  {journal} {Astrophys. J.}\ }\textbf {\bibinfo {volume} {875}},\ \bibinfo {pages} {160} (\bibinfo {year} {2019}{\natexlab{b}})},\ \Eprint {https://arxiv.org/abs/1810.02581} {arXiv:1810.02581 [gr-qc]} \BibitemShut {NoStop}%
\bibitem [{\citenamefont {Clark}\ \emph {et~al.}(2016)\citenamefont {Clark}, \citenamefont {Bauswein}, \citenamefont {Stergioulas},\ and\ \citenamefont {Shoemaker}}]{Clark:2015zxa}%
  \BibitemOpen
  \bibfield  {author} {\bibinfo {author} {\bibfnamefont {J.~A.}\ \bibnamefont {Clark}}, \bibinfo {author} {\bibfnamefont {A.}~\bibnamefont {Bauswein}}, \bibinfo {author} {\bibfnamefont {N.}~\bibnamefont {Stergioulas}},\ and\ \bibinfo {author} {\bibfnamefont {D.}~\bibnamefont {Shoemaker}},\ }\bibfield  {title} {\bibinfo {title} {{Observing Gravitational Waves From The Post-Merger Phase Of Binary Neutron Star Coalescence}},\ }\href {https://doi.org/10.1088/0264-9381/33/8/085003} {\bibfield  {journal} {\bibinfo  {journal} {Class. Quant. Grav.}\ }\textbf {\bibinfo {volume} {33}},\ \bibinfo {pages} {085003} (\bibinfo {year} {2016})},\ \Eprint {https://arxiv.org/abs/1509.08522} {arXiv:1509.08522 [astro-ph.HE]} \BibitemShut {NoStop}%
\bibitem [{\citenamefont {Torres-Rivas}\ \emph {et~al.}(2019)\citenamefont {Torres-Rivas}, \citenamefont {Chatziioannou}, \citenamefont {Bauswein},\ and\ \citenamefont {Clark}}]{Torres-Rivas:2018svp}%
  \BibitemOpen
  \bibfield  {author} {\bibinfo {author} {\bibfnamefont {A.}~\bibnamefont {Torres-Rivas}}, \bibinfo {author} {\bibfnamefont {K.}~\bibnamefont {Chatziioannou}}, \bibinfo {author} {\bibfnamefont {A.}~\bibnamefont {Bauswein}},\ and\ \bibinfo {author} {\bibfnamefont {J.~A.}\ \bibnamefont {Clark}},\ }\bibfield  {title} {\bibinfo {title} {{Observing the post-merger signal of GW170817-like events with improved gravitational-wave detectors}},\ }\href {https://doi.org/10.1103/PhysRevD.99.044014} {\bibfield  {journal} {\bibinfo  {journal} {Phys. Rev. D}\ }\textbf {\bibinfo {volume} {99}},\ \bibinfo {pages} {044014} (\bibinfo {year} {2019})},\ \Eprint {https://arxiv.org/abs/1811.08931} {arXiv:1811.08931 [gr-qc]} \BibitemShut {NoStop}%
\bibitem [{\citenamefont {Chan}\ \emph {et~al.}(2014)\citenamefont {Chan}, \citenamefont {Sham}, \citenamefont {Leung},\ and\ \citenamefont {Lin}}]{Chan:2014kua}%
  \BibitemOpen
  \bibfield  {author} {\bibinfo {author} {\bibfnamefont {T.~K.}\ \bibnamefont {Chan}}, \bibinfo {author} {\bibfnamefont {Y.~H.}\ \bibnamefont {Sham}}, \bibinfo {author} {\bibfnamefont {P.~T.}\ \bibnamefont {Leung}},\ and\ \bibinfo {author} {\bibfnamefont {L.~M.}\ \bibnamefont {Lin}},\ }\bibfield  {title} {\bibinfo {title} {{Multipolar universal relations between f-mode frequency and tidal deformability of compact stars}},\ }\href {https://doi.org/10.1103/PhysRevD.90.124023} {\bibfield  {journal} {\bibinfo  {journal} {Phys. Rev. D}\ }\textbf {\bibinfo {volume} {90}},\ \bibinfo {pages} {124023} (\bibinfo {year} {2014})},\ \Eprint {https://arxiv.org/abs/1408.3789} {arXiv:1408.3789 [gr-qc]} \BibitemShut {NoStop}%
\bibitem [{\citenamefont {Godzieba}\ \emph {et~al.}(2021)\citenamefont {Godzieba}, \citenamefont {Gamba}, \citenamefont {Radice},\ and\ \citenamefont {Bernuzzi}}]{Godzieba:2020bbz}%
  \BibitemOpen
  \bibfield  {author} {\bibinfo {author} {\bibfnamefont {D.~A.}\ \bibnamefont {Godzieba}}, \bibinfo {author} {\bibfnamefont {R.}~\bibnamefont {Gamba}}, \bibinfo {author} {\bibfnamefont {D.}~\bibnamefont {Radice}},\ and\ \bibinfo {author} {\bibfnamefont {S.}~\bibnamefont {Bernuzzi}},\ }\bibfield  {title} {\bibinfo {title} {{Updated universal relations for tidal deformabilities of neutron stars from phenomenological equations of state}},\ }\href {https://doi.org/10.1103/PhysRevD.103.063036} {\bibfield  {journal} {\bibinfo  {journal} {Phys. Rev. D}\ }\textbf {\bibinfo {volume} {103}},\ \bibinfo {pages} {063036} (\bibinfo {year} {2021})},\ \Eprint {https://arxiv.org/abs/2012.12151} {arXiv:2012.12151 [astro-ph.HE]} \BibitemShut {NoStop}%
\bibitem [{\citenamefont {Pradhan}\ \emph {et~al.}(2022)\citenamefont {Pradhan}, \citenamefont {Chatterjee}, \citenamefont {Lanoye},\ and\ \citenamefont {Jaikumar}}]{Pradhan:2022vdf}%
  \BibitemOpen
  \bibfield  {author} {\bibinfo {author} {\bibfnamefont {B.~K.}\ \bibnamefont {Pradhan}}, \bibinfo {author} {\bibfnamefont {D.}~\bibnamefont {Chatterjee}}, \bibinfo {author} {\bibfnamefont {M.}~\bibnamefont {Lanoye}},\ and\ \bibinfo {author} {\bibfnamefont {P.}~\bibnamefont {Jaikumar}},\ }\bibfield  {title} {\bibinfo {title} {{General relativistic treatment of f-mode oscillations of hyperonic stars}},\ }\href {https://doi.org/10.1103/PhysRevC.106.015805} {\bibfield  {journal} {\bibinfo  {journal} {Phys. Rev. C}\ }\textbf {\bibinfo {volume} {106}},\ \bibinfo {pages} {015805} (\bibinfo {year} {2022})},\ \Eprint {https://arxiv.org/abs/2203.03141} {arXiv:2203.03141 [astro-ph.HE]} \BibitemShut {NoStop}%
\bibitem [{\citenamefont {Pradhan}\ \emph {et~al.}(2024)\citenamefont {Pradhan}, \citenamefont {Chatterjee},\ and\ \citenamefont {Alvarez-Castillo}}]{Pradhan:2023zmg}%
  \BibitemOpen
  \bibfield  {author} {\bibinfo {author} {\bibfnamefont {B.~K.}\ \bibnamefont {Pradhan}}, \bibinfo {author} {\bibfnamefont {D.}~\bibnamefont {Chatterjee}},\ and\ \bibinfo {author} {\bibfnamefont {D.~E.}\ \bibnamefont {Alvarez-Castillo}},\ }\bibfield  {title} {\bibinfo {title} {{Probing hadron{\textendash}quark phase transition in twin stars using $f$-modes}},\ }\href {https://doi.org/10.1093/mnras/stae1425} {\bibfield  {journal} {\bibinfo  {journal} {Mon. Not. Roy. Astron. Soc.}\ }\textbf {\bibinfo {volume} {531}},\ \bibinfo {pages} {4640} (\bibinfo {year} {2024})},\ \Eprint {https://arxiv.org/abs/2309.08775} {arXiv:2309.08775 [nucl-th]} \BibitemShut {NoStop}%
\bibitem [{\citenamefont {Kantor}\ and\ \citenamefont {Gusakov}(2014)}]{Kantor:2014lja}%
  \BibitemOpen
  \bibfield  {author} {\bibinfo {author} {\bibfnamefont {E.~M.}\ \bibnamefont {Kantor}}\ and\ \bibinfo {author} {\bibfnamefont {M.~E.}\ \bibnamefont {Gusakov}},\ }\bibfield  {title} {\bibinfo {title} {{Composition temperature-dependent g-modes in superfluid neutron stars}},\ }\href {https://doi.org/10.1093/mnrasl/slu061} {\bibfield  {journal} {\bibinfo  {journal} {Mon. Not. Roy. Astron. Soc.}\ }\textbf {\bibinfo {volume} {442}},\ \bibinfo {pages} {90} (\bibinfo {year} {2014})},\ \Eprint {https://arxiv.org/abs/1404.6768} {arXiv:1404.6768 [astro-ph.SR]} \BibitemShut {NoStop}%
\bibitem [{\citenamefont {Passamonti}\ \emph {et~al.}(2016)\citenamefont {Passamonti}, \citenamefont {Andersson},\ and\ \citenamefont {Ho}}]{Passamonti:2015oia}%
  \BibitemOpen
  \bibfield  {author} {\bibinfo {author} {\bibfnamefont {A.}~\bibnamefont {Passamonti}}, \bibinfo {author} {\bibfnamefont {N.}~\bibnamefont {Andersson}},\ and\ \bibinfo {author} {\bibfnamefont {W.~C.~G.}\ \bibnamefont {Ho}},\ }\bibfield  {title} {\bibinfo {title} {{Buoyancy and g-modes in young superfluid neutron stars}},\ }\href {https://doi.org/10.1093/mnras/stv2149} {\bibfield  {journal} {\bibinfo  {journal} {Mon. Not. Roy. Astron. Soc.}\ }\textbf {\bibinfo {volume} {455}},\ \bibinfo {pages} {1489} (\bibinfo {year} {2016})},\ \Eprint {https://arxiv.org/abs/1504.07470} {arXiv:1504.07470 [astro-ph.SR]} \BibitemShut {NoStop}%
\bibitem [{\citenamefont {Constantinou}\ \emph {et~al.}(2021)\citenamefont {Constantinou}, \citenamefont {Han}, \citenamefont {Jaikumar},\ and\ \citenamefont {Prakash}}]{Constantinou:2021hba}%
  \BibitemOpen
  \bibfield  {author} {\bibinfo {author} {\bibfnamefont {C.}~\bibnamefont {Constantinou}}, \bibinfo {author} {\bibfnamefont {S.}~\bibnamefont {Han}}, \bibinfo {author} {\bibfnamefont {P.}~\bibnamefont {Jaikumar}},\ and\ \bibinfo {author} {\bibfnamefont {M.}~\bibnamefont {Prakash}},\ }\bibfield  {title} {\bibinfo {title} {{g modes of neutron stars with hadron-to-quark crossover transitions}},\ }\href {https://doi.org/10.1103/PhysRevD.104.123032} {\bibfield  {journal} {\bibinfo  {journal} {Phys. Rev. D}\ }\textbf {\bibinfo {volume} {104}},\ \bibinfo {pages} {123032} (\bibinfo {year} {2021})},\ \Eprint {https://arxiv.org/abs/2109.14091} {arXiv:2109.14091 [astro-ph.HE]} \BibitemShut {NoStop}%
\bibitem [{\citenamefont {Jaikumar}\ \emph {et~al.}(2021)\citenamefont {Jaikumar}, \citenamefont {Semposki}, \citenamefont {Prakash},\ and\ \citenamefont {Constantinou}}]{Jaikumar:2021jbw}%
  \BibitemOpen
  \bibfield  {author} {\bibinfo {author} {\bibfnamefont {P.}~\bibnamefont {Jaikumar}}, \bibinfo {author} {\bibfnamefont {A.}~\bibnamefont {Semposki}}, \bibinfo {author} {\bibfnamefont {M.}~\bibnamefont {Prakash}},\ and\ \bibinfo {author} {\bibfnamefont {C.}~\bibnamefont {Constantinou}},\ }\bibfield  {title} {\bibinfo {title} {{$g$-mode oscillations in hybrid stars: A tale of two sounds}},\ }\href {https://doi.org/10.1103/PhysRevD.103.123009} {\bibfield  {journal} {\bibinfo  {journal} {Phys. Rev. D}\ }\textbf {\bibinfo {volume} {103}},\ \bibinfo {pages} {123009} (\bibinfo {year} {2021})},\ \Eprint {https://arxiv.org/abs/2101.06349} {arXiv:2101.06349 [nucl-th]} \BibitemShut {NoStop}%
\bibitem [{\citenamefont {Zhao}\ \emph {et~al.}(2022)\citenamefont {Zhao}, \citenamefont {Constantinou}, \citenamefont {Jaikumar},\ and\ \citenamefont {Prakash}}]{Zhao:2022toc}%
  \BibitemOpen
  \bibfield  {author} {\bibinfo {author} {\bibfnamefont {T.}~\bibnamefont {Zhao}}, \bibinfo {author} {\bibfnamefont {C.}~\bibnamefont {Constantinou}}, \bibinfo {author} {\bibfnamefont {P.}~\bibnamefont {Jaikumar}},\ and\ \bibinfo {author} {\bibfnamefont {M.}~\bibnamefont {Prakash}},\ }\bibfield  {title} {\bibinfo {title} {{Quasinormal g modes of neutron stars with quarks}},\ }\href {https://doi.org/10.1103/PhysRevD.105.103025} {\bibfield  {journal} {\bibinfo  {journal} {Phys. Rev. D}\ }\textbf {\bibinfo {volume} {105}},\ \bibinfo {pages} {103025} (\bibinfo {year} {2022})},\ \Eprint {https://arxiv.org/abs/2202.01403} {arXiv:2202.01403 [gr-qc]} \BibitemShut {NoStop}%
\bibitem [{\citenamefont {Counsell}\ \emph {et~al.}(2024)\citenamefont {Counsell}, \citenamefont {Gittins}, \citenamefont {Andersson},\ and\ \citenamefont {Pnigouras}}]{Counsell:2024pua}%
  \BibitemOpen
  \bibfield  {author} {\bibinfo {author} {\bibfnamefont {R.}~\bibnamefont {Counsell}}, \bibinfo {author} {\bibfnamefont {F.}~\bibnamefont {Gittins}}, \bibinfo {author} {\bibfnamefont {N.}~\bibnamefont {Andersson}},\ and\ \bibinfo {author} {\bibfnamefont {P.}~\bibnamefont {Pnigouras}},\ }\bibfield  {title} {\bibinfo {title} {{Neutron star g modes in the relativistic Cowling approximation}},\ }\href {https://doi.org/10.1093/mnras/stae2721} {\bibfield  {journal} {\bibinfo  {journal} {Mon. Not. Roy. Astron. Soc.}\ }\textbf {\bibinfo {volume} {536}},\ \bibinfo {pages} {1967} (\bibinfo {year} {2024})},\ \Eprint {https://arxiv.org/abs/2409.20178} {arXiv:2409.20178 [gr-qc]} \BibitemShut {NoStop}%
\bibitem [{\citenamefont {Miniutti}\ \emph {et~al.}(2003)\citenamefont {Miniutti}, \citenamefont {Pons}, \citenamefont {Berti}, \citenamefont {Gualtieri},\ and\ \citenamefont {Ferrari}}]{Miniutti:2002bh}%
  \BibitemOpen
  \bibfield  {author} {\bibinfo {author} {\bibfnamefont {G.}~\bibnamefont {Miniutti}}, \bibinfo {author} {\bibfnamefont {J.~A.}\ \bibnamefont {Pons}}, \bibinfo {author} {\bibfnamefont {E.}~\bibnamefont {Berti}}, \bibinfo {author} {\bibfnamefont {L.}~\bibnamefont {Gualtieri}},\ and\ \bibinfo {author} {\bibfnamefont {V.}~\bibnamefont {Ferrari}},\ }\bibfield  {title} {\bibinfo {title} {{Non-radial oscillation modes as a probe of density discontinuities in neutron stars}},\ }\href {https://doi.org/10.1046/j.1365-8711.2003.06057.x} {\bibfield  {journal} {\bibinfo  {journal} {Mon. Not. Roy. Astron. Soc.}\ }\textbf {\bibinfo {volume} {338}},\ \bibinfo {pages} {389} (\bibinfo {year} {2003})},\ \Eprint {https://arxiv.org/abs/astro-ph/0206142} {arXiv:astro-ph/0206142} \BibitemShut {NoStop}%
\bibitem [{\citenamefont {Counsell}\ \emph {et~al.}(2025)\citenamefont {Counsell}, \citenamefont {Gittins}, \citenamefont {Andersson},\ and\ \citenamefont {Tews}}]{Counsell:2025hcv}%
  \BibitemOpen
  \bibfield  {author} {\bibinfo {author} {\bibfnamefont {A.~R.}\ \bibnamefont {Counsell}}, \bibinfo {author} {\bibfnamefont {F.}~\bibnamefont {Gittins}}, \bibinfo {author} {\bibfnamefont {N.}~\bibnamefont {Andersson}},\ and\ \bibinfo {author} {\bibfnamefont {I.}~\bibnamefont {Tews}},\ }\bibfield  {title} {\bibinfo {title} {{Interface Modes in Inspiralling Neutron Stars: A Gravitational-Wave Probe of First-Order Phase Transitions}},\ }\href {https://doi.org/10.1103/8hvq-6dy7} {\bibfield  {journal} {\bibinfo  {journal} {Phys. Rev. Lett.}\ }\textbf {\bibinfo {volume} {135}},\ \bibinfo {pages} {081402} (\bibinfo {year} {2025})},\ \Eprint {https://arxiv.org/abs/2504.06181} {arXiv:2504.06181 [gr-qc]} \BibitemShut {NoStop}%
\bibitem [{\citenamefont {Zhu}\ \emph {et~al.}(2023)\citenamefont {Zhu}, \citenamefont {Wang}, \citenamefont {Xia}, \citenamefont {Zhou},\ and\ \citenamefont {Ma}}]{Zhu:2022pja}%
  \BibitemOpen
  \bibfield  {author} {\bibinfo {author} {\bibfnamefont {J.}~\bibnamefont {Zhu}}, \bibinfo {author} {\bibfnamefont {C.}~\bibnamefont {Wang}}, \bibinfo {author} {\bibfnamefont {C.}~\bibnamefont {Xia}}, \bibinfo {author} {\bibfnamefont {E.}~\bibnamefont {Zhou}},\ and\ \bibinfo {author} {\bibfnamefont {Y.}~\bibnamefont {Ma}},\ }\bibfield  {title} {\bibinfo {title} {{Probing phase transitions in neutron stars via the crust-core interfacial mode}},\ }\href {https://doi.org/10.1103/PhysRevD.107.083023} {\bibfield  {journal} {\bibinfo  {journal} {Phys. Rev. D}\ }\textbf {\bibinfo {volume} {107}},\ \bibinfo {pages} {083023} (\bibinfo {year} {2023})},\ \Eprint {https://arxiv.org/abs/2211.11529} {arXiv:2211.11529 [astro-ph.HE]} \BibitemShut {NoStop}%
\bibitem [{\citenamefont {Lau}\ and\ \citenamefont {Yagi}(2021)}]{Lau:2020bfq}%
  \BibitemOpen
  \bibfield  {author} {\bibinfo {author} {\bibfnamefont {S.~Y.}\ \bibnamefont {Lau}}\ and\ \bibinfo {author} {\bibfnamefont {K.}~\bibnamefont {Yagi}},\ }\bibfield  {title} {\bibinfo {title} {{Probing hybrid stars with gravitational waves via interfacial modes}},\ }\href {https://doi.org/10.1103/PhysRevD.103.063015} {\bibfield  {journal} {\bibinfo  {journal} {Phys. Rev. D}\ }\textbf {\bibinfo {volume} {103}},\ \bibinfo {pages} {063015} (\bibinfo {year} {2021})},\ \Eprint {https://arxiv.org/abs/2012.13000} {arXiv:2012.13000 [astro-ph.HE]} \BibitemShut {NoStop}%
\bibitem [{\citenamefont {Rodriguez}\ \emph {et~al.}(2025)\citenamefont {Rodriguez}, \citenamefont {Jim{\'e}nez},\ and\ \citenamefont {Ranea-Sandoval}}]{Rodriguez:2025oes}%
  \BibitemOpen
  \bibfield  {author} {\bibinfo {author} {\bibfnamefont {M.~C.}\ \bibnamefont {Rodriguez}}, \bibinfo {author} {\bibfnamefont {J.~C.}\ \bibnamefont {Jim{\'e}nez}},\ and\ \bibinfo {author} {\bibfnamefont {I.~F.}\ \bibnamefont {Ranea-Sandoval}},\ }\bibfield  {title} {\bibinfo {title} {{Classification of $g$-modes for neutron stars with a strong transition: Novel universal relation including slow stable twin stars}},\ }\href@noop {} {\bibfield  {journal} {\bibinfo  {journal} {arXiv e-prints}\ } (\bibinfo {year} {2025})},\ \Eprint {https://arxiv.org/abs/2510.06463} {arXiv:2510.06463 [hep-ph]} \BibitemShut {NoStop}%
\bibitem [{\citenamefont {Pereira}\ \emph {et~al.}(2025)\citenamefont {Pereira}, \citenamefont {Tonetto}, \citenamefont {Bejger}, \citenamefont {Zdunik},\ and\ \citenamefont {Haensel}}]{Pereira:2025xsi}%
  \BibitemOpen
  \bibfield  {author} {\bibinfo {author} {\bibfnamefont {J.~P.}\ \bibnamefont {Pereira}}, \bibinfo {author} {\bibfnamefont {L.}~\bibnamefont {Tonetto}}, \bibinfo {author} {\bibfnamefont {M.}~\bibnamefont {Bejger}}, \bibinfo {author} {\bibfnamefont {J.~L.}\ \bibnamefont {Zdunik}},\ and\ \bibinfo {author} {\bibfnamefont {P.}~\bibnamefont {Haensel}},\ }\bibfield  {title} {\bibinfo {title} {{Dynamical Tides in Neutron Stars with First-Order Phase Transitions: The Role of the Discontinuity Mode}},\ }\href {https://doi.org/10.1103/k7l9-hw8g} {\bibfield  {journal} {\bibinfo  {journal} {Phys. Rev. Lett.}\ }\textbf {\bibinfo {volume} {135}},\ \bibinfo {pages} {231401} (\bibinfo {year} {2025})},\ \Eprint {https://arxiv.org/abs/2504.16911} {arXiv:2504.16911 [astro-ph.HE]} \BibitemShut {NoStop}%
\bibitem [{\citenamefont {Christian}\ and\ \citenamefont {Schaffner-Bielich}(2020)}]{Christian:2019qer}%
  \BibitemOpen
  \bibfield  {author} {\bibinfo {author} {\bibfnamefont {J.-E.}\ \bibnamefont {Christian}}\ and\ \bibinfo {author} {\bibfnamefont {J.}~\bibnamefont {Schaffner-Bielich}},\ }\bibfield  {title} {\bibinfo {title} {{Twin Stars and the Stiffness of the Nuclear Equation of State: Ruling Out Strong Phase Transitions below $1.7n_0$ with the New NICER Radius Measurements}},\ }\href {https://doi.org/10.3847/2041-8213/ab8af4} {\bibfield  {journal} {\bibinfo  {journal} {Astrophys. J. Lett.}\ }\textbf {\bibinfo {volume} {894}},\ \bibinfo {pages} {L8} (\bibinfo {year} {2020})},\ \Eprint {https://arxiv.org/abs/1912.09809} {arXiv:1912.09809 [astro-ph.HE]} \BibitemShut {NoStop}%
\bibitem [{\citenamefont {Christian}\ and\ \citenamefont {Schaffner-Bielich}(2021)}]{Christian:2020xwz}%
  \BibitemOpen
  \bibfield  {author} {\bibinfo {author} {\bibfnamefont {J.-E.}\ \bibnamefont {Christian}}\ and\ \bibinfo {author} {\bibfnamefont {J.}~\bibnamefont {Schaffner-Bielich}},\ }\bibfield  {title} {\bibinfo {title} {{Supermassive Neutron Stars Rule Out Twin Stars}},\ }\href {https://doi.org/10.1103/PhysRevD.103.063042} {\bibfield  {journal} {\bibinfo  {journal} {Phys. Rev. D}\ }\textbf {\bibinfo {volume} {103}},\ \bibinfo {pages} {063042} (\bibinfo {year} {2021})},\ \Eprint {https://arxiv.org/abs/2011.01001} {arXiv:2011.01001 [astro-ph.HE]} \BibitemShut {NoStop}%
\bibitem [{\citenamefont {Christian}\ and\ \citenamefont {Schaffner-Bielich}(2022)}]{Christian:2021uhd}%
  \BibitemOpen
  \bibfield  {author} {\bibinfo {author} {\bibfnamefont {J.-E.}\ \bibnamefont {Christian}}\ and\ \bibinfo {author} {\bibfnamefont {J.}~\bibnamefont {Schaffner-Bielich}},\ }\bibfield  {title} {\bibinfo {title} {{Confirming the Existence of Twin Stars in a NICER Way}},\ }\href {https://doi.org/10.3847/1538-4357/ac75cf} {\bibfield  {journal} {\bibinfo  {journal} {Astrophys. J.}\ }\textbf {\bibinfo {volume} {935}},\ \bibinfo {pages} {122} (\bibinfo {year} {2022})},\ \Eprint {https://arxiv.org/abs/2109.04191} {arXiv:2109.04191 [astro-ph.HE]} \BibitemShut {NoStop}%
\bibitem [{\citenamefont {Christian}\ \emph {et~al.}(2024)\citenamefont {Christian}, \citenamefont {Schaffner-Bielich},\ and\ \citenamefont {Rosswog}}]{Christian:2023hez}%
  \BibitemOpen
  \bibfield  {author} {\bibinfo {author} {\bibfnamefont {J.-E.}\ \bibnamefont {Christian}}, \bibinfo {author} {\bibfnamefont {J.}~\bibnamefont {Schaffner-Bielich}},\ and\ \bibinfo {author} {\bibfnamefont {S.}~\bibnamefont {Rosswog}},\ }\bibfield  {title} {\bibinfo {title} {{Which first order phase transitions to quark matter are possible in neutron stars?}},\ }\href {https://doi.org/10.1103/PhysRevD.109.063035} {\bibfield  {journal} {\bibinfo  {journal} {Phys. Rev. D}\ }\textbf {\bibinfo {volume} {109}},\ \bibinfo {pages} {063035} (\bibinfo {year} {2024})},\ \Eprint {https://arxiv.org/abs/2312.10148} {arXiv:2312.10148 [nucl-th]} \BibitemShut {NoStop}%
\bibitem [{\citenamefont {Blomqvist}\ \emph {et~al.}(2026)\citenamefont {Blomqvist}, \citenamefont {Ecker}, \citenamefont {Gorda},\ and\ \citenamefont {Vuorinen}}]{Blomqvist:2025cxe}%
  \BibitemOpen
  \bibfield  {author} {\bibinfo {author} {\bibfnamefont {S.}~\bibnamefont {Blomqvist}}, \bibinfo {author} {\bibfnamefont {C.}~\bibnamefont {Ecker}}, \bibinfo {author} {\bibfnamefont {T.}~\bibnamefont {Gorda}},\ and\ \bibinfo {author} {\bibfnamefont {A.}~\bibnamefont {Vuorinen}},\ }\bibfield  {title} {\bibinfo {title} {{Strong model-agnostic constraints for twin-star solutions}},\ }\href {https://doi.org/10.1103/f48x-c4zt} {\bibfield  {journal} {\bibinfo  {journal} {Phys. Rev. D}\ }\textbf {\bibinfo {volume} {113}},\ \bibinfo {pages} {L111504} (\bibinfo {year} {2026})},\ \Eprint {https://arxiv.org/abs/2512.19477} {arXiv:2512.19477 [astro-ph.HE]} \BibitemShut {NoStop}%
\bibitem [{\citenamefont {Sagert}\ \emph {et~al.}(2009)\citenamefont {Sagert}, \citenamefont {Fischer}, \citenamefont {Hempel}, \citenamefont {Pagliara}, \citenamefont {Schaffner-Bielich}, \citenamefont {Mezzacappa}, \citenamefont {Thielemann},\ and\ \citenamefont {Liebendorfer}}]{Sagert:2008ka}%
  \BibitemOpen
  \bibfield  {author} {\bibinfo {author} {\bibfnamefont {I.}~\bibnamefont {Sagert}}, \bibinfo {author} {\bibfnamefont {T.}~\bibnamefont {Fischer}}, \bibinfo {author} {\bibfnamefont {M.}~\bibnamefont {Hempel}}, \bibinfo {author} {\bibfnamefont {G.}~\bibnamefont {Pagliara}}, \bibinfo {author} {\bibfnamefont {J.}~\bibnamefont {Schaffner-Bielich}}, \bibinfo {author} {\bibfnamefont {A.}~\bibnamefont {Mezzacappa}}, \bibinfo {author} {\bibfnamefont {F.~K.}\ \bibnamefont {Thielemann}},\ and\ \bibinfo {author} {\bibfnamefont {M.}~\bibnamefont {Liebendorfer}},\ }\bibfield  {title} {\bibinfo {title} {{Signals of the QCD phase transition in core-collapse supernovae}},\ }\href {https://doi.org/10.1103/PhysRevLett.102.081101} {\bibfield  {journal} {\bibinfo  {journal} {Phys. Rev. Lett.}\ }\textbf {\bibinfo {volume} {102}},\ \bibinfo {pages} {081101} (\bibinfo {year} {2009})},\ \Eprint {https://arxiv.org/abs/0809.4225} {arXiv:0809.4225 [astro-ph]} \BibitemShut {NoStop}%
\bibitem [{\citenamefont {Fischer}\ \emph {et~al.}(2011)\citenamefont {Fischer}, \citenamefont {Sagert}, \citenamefont {Pagliara}, \citenamefont {Hempel}, \citenamefont {Schaffner-Bielich}, \citenamefont {Rauscher}, \citenamefont {Thielemann}, \citenamefont {Kappeli}, \citenamefont {Martinez-Pinedo},\ and\ \citenamefont {Liebendorfer}}]{Fischer:2010wp}%
  \BibitemOpen
  \bibfield  {author} {\bibinfo {author} {\bibfnamefont {T.}~\bibnamefont {Fischer}}, \bibinfo {author} {\bibfnamefont {I.}~\bibnamefont {Sagert}}, \bibinfo {author} {\bibfnamefont {G.}~\bibnamefont {Pagliara}}, \bibinfo {author} {\bibfnamefont {M.}~\bibnamefont {Hempel}}, \bibinfo {author} {\bibfnamefont {J.}~\bibnamefont {Schaffner-Bielich}}, \bibinfo {author} {\bibfnamefont {T.}~\bibnamefont {Rauscher}}, \bibinfo {author} {\bibfnamefont {F.~K.}\ \bibnamefont {Thielemann}}, \bibinfo {author} {\bibfnamefont {R.}~\bibnamefont {Kappeli}}, \bibinfo {author} {\bibfnamefont {G.}~\bibnamefont {Martinez-Pinedo}},\ and\ \bibinfo {author} {\bibfnamefont {M.}~\bibnamefont {Liebendorfer}},\ }\bibfield  {title} {\bibinfo {title} {{Core-collapse supernova explosions triggered by a quark-hadron phase transition during the early post-bounce phase}},\ }\href {https://doi.org/10.1088/0067-0049/194/2/39} {\bibfield  {journal} {\bibinfo  {journal} {Astrophys. J. Suppl.}\ }\textbf {\bibinfo {volume} {194}},\ \bibinfo {pages}
  {39} (\bibinfo {year} {2011})},\ \Eprint {https://arxiv.org/abs/1011.3409} {arXiv:1011.3409 [astro-ph.HE]} \BibitemShut {NoStop}%
\bibitem [{\citenamefont {Fischer}\ \emph {et~al.}(2018)\citenamefont {Fischer}, \citenamefont {Bastian}, \citenamefont {Wu}, \citenamefont {Baklanov}, \citenamefont {Sorokina}, \citenamefont {Blinnikov}, \citenamefont {Typel}, \citenamefont {Kl{\"a}hn},\ and\ \citenamefont {Blaschke}}]{Fischer:2017lag}%
  \BibitemOpen
  \bibfield  {author} {\bibinfo {author} {\bibfnamefont {T.}~\bibnamefont {Fischer}}, \bibinfo {author} {\bibfnamefont {N.-U.~F.}\ \bibnamefont {Bastian}}, \bibinfo {author} {\bibfnamefont {M.-R.}\ \bibnamefont {Wu}}, \bibinfo {author} {\bibfnamefont {P.}~\bibnamefont {Baklanov}}, \bibinfo {author} {\bibfnamefont {E.}~\bibnamefont {Sorokina}}, \bibinfo {author} {\bibfnamefont {S.}~\bibnamefont {Blinnikov}}, \bibinfo {author} {\bibfnamefont {S.}~\bibnamefont {Typel}}, \bibinfo {author} {\bibfnamefont {T.}~\bibnamefont {Kl{\"a}hn}},\ and\ \bibinfo {author} {\bibfnamefont {D.~B.}\ \bibnamefont {Blaschke}},\ }\bibfield  {title} {\bibinfo {title} {{Quark deconfinement as a supernova explosion engine for massive blue supergiant stars}},\ }\href {https://doi.org/10.1038/s41550-018-0583-0} {\bibfield  {journal} {\bibinfo  {journal} {Nature Astron.}\ }\textbf {\bibinfo {volume} {2}},\ \bibinfo {pages} {980} (\bibinfo {year} {2018})},\ \Eprint {https://arxiv.org/abs/1712.08788} {arXiv:1712.08788 [astro-ph.HE]}
  \BibitemShut {NoStop}%
\bibitem [{\citenamefont {Zha}\ \emph {et~al.}(2020)\citenamefont {Zha}, \citenamefont {O'Connor}, \citenamefont {Chu}, \citenamefont {Lin},\ and\ \citenamefont {Couch}}]{Zha:2020gjw}%
  \BibitemOpen
  \bibfield  {author} {\bibinfo {author} {\bibfnamefont {S.}~\bibnamefont {Zha}}, \bibinfo {author} {\bibfnamefont {E.~P.}\ \bibnamefont {O'Connor}}, \bibinfo {author} {\bibfnamefont {M.-c.}\ \bibnamefont {Chu}}, \bibinfo {author} {\bibfnamefont {L.-M.}\ \bibnamefont {Lin}},\ and\ \bibinfo {author} {\bibfnamefont {S.~M.}\ \bibnamefont {Couch}},\ }\bibfield  {title} {\bibinfo {title} {{Gravitational-Wave Signature of a First-Order Quantum Chromodynamics Phase Transition in Core-Collapse Supernovae}},\ }\href {https://doi.org/10.1103/PhysRevLett.127.219901} {\bibfield  {journal} {\bibinfo  {journal} {Phys. Rev. Lett.}\ }\textbf {\bibinfo {volume} {125}},\ \bibinfo {pages} {051102} (\bibinfo {year} {2020})},\ \bibinfo {note} {[Erratum: Phys.Rev.Lett. 127, 219901 (2021)]},\ \Eprint {https://arxiv.org/abs/2007.04716} {arXiv:2007.04716 [astro-ph.HE]} \BibitemShut {NoStop}%
\bibitem [{\citenamefont {Kuroda}\ \emph {et~al.}(2022)\citenamefont {Kuroda}, \citenamefont {Fischer}, \citenamefont {Takiwaki},\ and\ \citenamefont {Kotake}}]{Kuroda:2021eiv}%
  \BibitemOpen
  \bibfield  {author} {\bibinfo {author} {\bibfnamefont {T.}~\bibnamefont {Kuroda}}, \bibinfo {author} {\bibfnamefont {T.}~\bibnamefont {Fischer}}, \bibinfo {author} {\bibfnamefont {T.}~\bibnamefont {Takiwaki}},\ and\ \bibinfo {author} {\bibfnamefont {K.}~\bibnamefont {Kotake}},\ }\bibfield  {title} {\bibinfo {title} {{Core-collapse Supernova Simulations and the Formation of Neutron Stars, Hybrid Stars, and Black Holes}},\ }\href {https://doi.org/10.3847/1538-4357/ac31a8} {\bibfield  {journal} {\bibinfo  {journal} {Astrophys. J.}\ }\textbf {\bibinfo {volume} {924}},\ \bibinfo {pages} {38} (\bibinfo {year} {2022})},\ \Eprint {https://arxiv.org/abs/2109.01508} {arXiv:2109.01508 [astro-ph.HE]} \BibitemShut {NoStop}%
\bibitem [{\citenamefont {Bauswein}\ \emph {et~al.}(2019)\citenamefont {Bauswein}, \citenamefont {Bastian}, \citenamefont {Blaschke}, \citenamefont {Chatziioannou}, \citenamefont {Clark}, \citenamefont {Fischer},\ and\ \citenamefont {Oertel}}]{Bauswein:2018bma}%
  \BibitemOpen
  \bibfield  {author} {\bibinfo {author} {\bibfnamefont {A.}~\bibnamefont {Bauswein}}, \bibinfo {author} {\bibfnamefont {N.-U.~F.}\ \bibnamefont {Bastian}}, \bibinfo {author} {\bibfnamefont {D.~B.}\ \bibnamefont {Blaschke}}, \bibinfo {author} {\bibfnamefont {K.}~\bibnamefont {Chatziioannou}}, \bibinfo {author} {\bibfnamefont {J.~A.}\ \bibnamefont {Clark}}, \bibinfo {author} {\bibfnamefont {T.}~\bibnamefont {Fischer}},\ and\ \bibinfo {author} {\bibfnamefont {M.}~\bibnamefont {Oertel}},\ }\bibfield  {title} {\bibinfo {title} {{Identifying a first-order phase transition in neutron star mergers through gravitational waves}},\ }\href {https://doi.org/10.1103/PhysRevLett.122.061102} {\bibfield  {journal} {\bibinfo  {journal} {Phys. Rev. Lett.}\ }\textbf {\bibinfo {volume} {122}},\ \bibinfo {pages} {061102} (\bibinfo {year} {2019})},\ \Eprint {https://arxiv.org/abs/1809.01116} {arXiv:1809.01116 [astro-ph.HE]} \BibitemShut {NoStop}%
\bibitem [{\citenamefont {Most}\ \emph {et~al.}(2020)\citenamefont {Most}, \citenamefont {Jens~Papenfort}, \citenamefont {Dexheimer}, \citenamefont {Hanauske}, \citenamefont {Stoecker},\ and\ \citenamefont {Rezzolla}}]{Most:2019onn}%
  \BibitemOpen
  \bibfield  {author} {\bibinfo {author} {\bibfnamefont {E.~R.}\ \bibnamefont {Most}}, \bibinfo {author} {\bibfnamefont {L.}~\bibnamefont {Jens~Papenfort}}, \bibinfo {author} {\bibfnamefont {V.}~\bibnamefont {Dexheimer}}, \bibinfo {author} {\bibfnamefont {M.}~\bibnamefont {Hanauske}}, \bibinfo {author} {\bibfnamefont {H.}~\bibnamefont {Stoecker}},\ and\ \bibinfo {author} {\bibfnamefont {L.}~\bibnamefont {Rezzolla}},\ }\bibfield  {title} {\bibinfo {title} {{On the deconfinement phase transition in neutron-star mergers}},\ }\href {https://doi.org/10.1140/epja/s10050-020-00073-4} {\bibfield  {journal} {\bibinfo  {journal} {Eur. Phys. J. A}\ }\textbf {\bibinfo {volume} {56}},\ \bibinfo {pages} {59} (\bibinfo {year} {2020})},\ \Eprint {https://arxiv.org/abs/1910.13893} {arXiv:1910.13893 [astro-ph.HE]} \BibitemShut {NoStop}%
\bibitem [{\citenamefont {Gieg}\ \emph {et~al.}(2019)\citenamefont {Gieg}, \citenamefont {Dietrich},\ and\ \citenamefont {Ujevic}}]{Gieg:2019yzq}%
  \BibitemOpen
  \bibfield  {author} {\bibinfo {author} {\bibfnamefont {H.}~\bibnamefont {Gieg}}, \bibinfo {author} {\bibfnamefont {T.}~\bibnamefont {Dietrich}},\ and\ \bibinfo {author} {\bibfnamefont {M.}~\bibnamefont {Ujevic}},\ }\bibfield  {title} {\bibinfo {title} {{Simulating Binary Neutron Stars with Hybrid Equation of States: Gravitational Waves, Electromagnetic Signatures, and Challenges for Numerical Relativity}},\ }\href {https://doi.org/10.3390/particles2030023} {\bibfield  {journal} {\bibinfo  {journal} {Particles}\ }\textbf {\bibinfo {volume} {2}},\ \bibinfo {pages} {365} (\bibinfo {year} {2019})},\ \Eprint {https://arxiv.org/abs/1908.03135} {arXiv:1908.03135 [gr-qc]} \BibitemShut {NoStop}%
\bibitem [{\citenamefont {Blacker}\ \emph {et~al.}(2020)\citenamefont {Blacker}, \citenamefont {Bastian}, \citenamefont {Bauswein}, \citenamefont {Blaschke}, \citenamefont {Fischer}, \citenamefont {Oertel}, \citenamefont {Soultanis},\ and\ \citenamefont {Typel}}]{Blacker:2020nlq}%
  \BibitemOpen
  \bibfield  {author} {\bibinfo {author} {\bibfnamefont {S.}~\bibnamefont {Blacker}}, \bibinfo {author} {\bibfnamefont {N.-U.~F.}\ \bibnamefont {Bastian}}, \bibinfo {author} {\bibfnamefont {A.}~\bibnamefont {Bauswein}}, \bibinfo {author} {\bibfnamefont {D.~B.}\ \bibnamefont {Blaschke}}, \bibinfo {author} {\bibfnamefont {T.}~\bibnamefont {Fischer}}, \bibinfo {author} {\bibfnamefont {M.}~\bibnamefont {Oertel}}, \bibinfo {author} {\bibfnamefont {T.}~\bibnamefont {Soultanis}},\ and\ \bibinfo {author} {\bibfnamefont {S.}~\bibnamefont {Typel}},\ }\bibfield  {title} {\bibinfo {title} {{Constraining the onset density of the hadron-quark phase transition with gravitational-wave observations}},\ }\href {https://doi.org/10.1103/PhysRevD.102.123023} {\bibfield  {journal} {\bibinfo  {journal} {Phys. Rev. D}\ }\textbf {\bibinfo {volume} {102}},\ \bibinfo {pages} {123023} (\bibinfo {year} {2020})},\ \Eprint {https://arxiv.org/abs/2006.03789} {arXiv:2006.03789 [astro-ph.HE]} \BibitemShut {NoStop}%
\bibitem [{\citenamefont {{Sana}}\ \emph {et~al.}(2012)\citenamefont {{Sana}}, \citenamefont {{de Mink}}, \citenamefont {{de Koter}}, \citenamefont {{Langer}}, \citenamefont {{Evans}}, \citenamefont {{Gieles}}, \citenamefont {{Gosset}}, \citenamefont {{Izzard}}, \citenamefont {{Le Bouquin}},\ and\ \citenamefont {{Schneider}}}]{Sana:2012}%
  \BibitemOpen
  \bibfield  {author} {\bibinfo {author} {\bibfnamefont {H.}~\bibnamefont {{Sana}}}, \bibinfo {author} {\bibfnamefont {S.~E.}\ \bibnamefont {{de Mink}}}, \bibinfo {author} {\bibfnamefont {A.}~\bibnamefont {{de Koter}}}, \bibinfo {author} {\bibfnamefont {N.}~\bibnamefont {{Langer}}}, \bibinfo {author} {\bibfnamefont {C.~J.}\ \bibnamefont {{Evans}}}, \bibinfo {author} {\bibfnamefont {M.}~\bibnamefont {{Gieles}}}, \bibinfo {author} {\bibfnamefont {E.}~\bibnamefont {{Gosset}}}, \bibinfo {author} {\bibfnamefont {R.~G.}\ \bibnamefont {{Izzard}}}, \bibinfo {author} {\bibfnamefont {J.-B.}\ \bibnamefont {{Le Bouquin}}},\ and\ \bibinfo {author} {\bibfnamefont {F.~R.~N.}\ \bibnamefont {{Schneider}}},\ }\bibfield  {title} {\bibinfo {title} {{Binary Interaction Dominates the Evolution of Massive Stars}},\ }\href {https://doi.org/10.1126/science.1223344} {\bibfield  {journal} {\bibinfo  {journal} {Science}\ }\textbf {\bibinfo {volume} {337}},\ \bibinfo {pages} {444} (\bibinfo {year} {2012})},\ \Eprint
  {https://arxiv.org/abs/1207.6397} {arXiv:1207.6397 [astro-ph.SR]} \BibitemShut {NoStop}%
\bibitem [{\citenamefont {{Duch{\^e}ne}}\ and\ \citenamefont {{Kraus}}(2013)}]{Duchene:2013}%
  \BibitemOpen
  \bibfield  {author} {\bibinfo {author} {\bibfnamefont {G.}~\bibnamefont {{Duch{\^e}ne}}}\ and\ \bibinfo {author} {\bibfnamefont {A.}~\bibnamefont {{Kraus}}},\ }\bibfield  {title} {\bibinfo {title} {{Stellar Multiplicity}},\ }\href {https://doi.org/10.1146/annurev-astro-081710-102602} {\bibfield  {journal} {\bibinfo  {journal} {\araa}\ }\textbf {\bibinfo {volume} {51}},\ \bibinfo {pages} {269} (\bibinfo {year} {2013})},\ \Eprint {https://arxiv.org/abs/1303.3028} {arXiv:1303.3028 [astro-ph.SR]} \BibitemShut {NoStop}%
\bibitem [{\citenamefont {{Moe}}\ and\ \citenamefont {{Di Stefano}}(2017)}]{Moe:2017}%
  \BibitemOpen
  \bibfield  {author} {\bibinfo {author} {\bibfnamefont {M.}~\bibnamefont {{Moe}}}\ and\ \bibinfo {author} {\bibfnamefont {R.}~\bibnamefont {{Di Stefano}}},\ }\bibfield  {title} {\bibinfo {title} {{Mind Your Ps and Qs: The Interrelation between Period (P) and Mass-ratio (Q) Distributions of Binary Stars}},\ }\href {https://doi.org/10.3847/1538-4365/aa6fb6} {\bibfield  {journal} {\bibinfo  {journal} {\apjs}\ }\textbf {\bibinfo {volume} {230}},\ \bibinfo {eid} {15} (\bibinfo {year} {2017})},\ \Eprint {https://arxiv.org/abs/1606.05347} {arXiv:1606.05347 [astro-ph.SR]} \BibitemShut {NoStop}%
\bibitem [{\citenamefont {Postnov}\ and\ \citenamefont {Yungelson}(2014)}]{Postnov:2014tza}%
  \BibitemOpen
  \bibfield  {author} {\bibinfo {author} {\bibfnamefont {K.~A.}\ \bibnamefont {Postnov}}\ and\ \bibinfo {author} {\bibfnamefont {L.~R.}\ \bibnamefont {Yungelson}},\ }\bibfield  {title} {\bibinfo {title} {{The Evolution of Compact Binary Star Systems}},\ }\href {https://doi.org/10.12942/lrr-2014-3} {\bibfield  {journal} {\bibinfo  {journal} {Living Rev. Rel.}\ }\textbf {\bibinfo {volume} {17}},\ \bibinfo {pages} {3} (\bibinfo {year} {2014})},\ \Eprint {https://arxiv.org/abs/1403.4754} {arXiv:1403.4754 [astro-ph.HE]} \BibitemShut {NoStop}%
\bibitem [{\citenamefont {Tauris}\ \emph {et~al.}(2017)\citenamefont {Tauris} \emph {et~al.}}]{Tauris:2017omb}%
  \BibitemOpen
  \bibfield  {author} {\bibinfo {author} {\bibfnamefont {T.~M.}\ \bibnamefont {Tauris}} \emph {et~al.},\ }\bibfield  {title} {\bibinfo {title} {{Formation of Double Neutron Star Systems}},\ }\href {https://doi.org/10.3847/1538-4357/aa7e89} {\bibfield  {journal} {\bibinfo  {journal} {Astrophys. J.}\ }\textbf {\bibinfo {volume} {846}},\ \bibinfo {pages} {170} (\bibinfo {year} {2017})},\ \Eprint {https://arxiv.org/abs/1706.09438} {arXiv:1706.09438 [astro-ph.HE]} \BibitemShut {NoStop}%
\bibitem [{\citenamefont {Vigna-G{\'o}mez}\ \emph {et~al.}(2018)\citenamefont {Vigna-G{\'o}mez} \emph {et~al.}}]{Vigna-Gomez:2018dza}%
  \BibitemOpen
  \bibfield  {author} {\bibinfo {author} {\bibfnamefont {A.}~\bibnamefont {Vigna-G{\'o}mez}} \emph {et~al.},\ }\bibfield  {title} {\bibinfo {title} {{On the formation history of Galactic double neutron stars}},\ }\href {https://doi.org/10.1093/mnras/sty2463} {\bibfield  {journal} {\bibinfo  {journal} {Mon. Not. Roy. Astron. Soc.}\ }\textbf {\bibinfo {volume} {481}},\ \bibinfo {pages} {4009} (\bibinfo {year} {2018})},\ \Eprint {https://arxiv.org/abs/1805.07974} {arXiv:1805.07974 [astro-ph.SR]} \BibitemShut {NoStop}%
\bibitem [{\citenamefont {{Bhattacharya}}\ and\ \citenamefont {{van den Heuvel}}(1991)}]{Bhattacharya:1991}%
  \BibitemOpen
  \bibfield  {author} {\bibinfo {author} {\bibfnamefont {D.}~\bibnamefont {{Bhattacharya}}}\ and\ \bibinfo {author} {\bibfnamefont {E.~P.~J.}\ \bibnamefont {{van den Heuvel}}},\ }\bibfield  {title} {\bibinfo {title} {{Formation and evolution of binary and millisecond radio pulsars}},\ }\href {https://doi.org/10.1016/0370-1573(91)90064-S} {\bibfield  {journal} {\bibinfo  {journal} {\physrep}\ }\textbf {\bibinfo {volume} {203}},\ \bibinfo {pages} {1} (\bibinfo {year} {1991})}\BibitemShut {NoStop}%
\bibitem [{\citenamefont {{Tauris}}\ \emph {et~al.}(2012)\citenamefont {{Tauris}}, \citenamefont {{Langer}},\ and\ \citenamefont {{Kramer}}}]{Tauris:2012}%
  \BibitemOpen
  \bibfield  {author} {\bibinfo {author} {\bibfnamefont {T.~M.}\ \bibnamefont {{Tauris}}}, \bibinfo {author} {\bibfnamefont {N.}~\bibnamefont {{Langer}}},\ and\ \bibinfo {author} {\bibfnamefont {M.}~\bibnamefont {{Kramer}}},\ }\bibfield  {title} {\bibinfo {title} {{Formation of millisecond pulsars with CO white dwarf companions - II. Accretion, spin-up, true ages and comparison to MSPs with He white dwarf companions}},\ }\href {https://doi.org/10.1111/j.1365-2966.2012.21446.x} {\bibfield  {journal} {\bibinfo  {journal} {\mnras}\ }\textbf {\bibinfo {volume} {425}},\ \bibinfo {pages} {1601} (\bibinfo {year} {2012})},\ \Eprint {https://arxiv.org/abs/1206.1862} {arXiv:1206.1862 [astro-ph.SR]} \BibitemShut {NoStop}%
\bibitem [{\citenamefont {Largani}\ \emph {et~al.}(2024)\citenamefont {Largani}, \citenamefont {Fischer}, \citenamefont {Shibagaki}, \citenamefont {Cerd{\'a}-Dur{\'a}n},\ and\ \citenamefont {Torres-Forn{\'e}}}]{Largani:2023kjx}%
  \BibitemOpen
  \bibfield  {author} {\bibinfo {author} {\bibfnamefont {N.~K.}\ \bibnamefont {Largani}}, \bibinfo {author} {\bibfnamefont {T.}~\bibnamefont {Fischer}}, \bibinfo {author} {\bibfnamefont {S.}~\bibnamefont {Shibagaki}}, \bibinfo {author} {\bibfnamefont {P.}~\bibnamefont {Cerd{\'a}-Dur{\'a}n}},\ and\ \bibinfo {author} {\bibfnamefont {A.}~\bibnamefont {Torres-Forn{\'e}}},\ }\bibfield  {title} {\bibinfo {title} {{Neutron stars in accreting systems {\textendash} Signatures of the QCD phase transition}},\ }\href {https://doi.org/10.1051/0004-6361/202348742} {\bibfield  {journal} {\bibinfo  {journal} {Astron. Astrophys.}\ }\textbf {\bibinfo {volume} {687}},\ \bibinfo {pages} {A245} (\bibinfo {year} {2024})},\ \Eprint {https://arxiv.org/abs/2311.15992} {arXiv:2311.15992 [astro-ph.HE]} \BibitemShut {NoStop}%
\bibitem [{\citenamefont {Haque}\ \emph {et~al.}(2026)\citenamefont {Haque}, \citenamefont {Rezzolla},\ and\ \citenamefont {Mallick}}]{Haque:2026ero}%
  \BibitemOpen
  \bibfield  {author} {\bibinfo {author} {\bibfnamefont {S.}~\bibnamefont {Haque}}, \bibinfo {author} {\bibfnamefont {L.}~\bibnamefont {Rezzolla}},\ and\ \bibinfo {author} {\bibfnamefont {R.}~\bibnamefont {Mallick}},\ }\bibfield  {title} {\bibinfo {title} {{Dynamical response of twin stars to perturbations}},\ }\href {https://doi.org/10.1103/99c1-2p2y} {\bibfield  {journal} {\bibinfo  {journal} {Phys. Rev. D}\ }\textbf {\bibinfo {volume} {113}},\ \bibinfo {pages} {103044} (\bibinfo {year} {2026})},\ \Eprint {https://arxiv.org/abs/2602.02654} {arXiv:2602.02654 [gr-qc]} \BibitemShut {NoStop}%
\bibitem [{\citenamefont {Naseri}\ and\ \citenamefont {Paschalidis}(2026)}]{Naseri:2026xrq}%
  \BibitemOpen
  \bibfield  {author} {\bibinfo {author} {\bibfnamefont {M.}~\bibnamefont {Naseri}}\ and\ \bibinfo {author} {\bibfnamefont {V.}~\bibnamefont {Paschalidis}},\ }\bibfield  {title} {\bibinfo {title} {{On the dynamical accessibility of twin stars}},\ }\href@noop {} {\bibfield  {journal} {\bibinfo  {journal} {arXiv e-prints}\ } (\bibinfo {year} {2026})},\ \Eprint {https://arxiv.org/abs/2609.16126} {arXiv:2609.16126 [astro-ph.HE]} \BibitemShut {NoStop}%
\bibitem [{\citenamefont {Ecker}\ \emph {et~al.}(2026)\citenamefont {Ecker}, \citenamefont {Giliberti},\ and\ \citenamefont {Rezzolla}}]{Ecker:2026ixk}%
  \BibitemOpen
  \bibfield  {author} {\bibinfo {author} {\bibfnamefont {C.}~\bibnamefont {Ecker}}, \bibinfo {author} {\bibfnamefont {M.}~\bibnamefont {Giliberti}},\ and\ \bibinfo {author} {\bibfnamefont {L.}~\bibnamefont {Rezzolla}},\ }\bibfield  {title} {\bibinfo {title} {{Multi-messenger and multi-band signal from first-order phase transitions in proto-neutron stars}},\ }\href@noop {} {\bibfield  {journal} {\bibinfo  {journal} {arXiv e-prints}\ } (\bibinfo {year} {2026})},\ \Eprint {https://arxiv.org/abs/2609.29761} {arXiv:2609.29761 [gr-qc]} \BibitemShut {NoStop}%
\bibitem [{\citenamefont {{Gerlach}}(1968)}]{Gerlach:1968}%
  \BibitemOpen
  \bibfield  {author} {\bibinfo {author} {\bibfnamefont {U.~H.}\ \bibnamefont {{Gerlach}}},\ }\bibfield  {title} {\bibinfo {title} {{Equation of State at Supranuclear Densities and the Existence of a Third Family of Superdense Stars}},\ }\href {https://doi.org/10.1103/PhysRev.172.1325} {\bibfield  {journal} {\bibinfo  {journal} {Physical Review}\ }\textbf {\bibinfo {volume} {172}},\ \bibinfo {pages} {1325} (\bibinfo {year} {1968})}\BibitemShut {NoStop}%
\bibitem [{\citenamefont {Glendenning}\ and\ \citenamefont {Kettner}(2000)}]{Glendenning:1998ag}%
  \BibitemOpen
  \bibfield  {author} {\bibinfo {author} {\bibfnamefont {N.~K.}\ \bibnamefont {Glendenning}}\ and\ \bibinfo {author} {\bibfnamefont {C.}~\bibnamefont {Kettner}},\ }\bibfield  {title} {\bibinfo {title} {{Nonidentical neutron star twins}},\ }\href@noop {} {\bibfield  {journal} {\bibinfo  {journal} {Astron. Astrophys.}\ }\textbf {\bibinfo {volume} {353}},\ \bibinfo {pages} {L9} (\bibinfo {year} {2000})},\ \Eprint {https://arxiv.org/abs/astro-ph/9807155} {arXiv:astro-ph/9807155} \BibitemShut {NoStop}%
\bibitem [{\citenamefont {Bastian}(2021)}]{Bastian:2020unt}%
  \BibitemOpen
  \bibfield  {author} {\bibinfo {author} {\bibfnamefont {N.-U.~F.}\ \bibnamefont {Bastian}},\ }\bibfield  {title} {\bibinfo {title} {{Phenomenological quark-hadron equations of state with first-order phase transitions for astrophysical applications}},\ }\href {https://doi.org/10.1103/PhysRevD.103.023001} {\bibfield  {journal} {\bibinfo  {journal} {Phys. Rev. D}\ }\textbf {\bibinfo {volume} {103}},\ \bibinfo {pages} {023001} (\bibinfo {year} {2021})},\ \Eprint {https://arxiv.org/abs/2009.10846} {arXiv:2009.10846 [nucl-th]} \BibitemShut {NoStop}%
\bibitem [{\citenamefont {Typel}\ \emph {et~al.}(2010)\citenamefont {Typel}, \citenamefont {Ropke}, \citenamefont {Klahn}, \citenamefont {Blaschke},\ and\ \citenamefont {Wolter}}]{Typel2010}%
  \BibitemOpen
  \bibfield  {author} {\bibinfo {author} {\bibfnamefont {S.}~\bibnamefont {Typel}}, \bibinfo {author} {\bibfnamefont {G.}~\bibnamefont {Ropke}}, \bibinfo {author} {\bibfnamefont {T.}~\bibnamefont {Klahn}}, \bibinfo {author} {\bibfnamefont {D.}~\bibnamefont {Blaschke}},\ and\ \bibinfo {author} {\bibfnamefont {H.}~\bibnamefont {Wolter}},\ }\bibfield  {title} {\bibinfo {title} {{Composition and thermodynamics of nuclear matter with light clusters}},\ }\href {https://doi.org/10.1103/PhysRevC.81.015803} {\bibfield  {journal} {\bibinfo  {journal} {Phys. Rev. C}\ }\textbf {\bibinfo {volume} {81}},\ \bibinfo {pages} {015803} (\bibinfo {year} {2010})}\BibitemShut {NoStop}%
\bibitem [{\citenamefont {Hempel}\ and\ \citenamefont {Schaffner-Bielich}(2010)}]{Hempel2009}%
  \BibitemOpen
  \bibfield  {author} {\bibinfo {author} {\bibfnamefont {M.}~\bibnamefont {Hempel}}\ and\ \bibinfo {author} {\bibfnamefont {J.}~\bibnamefont {Schaffner-Bielich}},\ }\bibfield  {title} {\bibinfo {title} {{Statistical Model for a Complete Supernova Equation of State}},\ }\href {https://doi.org/10.1016/j.nuclphysa.2010.02.010} {\bibfield  {journal} {\bibinfo  {journal} {Nucl. Phys. A}\ }\textbf {\bibinfo {volume} {837}},\ \bibinfo {pages} {210} (\bibinfo {year} {2010})}\BibitemShut {NoStop}%
\bibitem [{\citenamefont {Kaltenborn}\ \emph {et~al.}(2017)\citenamefont {Kaltenborn}, \citenamefont {Bastian},\ and\ \citenamefont {Blaschke}}]{Kaltenborn:2017hus}%
  \BibitemOpen
  \bibfield  {author} {\bibinfo {author} {\bibfnamefont {M.~A.~R.}\ \bibnamefont {Kaltenborn}}, \bibinfo {author} {\bibfnamefont {N.-U.~F.}\ \bibnamefont {Bastian}},\ and\ \bibinfo {author} {\bibfnamefont {D.~B.}\ \bibnamefont {Blaschke}},\ }\bibfield  {title} {\bibinfo {title} {{Quark-nuclear hybrid star equation of state with excluded volume effects}},\ }\href {https://doi.org/10.1103/PhysRevD.96.056024} {\bibfield  {journal} {\bibinfo  {journal} {Phys. Rev. D}\ }\textbf {\bibinfo {volume} {96}},\ \bibinfo {pages} {056024} (\bibinfo {year} {2017})},\ \Eprint {https://arxiv.org/abs/1701.04400} {arXiv:1701.04400 [astro-ph.HE]} \BibitemShut {NoStop}%
\bibitem [{\citenamefont {Schaffner-Bielich}(2020)}]{Schaffner-Bielich:2020psc}%
  \BibitemOpen
  \bibfield  {author} {\bibinfo {author} {\bibfnamefont {J.}~\bibnamefont {Schaffner-Bielich}},\ }\href {https://doi.org/10.1017/9781316848357} {\emph {\bibinfo {title} {{Compact Star Physics}}}}\ (\bibinfo  {publisher} {Cambridge University Press},\ \bibinfo {year} {2020})\BibitemShut {NoStop}%
\bibitem [{\citenamefont {Bruegmann}\ \emph {et~al.}(2008)\citenamefont {Bruegmann}, \citenamefont {Gonzalez}, \citenamefont {Hannam}, \citenamefont {Husa}, \citenamefont {Sperhake},\ and\ \citenamefont {Tichy}}]{Bruegmann:2006ulg}%
  \BibitemOpen
  \bibfield  {author} {\bibinfo {author} {\bibfnamefont {B.}~\bibnamefont {Bruegmann}}, \bibinfo {author} {\bibfnamefont {J.~A.}\ \bibnamefont {Gonzalez}}, \bibinfo {author} {\bibfnamefont {M.}~\bibnamefont {Hannam}}, \bibinfo {author} {\bibfnamefont {S.}~\bibnamefont {Husa}}, \bibinfo {author} {\bibfnamefont {U.}~\bibnamefont {Sperhake}},\ and\ \bibinfo {author} {\bibfnamefont {W.}~\bibnamefont {Tichy}},\ }\bibfield  {title} {\bibinfo {title} {{Calibration of Moving Puncture Simulations}},\ }\href {https://doi.org/10.1103/PhysRevD.77.024027} {\bibfield  {journal} {\bibinfo  {journal} {Phys. Rev. D}\ }\textbf {\bibinfo {volume} {77}},\ \bibinfo {pages} {024027} (\bibinfo {year} {2008})},\ \Eprint {https://arxiv.org/abs/gr-qc/0610128} {arXiv:gr-qc/0610128} \BibitemShut {NoStop}%
\bibitem [{\citenamefont {Thierfelder}\ \emph {et~al.}(2011)\citenamefont {Thierfelder}, \citenamefont {Bernuzzi},\ and\ \citenamefont {Bruegmann}}]{Thierfelder:2011yi}%
  \BibitemOpen
  \bibfield  {author} {\bibinfo {author} {\bibfnamefont {M.}~\bibnamefont {Thierfelder}}, \bibinfo {author} {\bibfnamefont {S.}~\bibnamefont {Bernuzzi}},\ and\ \bibinfo {author} {\bibfnamefont {B.}~\bibnamefont {Bruegmann}},\ }\bibfield  {title} {\bibinfo {title} {{Numerical relativity simulations of binary neutron stars}},\ }\href {https://doi.org/10.1103/PhysRevD.84.044012} {\bibfield  {journal} {\bibinfo  {journal} {Phys. Rev. D}\ }\textbf {\bibinfo {volume} {84}},\ \bibinfo {pages} {044012} (\bibinfo {year} {2011})},\ \Eprint {https://arxiv.org/abs/1104.4751} {arXiv:1104.4751 [gr-qc]} \BibitemShut {NoStop}%
\bibitem [{\citenamefont {Bernuzzi}\ and\ \citenamefont {Dietrich}(2016)}]{Bernuzzi:2016pie}%
  \BibitemOpen
  \bibfield  {author} {\bibinfo {author} {\bibfnamefont {S.}~\bibnamefont {Bernuzzi}}\ and\ \bibinfo {author} {\bibfnamefont {T.}~\bibnamefont {Dietrich}},\ }\bibfield  {title} {\bibinfo {title} {{Gravitational waveforms from binary neutron star mergers with high-order weighted-essentially-nonoscillatory schemes in numerical relativity}},\ }\href {https://doi.org/10.1103/PhysRevD.94.064062} {\bibfield  {journal} {\bibinfo  {journal} {Phys. Rev. D}\ }\textbf {\bibinfo {volume} {94}},\ \bibinfo {pages} {064062} (\bibinfo {year} {2016})},\ \Eprint {https://arxiv.org/abs/1604.07999} {arXiv:1604.07999 [gr-qc]} \BibitemShut {NoStop}%
\bibitem [{\citenamefont {Dietrich}\ \emph {et~al.}(2015)\citenamefont {Dietrich}, \citenamefont {Bernuzzi}, \citenamefont {Ujevic},\ and\ \citenamefont {Br{\"u}gmann}}]{Dietrich:2015iva}%
  \BibitemOpen
  \bibfield  {author} {\bibinfo {author} {\bibfnamefont {T.}~\bibnamefont {Dietrich}}, \bibinfo {author} {\bibfnamefont {S.}~\bibnamefont {Bernuzzi}}, \bibinfo {author} {\bibfnamefont {M.}~\bibnamefont {Ujevic}},\ and\ \bibinfo {author} {\bibfnamefont {B.}~\bibnamefont {Br{\"u}gmann}},\ }\bibfield  {title} {\bibinfo {title} {{Numerical relativity simulations of neutron star merger remnants using conservative mesh refinement}},\ }\href {https://doi.org/10.1103/PhysRevD.91.124041} {\bibfield  {journal} {\bibinfo  {journal} {Phys. Rev. D}\ }\textbf {\bibinfo {volume} {91}},\ \bibinfo {pages} {124041} (\bibinfo {year} {2015})},\ \Eprint {https://arxiv.org/abs/1504.01266} {arXiv:1504.01266 [gr-qc]} \BibitemShut {NoStop}%
\bibitem [{\citenamefont {Bernuzzi}\ and\ \citenamefont {Hilditch}(2010)}]{Bernuzzi:2009ex}%
  \BibitemOpen
  \bibfield  {author} {\bibinfo {author} {\bibfnamefont {S.}~\bibnamefont {Bernuzzi}}\ and\ \bibinfo {author} {\bibfnamefont {D.}~\bibnamefont {Hilditch}},\ }\bibfield  {title} {\bibinfo {title} {{Constraint violation in free evolution schemes: Comparing BSSNOK with a conformal decomposition of Z4}},\ }\href {https://doi.org/10.1103/PhysRevD.81.084003} {\bibfield  {journal} {\bibinfo  {journal} {Phys. Rev. D}\ }\textbf {\bibinfo {volume} {81}},\ \bibinfo {pages} {084003} (\bibinfo {year} {2010})},\ \Eprint {https://arxiv.org/abs/0912.2920} {arXiv:0912.2920 [gr-qc]} \BibitemShut {NoStop}%
\bibitem [{\citenamefont {Hilditch}\ \emph {et~al.}(2013)\citenamefont {Hilditch}, \citenamefont {Bernuzzi}, \citenamefont {Thierfelder}, \citenamefont {Cao}, \citenamefont {Tichy},\ and\ \citenamefont {Bruegmann}}]{Hilditch:2012fp}%
  \BibitemOpen
  \bibfield  {author} {\bibinfo {author} {\bibfnamefont {D.}~\bibnamefont {Hilditch}}, \bibinfo {author} {\bibfnamefont {S.}~\bibnamefont {Bernuzzi}}, \bibinfo {author} {\bibfnamefont {M.}~\bibnamefont {Thierfelder}}, \bibinfo {author} {\bibfnamefont {Z.}~\bibnamefont {Cao}}, \bibinfo {author} {\bibfnamefont {W.}~\bibnamefont {Tichy}},\ and\ \bibinfo {author} {\bibfnamefont {B.}~\bibnamefont {Bruegmann}},\ }\bibfield  {title} {\bibinfo {title} {{Compact binary evolutions with the Z4c formulation}},\ }\href {https://doi.org/10.1103/PhysRevD.88.084057} {\bibfield  {journal} {\bibinfo  {journal} {Phys. Rev. D}\ }\textbf {\bibinfo {volume} {88}},\ \bibinfo {pages} {084057} (\bibinfo {year} {2013})},\ \Eprint {https://arxiv.org/abs/1212.2901} {arXiv:1212.2901 [gr-qc]} \BibitemShut {NoStop}%
\bibitem [{\citenamefont {Bona}\ \emph {et~al.}(1995)\citenamefont {Bona}, \citenamefont {Masso}, \citenamefont {Seidel},\ and\ \citenamefont {Stela}}]{Bona:1994dr}%
  \BibitemOpen
  \bibfield  {author} {\bibinfo {author} {\bibfnamefont {C.}~\bibnamefont {Bona}}, \bibinfo {author} {\bibfnamefont {J.}~\bibnamefont {Masso}}, \bibinfo {author} {\bibfnamefont {E.}~\bibnamefont {Seidel}},\ and\ \bibinfo {author} {\bibfnamefont {J.}~\bibnamefont {Stela}},\ }\bibfield  {title} {\bibinfo {title} {{A New formalism for numerical relativity}},\ }\href {https://doi.org/10.1103/PhysRevLett.75.600} {\bibfield  {journal} {\bibinfo  {journal} {Phys. Rev. Lett.}\ }\textbf {\bibinfo {volume} {75}},\ \bibinfo {pages} {600} (\bibinfo {year} {1995})},\ \Eprint {https://arxiv.org/abs/gr-qc/9412071} {arXiv:gr-qc/9412071} \BibitemShut {NoStop}%
\bibitem [{\citenamefont {Alcubierre}\ \emph {et~al.}(2003)\citenamefont {Alcubierre}, \citenamefont {Bruegmann}, \citenamefont {Diener}, \citenamefont {Koppitz}, \citenamefont {Pollney}, \citenamefont {Seidel},\ and\ \citenamefont {Takahashi}}]{Alcubierre:2002kk}%
  \BibitemOpen
  \bibfield  {author} {\bibinfo {author} {\bibfnamefont {M.}~\bibnamefont {Alcubierre}}, \bibinfo {author} {\bibfnamefont {B.}~\bibnamefont {Bruegmann}}, \bibinfo {author} {\bibfnamefont {P.}~\bibnamefont {Diener}}, \bibinfo {author} {\bibfnamefont {M.}~\bibnamefont {Koppitz}}, \bibinfo {author} {\bibfnamefont {D.}~\bibnamefont {Pollney}}, \bibinfo {author} {\bibfnamefont {E.}~\bibnamefont {Seidel}},\ and\ \bibinfo {author} {\bibfnamefont {R.}~\bibnamefont {Takahashi}},\ }\bibfield  {title} {\bibinfo {title} {{Gauge conditions for long term numerical black hole evolutions without excision}},\ }\href {https://doi.org/10.1103/PhysRevD.67.084023} {\bibfield  {journal} {\bibinfo  {journal} {Phys. Rev. D}\ }\textbf {\bibinfo {volume} {67}},\ \bibinfo {pages} {084023} (\bibinfo {year} {2003})},\ \Eprint {https://arxiv.org/abs/gr-qc/0206072} {arXiv:gr-qc/0206072} \BibitemShut {NoStop}%
\bibitem [{\citenamefont {Banyuls}\ \emph {et~al.}(1997)\citenamefont {Banyuls}, \citenamefont {Font}, \citenamefont {Ibanez}, \citenamefont {Marti},\ and\ \citenamefont {Miralles}}]{Banyuls:1997zz}%
  \BibitemOpen
  \bibfield  {author} {\bibinfo {author} {\bibfnamefont {F.}~\bibnamefont {Banyuls}}, \bibinfo {author} {\bibfnamefont {J.~A.}\ \bibnamefont {Font}}, \bibinfo {author} {\bibfnamefont {J.~M.~A.}\ \bibnamefont {Ibanez}}, \bibinfo {author} {\bibfnamefont {J.~M.~A.}\ \bibnamefont {Marti}},\ and\ \bibinfo {author} {\bibfnamefont {J.~A.}\ \bibnamefont {Miralles}},\ }\bibfield  {title} {\bibinfo {title} {{Numerical {3+1} General Relativistic Hydrodynamics: A Local Characteristic Approach}},\ }\href@noop {} {\bibfield  {journal} {\bibinfo  {journal} {Astrophys. J.}\ }\textbf {\bibinfo {volume} {476}},\ \bibinfo {pages} {221} (\bibinfo {year} {1997})}\BibitemShut {NoStop}%
\bibitem [{\citenamefont {Schianchi}\ \emph {et~al.}(2024)\citenamefont {Schianchi}, \citenamefont {Gieg}, \citenamefont {Nedora}, \citenamefont {Neuweiler}, \citenamefont {Ujevic}, \citenamefont {Bulla},\ and\ \citenamefont {Dietrich}}]{Schianchi:2023uky}%
  \BibitemOpen
  \bibfield  {author} {\bibinfo {author} {\bibfnamefont {F.}~\bibnamefont {Schianchi}}, \bibinfo {author} {\bibfnamefont {H.}~\bibnamefont {Gieg}}, \bibinfo {author} {\bibfnamefont {V.}~\bibnamefont {Nedora}}, \bibinfo {author} {\bibfnamefont {A.}~\bibnamefont {Neuweiler}}, \bibinfo {author} {\bibfnamefont {M.}~\bibnamefont {Ujevic}}, \bibinfo {author} {\bibfnamefont {M.}~\bibnamefont {Bulla}},\ and\ \bibinfo {author} {\bibfnamefont {T.}~\bibnamefont {Dietrich}},\ }\bibfield  {title} {\bibinfo {title} {{M1 neutrino transport within the numerical-relativistic code BAM with application to low mass binary neutron star mergers}},\ }\href {https://doi.org/10.1103/PhysRevD.109.044012} {\bibfield  {journal} {\bibinfo  {journal} {Phys. Rev. D}\ }\textbf {\bibinfo {volume} {109}},\ \bibinfo {pages} {044012} (\bibinfo {year} {2024})},\ \Eprint {https://arxiv.org/abs/2307.04572} {arXiv:2307.04572 [gr-qc]} \BibitemShut {NoStop}%
\bibitem [{\citenamefont {Izquierdo}\ \emph {et~al.}(2023)\citenamefont {Izquierdo}, \citenamefont {Pareschi}, \citenamefont {Mi{\~n}ano}, \citenamefont {Mass{\'o}},\ and\ \citenamefont {Palenzuela}}]{Izquierdo:2022eaz}%
  \BibitemOpen
  \bibfield  {author} {\bibinfo {author} {\bibfnamefont {M.~R.}\ \bibnamefont {Izquierdo}}, \bibinfo {author} {\bibfnamefont {L.}~\bibnamefont {Pareschi}}, \bibinfo {author} {\bibfnamefont {B.}~\bibnamefont {Mi{\~n}ano}}, \bibinfo {author} {\bibfnamefont {J.}~\bibnamefont {Mass{\'o}}},\ and\ \bibinfo {author} {\bibfnamefont {C.}~\bibnamefont {Palenzuela}},\ }\bibfield  {title} {\bibinfo {title} {{Global high-order numerical schemes for the time evolution of the general relativistic radiation magneto-hydrodynamics equations}},\ }\href {https://doi.org/10.1088/1361-6382/acd97f} {\bibfield  {journal} {\bibinfo  {journal} {Class. Quant. Grav.}\ }\textbf {\bibinfo {volume} {40}},\ \bibinfo {pages} {145014} (\bibinfo {year} {2023})},\ \Eprint {https://arxiv.org/abs/2211.00027} {arXiv:2211.00027 [gr-qc]} \BibitemShut {NoStop}%
\bibitem [{\citenamefont {Jaeger}\ \emph {et~al.}(2026)\citenamefont {Jaeger}, \citenamefont {Gieg},\ and\ \citenamefont {Dietrich}}]{Jaeger:2026lnh}%
  \BibitemOpen
  \bibfield  {author} {\bibinfo {author} {\bibfnamefont {R.}~\bibnamefont {Jaeger}}, \bibinfo {author} {\bibfnamefont {H.}~\bibnamefont {Gieg}},\ and\ \bibinfo {author} {\bibfnamefont {T.}~\bibnamefont {Dietrich}},\ }\bibfield  {title} {\bibinfo {title} {{Investigating the influence of the full-kinematics treatment for charged-current processes in binary neutron star mergers}},\ }\href@noop {} {\bibfield  {journal} {\bibinfo  {journal} {arXiv e-prints}\ } (\bibinfo {year} {2026})},\ \Eprint {https://arxiv.org/abs/2609.19922} {arXiv:2609.19922 [astro-ph.HE]} \BibitemShut {NoStop}%
\bibitem [{\citenamefont {{Borges}}\ \emph {et~al.}(2008)\citenamefont {{Borges}}, \citenamefont {{Carmona}}, \citenamefont {{Costa}},\ and\ \citenamefont {{Don}}}]{Borges:2008}%
  \BibitemOpen
  \bibfield  {author} {\bibinfo {author} {\bibfnamefont {R.}~\bibnamefont {{Borges}}}, \bibinfo {author} {\bibfnamefont {M.}~\bibnamefont {{Carmona}}}, \bibinfo {author} {\bibfnamefont {B.}~\bibnamefont {{Costa}}},\ and\ \bibinfo {author} {\bibfnamefont {W.~S.}\ \bibnamefont {{Don}}},\ }\bibfield  {title} {\bibinfo {title} {{An improved weighted essentially non-oscillatory scheme for hyperbolic conservation laws}},\ }\href {https://doi.org/10.1016/j.jcp.2007.11.038} {\bibfield  {journal} {\bibinfo  {journal} {Journal of Computational Physics}\ }\textbf {\bibinfo {volume} {227}},\ \bibinfo {pages} {3191} (\bibinfo {year} {2008})}\BibitemShut {NoStop}%
\bibitem [{\citenamefont {Einfeldt}(1988)}]{Einfeldt:1988}%
  \BibitemOpen
  \bibfield  {author} {\bibinfo {author} {\bibfnamefont {B.}~\bibnamefont {Einfeldt}},\ }\bibfield  {title} {\bibinfo {title} {On godunov-type methods for gas dynamics},\ }\href {https://doi.org/10.1137/0725021} {\bibfield  {journal} {\bibinfo  {journal} {SIAM Journal on Numerical Analysis}\ }\textbf {\bibinfo {volume} {25}},\ \bibinfo {pages} {294} (\bibinfo {year} {1988})},\ \Eprint {https://arxiv.org/abs/https://doi.org/10.1137/0725021} {https://doi.org/10.1137/0725021} \BibitemShut {NoStop}%
\bibitem [{\citenamefont {Amiram}\ \emph {et~al.}(2006)\citenamefont {Amiram}, \citenamefont {Peter~D.},\ and\ \citenamefont {Bram~van}}]{Amiram:2006zjz}%
  \BibitemOpen
  \bibfield  {author} {\bibinfo {author} {\bibfnamefont {H.}~\bibnamefont {Amiram}}, \bibinfo {author} {\bibfnamefont {L.}~\bibnamefont {Peter~D.}},\ and\ \bibinfo {author} {\bibfnamefont {L.}~\bibnamefont {Bram~van}},\ }\bibfield  {title} {\bibinfo {title} {{On Upstream Differencing and Godunov-Type Schemes for Hyperbolic Conservation Laws}},\ }\href {https://doi.org/10.1137/1025002} {\bibfield  {journal} {\bibinfo  {journal} {SIAM Rev.}\ }\textbf {\bibinfo {volume} {25}},\ \bibinfo {pages} {35} (\bibinfo {year} {2006})}\BibitemShut {NoStop}%
\bibitem [{\citenamefont {Guo}\ \emph {et~al.}(2020)\citenamefont {Guo}, \citenamefont {Mart\'\i{}nez-Pinedo}, \citenamefont {Lohs},\ and\ \citenamefont {Fischer}}]{Guo:2020tgx}%
  \BibitemOpen
  \bibfield  {author} {\bibinfo {author} {\bibfnamefont {G.}~\bibnamefont {Guo}}, \bibinfo {author} {\bibfnamefont {G.}~\bibnamefont {Mart\'\i{}nez-Pinedo}}, \bibinfo {author} {\bibfnamefont {A.}~\bibnamefont {Lohs}},\ and\ \bibinfo {author} {\bibfnamefont {T.}~\bibnamefont {Fischer}},\ }\bibfield  {title} {\bibinfo {title} {{Charged-Current Muonic Reactions in Core-Collapse Supernovae}},\ }\href {https://doi.org/10.1103/PhysRevD.102.023037} {\bibfield  {journal} {\bibinfo  {journal} {Phys. Rev. D}\ }\textbf {\bibinfo {volume} {102}},\ \bibinfo {pages} {023037} (\bibinfo {year} {2020})},\ \Eprint {https://arxiv.org/abs/2006.12051} {arXiv:2006.12051 [hep-ph]} \BibitemShut {NoStop}%
\bibitem [{\citenamefont {Steiner}\ \emph {et~al.}(2001)\citenamefont {Steiner}, \citenamefont {Prakash},\ and\ \citenamefont {Lattimer}}]{Steiner:2001rp}%
  \BibitemOpen
  \bibfield  {author} {\bibinfo {author} {\bibfnamefont {A.~W.}\ \bibnamefont {Steiner}}, \bibinfo {author} {\bibfnamefont {M.}~\bibnamefont {Prakash}},\ and\ \bibinfo {author} {\bibfnamefont {J.~M.}\ \bibnamefont {Lattimer}},\ }\bibfield  {title} {\bibinfo {title} {{Diffusion of neutrinos in protoneutron star matter with quarks}},\ }\href {https://doi.org/10.1016/S0370-2693(01)00434-8} {\bibfield  {journal} {\bibinfo  {journal} {Phys. Lett. B}\ }\textbf {\bibinfo {volume} {509}},\ \bibinfo {pages} {10} (\bibinfo {year} {2001})},\ \Eprint {https://arxiv.org/abs/astro-ph/0101566} {arXiv:astro-ph/0101566} \BibitemShut {NoStop}%
\bibitem [{\citenamefont {Colvero}\ and\ \citenamefont {Lugones}(2014)}]{Colvero:2014zfa}%
  \BibitemOpen
  \bibfield  {author} {\bibinfo {author} {\bibfnamefont {G.~C.}\ \bibnamefont {Colvero}}\ and\ \bibinfo {author} {\bibfnamefont {G.}~\bibnamefont {Lugones}},\ }\bibfield  {title} {\bibinfo {title} {{Neutrino diffusive transport in hot quark matter: a detailed analysis}},\ }\href {https://doi.org/10.1103/PhysRevC.89.055803} {\bibfield  {journal} {\bibinfo  {journal} {Phys. Rev. C}\ }\textbf {\bibinfo {volume} {89}},\ \bibinfo {pages} {055803} (\bibinfo {year} {2014})},\ \Eprint {https://arxiv.org/abs/1405.3294} {arXiv:1405.3294 [astro-ph.SR]} \BibitemShut {NoStop}%
\bibitem [{\citenamefont {Chiesa}\ \emph {et~al.}(2025)\citenamefont {Chiesa}, \citenamefont {Bhattacharyya}, \citenamefont {Mazzini}, \citenamefont {Guercilena}, \citenamefont {Perego},\ and\ \citenamefont {Radice}}]{Chiesa:2024lnu}%
  \BibitemOpen
  \bibfield  {author} {\bibinfo {author} {\bibfnamefont {L.}~\bibnamefont {Chiesa}}, \bibinfo {author} {\bibfnamefont {M.}~\bibnamefont {Bhattacharyya}}, \bibinfo {author} {\bibfnamefont {F.}~\bibnamefont {Mazzini}}, \bibinfo {author} {\bibfnamefont {F.~M.}\ \bibnamefont {Guercilena}}, \bibinfo {author} {\bibfnamefont {A.}~\bibnamefont {Perego}},\ and\ \bibinfo {author} {\bibfnamefont {D.}~\bibnamefont {Radice}},\ }\bibfield  {title} {\bibinfo {title} {{Open-source library for performance-portable neutrino reaction rates: Application to neutron star mergers}},\ }\href {https://doi.org/10.1103/PhysRevD.111.063053} {\bibfield  {journal} {\bibinfo  {journal} {Phys. Rev. D}\ }\textbf {\bibinfo {volume} {111}},\ \bibinfo {pages} {063053} (\bibinfo {year} {2025})},\ \Eprint {https://arxiv.org/abs/2412.04570} {arXiv:2412.04570 [astro-ph.HE]} \BibitemShut {NoStop}%
\bibitem [{\citenamefont {Cheong}\ \emph {et~al.}(2025)\citenamefont {Cheong}, \citenamefont {Foucart}, \citenamefont {Ng}, \citenamefont {Offermans}, \citenamefont {Duez}, \citenamefont {Muhammed},\ and\ \citenamefont {Chawhan}}]{Cheong:2024cnb}%
  \BibitemOpen
  \bibfield  {author} {\bibinfo {author} {\bibfnamefont {P.~C.-K.}\ \bibnamefont {Cheong}}, \bibinfo {author} {\bibfnamefont {F.}~\bibnamefont {Foucart}}, \bibinfo {author} {\bibfnamefont {H.~H.-Y.}\ \bibnamefont {Ng}}, \bibinfo {author} {\bibfnamefont {A.}~\bibnamefont {Offermans}}, \bibinfo {author} {\bibfnamefont {M.~D.}\ \bibnamefont {Duez}}, \bibinfo {author} {\bibfnamefont {N.}~\bibnamefont {Muhammed}},\ and\ \bibinfo {author} {\bibfnamefont {P.}~\bibnamefont {Chawhan}},\ }\bibfield  {title} {\bibinfo {title} {{Influence of neutrino-electron scattering and neutrino-pair annihilation on hypermassive neutron star}},\ }\href {https://doi.org/10.1103/PhysRevD.111.043036} {\bibfield  {journal} {\bibinfo  {journal} {Phys. Rev. D}\ }\textbf {\bibinfo {volume} {111}},\ \bibinfo {pages} {043036} (\bibinfo {year} {2025})},\ \Eprint {https://arxiv.org/abs/2410.20681} {arXiv:2410.20681 [astro-ph.HE]} \BibitemShut {NoStop}%
\bibitem [{\citenamefont {Foucart}\ \emph {et~al.}(2026)\citenamefont {Foucart}, \citenamefont {Rath}, \citenamefont {Davidson}, \citenamefont {Cheong}, \citenamefont {Duez}, \citenamefont {Kidder}, \citenamefont {Pfeiffer},\ and\ \citenamefont {Scheel}}]{Foucart:2026kpd}%
  \BibitemOpen
  \bibfield  {author} {\bibinfo {author} {\bibfnamefont {F.}~\bibnamefont {Foucart}}, \bibinfo {author} {\bibfnamefont {S.}~\bibnamefont {Rath}}, \bibinfo {author} {\bibfnamefont {R.}~\bibnamefont {Davidson}}, \bibinfo {author} {\bibfnamefont {P.~C.-K.}\ \bibnamefont {Cheong}}, \bibinfo {author} {\bibfnamefont {M.~D.}\ \bibnamefont {Duez}}, \bibinfo {author} {\bibfnamefont {L.}~\bibnamefont {Kidder}}, \bibinfo {author} {\bibfnamefont {H.}~\bibnamefont {Pfeiffer}},\ and\ \bibinfo {author} {\bibfnamefont {M.}~\bibnamefont {Scheel}},\ }\bibfield  {title} {\bibinfo {title} {{Impact of neutrino-electron scattering and an improved treatment of pair processes on binary neutron star mergers}},\ }\href@noop {} {\bibfield  {journal} {\bibinfo  {journal} {arXiv e-prints}\ } (\bibinfo {year} {2026})},\ \Eprint {https://arxiv.org/abs/2606.27425} {arXiv:2606.27425 [astro-ph.HE]} \BibitemShut {NoStop}%
\bibitem [{\citenamefont {Alford}\ \emph {et~al.}(2025)\citenamefont {Alford}, \citenamefont {Brodie}, \citenamefont {Buballa}, \citenamefont {Gholami}, \citenamefont {Haber},\ and\ \citenamefont {Hofmann}}]{Alford:2025jtm}%
  \BibitemOpen
  \bibfield  {author} {\bibinfo {author} {\bibfnamefont {M.~G.}\ \bibnamefont {Alford}}, \bibinfo {author} {\bibfnamefont {L.}~\bibnamefont {Brodie}}, \bibinfo {author} {\bibfnamefont {M.}~\bibnamefont {Buballa}}, \bibinfo {author} {\bibfnamefont {H.}~\bibnamefont {Gholami}}, \bibinfo {author} {\bibfnamefont {A.}~\bibnamefont {Haber}},\ and\ \bibinfo {author} {\bibfnamefont {M.}~\bibnamefont {Hofmann}},\ }\bibfield  {title} {\bibinfo {title} {{Neutrino absorption in two-flavor color-superconducting quark matter}},\ }\href@noop {} {\bibfield  {journal} {\bibinfo  {journal} {arXiv preprint}\ } (\bibinfo {year} {2025})},\ \Eprint {https://arxiv.org/abs/2509.04240} {arXiv:2509.04240 [nucl-th]} \BibitemShut {NoStop}%
\bibitem [{\citenamefont {Burrows}\ \emph {et~al.}(2006)\citenamefont {Burrows}, \citenamefont {Reddy},\ and\ \citenamefont {Thompson}}]{Burrows:2004vq}%
  \BibitemOpen
  \bibfield  {author} {\bibinfo {author} {\bibfnamefont {A.}~\bibnamefont {Burrows}}, \bibinfo {author} {\bibfnamefont {S.}~\bibnamefont {Reddy}},\ and\ \bibinfo {author} {\bibfnamefont {T.~A.}\ \bibnamefont {Thompson}},\ }\bibfield  {title} {\bibinfo {title} {{Neutrino opacities in nuclear matter}},\ }\href {https://doi.org/10.1016/j.nuclphysa.2004.06.012} {\bibfield  {journal} {\bibinfo  {journal} {Nucl. Phys. A}\ }\textbf {\bibinfo {volume} {777}},\ \bibinfo {pages} {356} (\bibinfo {year} {2006})},\ \Eprint {https://arxiv.org/abs/astro-ph/0404432} {arXiv:astro-ph/0404432} \BibitemShut {NoStop}%
\bibitem [{\citenamefont {Perego}\ \emph {et~al.}(2019)\citenamefont {Perego}, \citenamefont {Bernuzzi},\ and\ \citenamefont {Radice}}]{Perego:2019adq}%
  \BibitemOpen
  \bibfield  {author} {\bibinfo {author} {\bibfnamefont {A.}~\bibnamefont {Perego}}, \bibinfo {author} {\bibfnamefont {S.}~\bibnamefont {Bernuzzi}},\ and\ \bibinfo {author} {\bibfnamefont {D.}~\bibnamefont {Radice}},\ }\bibfield  {title} {\bibinfo {title} {{Thermodynamics conditions of matter in neutron star mergers}},\ }\href {https://doi.org/10.1140/epja/i2019-12810-7} {\bibfield  {journal} {\bibinfo  {journal} {Eur. Phys. J. A}\ }\textbf {\bibinfo {volume} {55}},\ \bibinfo {pages} {124} (\bibinfo {year} {2019})},\ \Eprint {https://arxiv.org/abs/1903.07898} {arXiv:1903.07898 [gr-qc]} \BibitemShut {NoStop}%
\bibitem [{\citenamefont {Radice}\ \emph {et~al.}(2022)\citenamefont {Radice}, \citenamefont {Bernuzzi}, \citenamefont {Perego},\ and\ \citenamefont {Haas}}]{Radice:2021jtw}%
  \BibitemOpen
  \bibfield  {author} {\bibinfo {author} {\bibfnamefont {D.}~\bibnamefont {Radice}}, \bibinfo {author} {\bibfnamefont {S.}~\bibnamefont {Bernuzzi}}, \bibinfo {author} {\bibfnamefont {A.}~\bibnamefont {Perego}},\ and\ \bibinfo {author} {\bibfnamefont {R.}~\bibnamefont {Haas}},\ }\bibfield  {title} {\bibinfo {title} {{A new moment-based general-relativistic neutrino-radiation transport code: Methods and first applications to neutron star mergers}},\ }\href {https://doi.org/10.1093/mnras/stac589} {\bibfield  {journal} {\bibinfo  {journal} {Mon. Not. Roy. Astron. Soc.}\ }\textbf {\bibinfo {volume} {512}},\ \bibinfo {pages} {1499} (\bibinfo {year} {2022})},\ \Eprint {https://arxiv.org/abs/2111.14858} {arXiv:2111.14858 [astro-ph.HE]} \BibitemShut {NoStop}%
\bibitem [{\citenamefont {Musolino}\ and\ \citenamefont {Rezzolla}(2024)}]{Musolino:2023pao}%
  \BibitemOpen
  \bibfield  {author} {\bibinfo {author} {\bibfnamefont {C.}~\bibnamefont {Musolino}}\ and\ \bibinfo {author} {\bibfnamefont {L.}~\bibnamefont {Rezzolla}},\ }\bibfield  {title} {\bibinfo {title} {{A practical guide to a moment approach for neutrino transport in numerical relativity}},\ }\href {https://doi.org/10.1093/mnras/stae224} {\bibfield  {journal} {\bibinfo  {journal} {Mon. Not. Roy. Astron. Soc.}\ }\textbf {\bibinfo {volume} {528}},\ \bibinfo {pages} {5952} (\bibinfo {year} {2024})},\ \Eprint {https://arxiv.org/abs/2304.09168} {arXiv:2304.09168 [gr-qc]} \BibitemShut {NoStop}%
\bibitem [{\citenamefont {Foucart}\ \emph {et~al.}(2024)\citenamefont {Foucart}, \citenamefont {Cheong}, \citenamefont {Duez}, \citenamefont {Kidder}, \citenamefont {Pfeiffer},\ and\ \citenamefont {Scheel}}]{Foucart:2024npn}%
  \BibitemOpen
  \bibfield  {author} {\bibinfo {author} {\bibfnamefont {F.}~\bibnamefont {Foucart}}, \bibinfo {author} {\bibfnamefont {P.~C.-K.}\ \bibnamefont {Cheong}}, \bibinfo {author} {\bibfnamefont {M.~D.}\ \bibnamefont {Duez}}, \bibinfo {author} {\bibfnamefont {L.~E.}\ \bibnamefont {Kidder}}, \bibinfo {author} {\bibfnamefont {H.~P.}\ \bibnamefont {Pfeiffer}},\ and\ \bibinfo {author} {\bibfnamefont {M.~A.}\ \bibnamefont {Scheel}},\ }\bibfield  {title} {\bibinfo {title} {{Robustness of neutron star merger simulations to changes in neutrino transport and neutrino-matter interactions}},\ }\href {https://doi.org/10.1103/PhysRevD.110.083028} {\bibfield  {journal} {\bibinfo  {journal} {Phys. Rev. D}\ }\textbf {\bibinfo {volume} {110}},\ \bibinfo {pages} {083028} (\bibinfo {year} {2024})},\ \Eprint {https://arxiv.org/abs/2407.15989} {arXiv:2407.15989 [astro-ph.HE]} \BibitemShut {NoStop}%
\bibitem [{\citenamefont {Foucart}\ \emph {et~al.}(2016)\citenamefont {Foucart}, \citenamefont {O'Connor}, \citenamefont {Roberts}, \citenamefont {Kidder}, \citenamefont {Pfeiffer},\ and\ \citenamefont {Scheel}}]{Foucart:2016rxm}%
  \BibitemOpen
  \bibfield  {author} {\bibinfo {author} {\bibfnamefont {F.}~\bibnamefont {Foucart}}, \bibinfo {author} {\bibfnamefont {E.}~\bibnamefont {O'Connor}}, \bibinfo {author} {\bibfnamefont {L.}~\bibnamefont {Roberts}}, \bibinfo {author} {\bibfnamefont {L.~E.}\ \bibnamefont {Kidder}}, \bibinfo {author} {\bibfnamefont {H.~P.}\ \bibnamefont {Pfeiffer}},\ and\ \bibinfo {author} {\bibfnamefont {M.~A.}\ \bibnamefont {Scheel}},\ }\bibfield  {title} {\bibinfo {title} {{Impact of an improved neutrino energy estimate on outflows in neutron star merger simulations}},\ }\href {https://doi.org/10.1103/PhysRevD.94.123016} {\bibfield  {journal} {\bibinfo  {journal} {Phys. Rev. D}\ }\textbf {\bibinfo {volume} {94}},\ \bibinfo {pages} {123016} (\bibinfo {year} {2016})},\ \Eprint {https://arxiv.org/abs/1607.07450} {arXiv:1607.07450 [astro-ph.HE]} \BibitemShut {NoStop}%
\bibitem [{\citenamefont {Ng}\ \emph {et~al.}(2024)\citenamefont {Ng}, \citenamefont {Cheong}, \citenamefont {Lam},\ and\ \citenamefont {Li}}]{Ng:2023syk}%
  \BibitemOpen
  \bibfield  {author} {\bibinfo {author} {\bibfnamefont {H.~H.-Y.}\ \bibnamefont {Ng}}, \bibinfo {author} {\bibfnamefont {P.~C.-K.}\ \bibnamefont {Cheong}}, \bibinfo {author} {\bibfnamefont {A.~T.-L.}\ \bibnamefont {Lam}},\ and\ \bibinfo {author} {\bibfnamefont {T.~G.~F.}\ \bibnamefont {Li}},\ }\bibfield  {title} {\bibinfo {title} {{General-relativistic Radiation Transport Scheme in Gmunu. II. Implementation of Novel Microphysical Library for Neutrino Radiation{\textemdash}Weakhub}},\ }\href {https://doi.org/10.3847/1538-4365/ad2fbd} {\bibfield  {journal} {\bibinfo  {journal} {Astrophys. J. Suppl.}\ }\textbf {\bibinfo {volume} {272}},\ \bibinfo {pages} {9} (\bibinfo {year} {2024})},\ \Eprint {https://arxiv.org/abs/2309.03526} {arXiv:2309.03526 [astro-ph.HE]} \BibitemShut {NoStop}%
\bibitem [{\citenamefont {Dimmelmeier}\ \emph {et~al.}(2009)\citenamefont {Dimmelmeier}, \citenamefont {Bejger}, \citenamefont {Haensel},\ and\ \citenamefont {Zdunik}}]{Dimmelmeier:2009vw}%
  \BibitemOpen
  \bibfield  {author} {\bibinfo {author} {\bibfnamefont {H.}~\bibnamefont {Dimmelmeier}}, \bibinfo {author} {\bibfnamefont {M.}~\bibnamefont {Bejger}}, \bibinfo {author} {\bibfnamefont {P.}~\bibnamefont {Haensel}},\ and\ \bibinfo {author} {\bibfnamefont {J.~L.}\ \bibnamefont {Zdunik}},\ }\bibfield  {title} {\bibinfo {title} {{Dynamic migration of rotating neutron stars due to a phase transition instability}},\ }\href {https://doi.org/10.1111/j.1365-2966.2009.14891.x} {\bibfield  {journal} {\bibinfo  {journal} {Mon. Not. Roy. Astron. Soc.}\ }\textbf {\bibinfo {volume} {396}},\ \bibinfo {pages} {2269} (\bibinfo {year} {2009})},\ \Eprint {https://arxiv.org/abs/0901.3819} {arXiv:0901.3819 [astro-ph.SR]} \BibitemShut {NoStop}%
\bibitem [{\citenamefont {Gieg}\ \emph {et~al.}(2026)\citenamefont {Gieg}, \citenamefont {Jaeger}, \citenamefont {Ujevic},\ and\ \citenamefont {Dietrich}}]{Gieg:2026beb}%
  \BibitemOpen
  \bibfield  {author} {\bibinfo {author} {\bibfnamefont {H.}~\bibnamefont {Gieg}}, \bibinfo {author} {\bibfnamefont {R.}~\bibnamefont {Jaeger}}, \bibinfo {author} {\bibfnamefont {M.}~\bibnamefont {Ujevic}},\ and\ \bibinfo {author} {\bibfnamefont {T.}~\bibnamefont {Dietrich}},\ }\bibfield  {title} {\bibinfo {title} {{Consistent treatment of muons in binary neutron star mergers}},\ }\href {https://doi.org/10.1103/qmb8-q3cx} {\bibfield  {journal} {\bibinfo  {journal} {Phys. Rev. D}\ }\textbf {\bibinfo {volume} {114}},\ \bibinfo {pages} {043036} (\bibinfo {year} {2026})},\ \Eprint {https://arxiv.org/abs/2604.14225} {arXiv:2604.14225 [astro-ph.HE]} \BibitemShut {NoStop}%
\bibitem [{\citenamefont {Thompson}\ \emph {et~al.}(2000)\citenamefont {Thompson}, \citenamefont {Burrows},\ and\ \citenamefont {Horvath}}]{Thompson:2000gv}%
  \BibitemOpen
  \bibfield  {author} {\bibinfo {author} {\bibfnamefont {T.~A.}\ \bibnamefont {Thompson}}, \bibinfo {author} {\bibfnamefont {A.}~\bibnamefont {Burrows}},\ and\ \bibinfo {author} {\bibfnamefont {J.~E.}\ \bibnamefont {Horvath}},\ }\bibfield  {title} {\bibinfo {title} {{Mu and tau-neutrino thermalization and production in supernovae: Processes and time scales}},\ }\href {https://doi.org/10.1103/PhysRevC.62.035802} {\bibfield  {journal} {\bibinfo  {journal} {Phys. Rev. C}\ }\textbf {\bibinfo {volume} {62}},\ \bibinfo {pages} {035802} (\bibinfo {year} {2000})},\ \Eprint {https://arxiv.org/abs/astro-ph/0003054} {arXiv:astro-ph/0003054} \BibitemShut {NoStop}%
\bibitem [{\citenamefont {Shapiro}\ and\ \citenamefont {Teukolsky}(1983)}]{Shapiro:1983du}%
  \BibitemOpen
  \bibfield  {author} {\bibinfo {author} {\bibfnamefont {S.~L.}\ \bibnamefont {Shapiro}}\ and\ \bibinfo {author} {\bibfnamefont {S.~A.}\ \bibnamefont {Teukolsky}},\ }\href {https://doi.org/10.1002/9783527617661} {\emph {\bibinfo {title} {{Black holes, white dwarfs, and neutron stars: The physics of compact objects}}}}\ (\bibinfo {year} {1983})\BibitemShut {NoStop}%
\bibitem [{\citenamefont {Prakash}\ \emph {et~al.}(2001)\citenamefont {Prakash}, \citenamefont {Lattimer}, \citenamefont {Pons}, \citenamefont {Steiner},\ and\ \citenamefont {Reddy}}]{Prakash:2000jr}%
  \BibitemOpen
  \bibfield  {author} {\bibinfo {author} {\bibfnamefont {M.}~\bibnamefont {Prakash}}, \bibinfo {author} {\bibfnamefont {J.~M.}\ \bibnamefont {Lattimer}}, \bibinfo {author} {\bibfnamefont {J.~A.}\ \bibnamefont {Pons}}, \bibinfo {author} {\bibfnamefont {A.~W.}\ \bibnamefont {Steiner}},\ and\ \bibinfo {author} {\bibfnamefont {S.}~\bibnamefont {Reddy}},\ }\bibfield  {title} {\bibinfo {title} {{Evolution of a neutron star from its birth to old age}},\ }\href@noop {} {\bibfield  {journal} {\bibinfo  {journal} {Lect. Notes Phys.}\ }\textbf {\bibinfo {volume} {578}},\ \bibinfo {pages} {364} (\bibinfo {year} {2001})},\ \Eprint {https://arxiv.org/abs/astro-ph/0012136} {arXiv:astro-ph/0012136} \BibitemShut {NoStop}%
\bibitem [{\citenamefont {Yakovlev}\ and\ \citenamefont {Pethick}(2004)}]{Yakovlev:2004iq}%
  \BibitemOpen
  \bibfield  {author} {\bibinfo {author} {\bibfnamefont {D.~G.}\ \bibnamefont {Yakovlev}}\ and\ \bibinfo {author} {\bibfnamefont {C.~J.}\ \bibnamefont {Pethick}},\ }\bibfield  {title} {\bibinfo {title} {{Neutron star cooling}},\ }\href {https://doi.org/10.1146/annurev.astro.42.053102.134013} {\bibfield  {journal} {\bibinfo  {journal} {Ann. Rev. Astron. Astrophys.}\ }\textbf {\bibinfo {volume} {42}},\ \bibinfo {pages} {169} (\bibinfo {year} {2004})},\ \Eprint {https://arxiv.org/abs/astro-ph/0402143} {arXiv:astro-ph/0402143} \BibitemShut {NoStop}%
\end{thebibliography}%
%\bibliographystyle{aasjournal}

% Produces the bibliography via BibTeX.

\end{document}